\documentclass[12pt,a4paper,notitlepage]{article}

\usepackage{graphicx}
\usepackage{amssymb}
\usepackage{amsmath}

\usepackage{amsthm}
\usepackage{amsfonts}
\usepackage{times}

\usepackage[format=hang]{subcaption}
\usepackage{color}
\usepackage[T1]{fontenc}
\usepackage{hyperref}
\usepackage[square]{natbib}
\usepackage{upgreek}

\usepackage{pgfplots}
\usepackage{todonotes}
\usepackage{bm}

\usepackage{float}

\usepackage{subcaption} 
\usepackage{siunitx}
\usepackage{tikz}
\usepackage{tikz-3dplot}

\defcitealias{KlockeZeisEtAl2014}{Klocke, Zeis et~al.}
\defcitealias{KlockeKlinkEtAl2014}{Klocke, Klink et~al.}

\pgfrealjobname{paper} %Terminal: pdflatex --jobname 'XYZ beginpgfgraphicnamed' paper.tex
\pgfplotsset{compat=1.14}

\definecolor{rwth1}{RGB}{0,84,159}      % RWTH-Blau
\definecolor{rwth2}{RGB}{142,186,229}   % RWTH-Hellblau
\definecolor{rwth3}{RGB}{0,97,101}      % Petrol 
\definecolor{rwth4}{RGB}{0,152,161}     % Türkis
\definecolor{rwth5}{RGB}{87,171,39}     % Grün
\definecolor{rwth6}{RGB}{189,205,0}     % Maigrün
\definecolor{rwth7}{RGB}{255,237,0}     % Gelb
\definecolor{rwth8}{RGB}{246,168,0}     % Orange
\definecolor{rwth9}{RGB}{227,0,102}     % Magenta
\definecolor{rwth10}{RGB}{204,7,30}     % Rot
\definecolor{rwth11}{RGB}{161,16,53}    % Bordeaux
\definecolor{rwth12}{RGB}{97,33,88}     % Violett
\definecolor{rwth13}{RGB}{122,111,172}  % Lila

\definecolor{sfb1}{RGB}{0,84,165}      % Blau
\definecolor{sfb2}{RGB}{201,0,35}      % Rot
\definecolor{sfb3}{RGB}{231,95,1}      % Orange
\definecolor{sfb4}{RGB}{127,127,127}   % Grau
\definecolor{sfb5}{RGB}{217,217,217}   % Hellgrau

\tikzstyle{dashpattern0} = [dash pattern = ]
\tikzstyle{dashpattern1} = [dash pattern = on 4.25pt off 0.75pt]
\tikzstyle{dashpattern2} = [dash pattern = on 1.5pt off 0.5pt]
\tikzstyle{dashpattern3} = [dash pattern = on 0.75pt off 0.4pt]
\tikzstyle{dashpattern4} = [dash pattern = on 3pt off 1pt on 1pt off 1pt]
\tikzstyle{dashpattern5} = [dash pattern = on 3.75pt off 0.5pt on 0.75pt off 0.5pt on 0.75pt off 0.5pt]
\tikzstyle{dashpattern6} = [dash pattern = on 3.25pt off 0.5pt on 0.75pt off 0.5pt on 0.75pt off 0.5pt on 0.75pt off 0.5pt]
\tikzstyle{dashpattern7} = [dash pattern = on 3.25pt off 0.5pt on 0.75pt off 0.5pt on 0.75pt off 0.5pt on 0.75pt off 0.5pt on 0.75pt off 0.5pt]
\tikzstyle{dashpattern8} = [line cap=round, dash pattern = on 3.25pt off 2.75pt]
\tikzstyle{dashpattern9} = [line cap=round, dash pattern = on 0.01pt off 2pt]
\tikzstyle{dashpattern10}= [line cap=round, dash pattern = on 3.25pt off 2pt on 0.01pt off 2pt]
\tikzstyle{dashpattern11}= [line cap=round, dash pattern = on 3.5pt off 1.75pt on 0.01pt off 1.75pt on 0.01pt off 1.75pt]
\tikzstyle{dashpattern12}= [line cap=round, dash pattern = on 3.5pt off 1.75pt on 0.01pt off 1.75pt on 0.01pt off 1.75pt on 0.01pt off 1.75pt]
\tikzstyle{dashpattern13}= [line cap=round, dash pattern = on 3.5pt off 1.75pt on 0.01pt off 1.75pt on 0.01pt off 1.75pt on 0.01pt off 1.75pt on 0.01pt off 1.75pt]

\newcommand{\grad}[1]{{\rm grad} \hspace{-0.5mm} \left( #1 \right)}
\renewcommand{\div}[1]{{\rm div} \hspace{-0.5mm} \left( #1 \right)}
\newcommand{\T}{^{\mathrm{T}}}
\newcommand{\dV}{\mathrm{d}V}
\newcommand{\dVe}{\mathrm{d}V^e}
\newcommand{\dt}{\Delta t}
\newcommand{\intO}{\int_\Omega}
\newcommand{\intOe}{\int_{\Omega^e}}
\newcommand{\pd}[2]{\displaystyle\frac{\partial #1}{\partial #2}}
\newcommand{\td}[2]{\frac{{\rm d} #1}{{\rm d} #2}}

\newcommand{\coloneq}{\mathrel{\resizebox{\widthof{$\mathord{=}$}}{\height}{ $\!\!\resizebox{1.2\width}{0.8\height}{\raisebox{0.23ex}{$\mathop{:}$}}\!\!=\!\!$ }}}
\newcommand{\ve}{\bm{v}^e}
\newcommand{\varve}{\delta \bm{v}^e}
\newcommand{\varveT}{\delta \bm{v}^{e \mathrm{T}}}
\newcommand{\dve}{\Delta \bm{v}_{n+1}^e}
\newcommand{\dv}{\Delta \bm{v}_{n+1}}
\newcommand{\delv}{\Delta v}
\newcommand{\delvpol}{\Delta v_\mathrm{pol}}
\newcommand{\Nv}{\bm{N}_v^e}

\newcommand{\Te}{\bm{\theta}^e}
\newcommand{\varTe}{\delta \bm{\theta}^e}
\newcommand{\varTeT}{\delta \bm{\theta}^{e \mathrm{T}}}
\newcommand{\dTe}{\Delta \bm{\theta}_{n+1}^e}
\newcommand{\dT}{\Delta \bm{\theta}_{n+1}}
\newcommand{\NT}{\bm{N}_\theta^e}
\newcommand{\NTT}{\bm{N}_\theta^{e \mathrm{T}}}
\newcommand{\Bv}{\mathbf{B}_{v}^e}
\newcommand{\BvT}{\mathbf{B}_{v}^{e \mathrm{T}}}
\newcommand{\BT}{\mathbf{B}_{\theta}^e}
\newcommand{\BTT}{\mathbf{B}_{\theta}^{e \mathrm{T}}}
\newcommand{\rv}{\bm{r}_{v}^e}
\newcommand{\Rv}{\bm{R}_{v}}
\newcommand{\rT}{\bm{r}_{\theta}^e}

\newcommand{\RT}{\bm{R}_{\theta}}
\newcommand{\kvv}{\mathbf{k}_{vv}^e}
\newcommand{\kvT}{\mathbf{k}_{v \theta}^e}
\newcommand{\kTv}{\mathbf{k}_{\theta v}^e}
\newcommand{\kTT}{\mathbf{k}_{\theta \theta}^e}

\newcommand{\Kvv}{\mathbf{K}_{vv}}
\newcommand{\KvT}{\mathbf{K}_{v \theta}}
\newcommand{\KTv}{\mathbf{K}_{\theta v}}
\newcommand{\KTT}{\mathbf{K}_{\theta \theta}}
\newcommand{\cvv}{\mathbf{c}_{vv}^e}

\newcommand{\cTv}{\mathbf{c}_{\theta v}^e}
\newcommand{\cTT}{\mathbf{c}_{\theta \theta}^e}
\newcommand{\Cvv}{\mathbf{C}_{vv}}

\newcommand{\CTv}{\mathbf{C}_{\theta v}}
\newcommand{\CTT}{\mathbf{C}_{\theta \theta}}

\newcommand{\nel}{n_{\mathrm{el}}}
\newcommand{\ngp}{n_{\mathrm{gp}}}
\newcommand{\ofx}{\hspace{-0.6mm}\left(\bm{x}\right)}
\newcommand{\assmbl}{\text{\raisebox{-1.5mm}{$\overset{\nel}{\underset{e=1}{\textbf{\textsf{\LARGE {\fontfamily{lmss}\selectfont A}}}}}$}}}
\newcommand{\alphabar}{\bar{\alpha}}
\newcommand{\alphabarn}{\bar{\alpha} \left( d_n, \Tn \right)}
\newcommand{\cTbar}{\bar{c}_\theta}
\newcommand{\cT}{c_\theta}
\newcommand{\epsnull}{\epsilon_\mathrm{0}}
\newcommand{\epsrbar}{\bar{\epsilon}_\mathrm{r}}
\newcommand{\epsr}{\epsilon_\mathrm{r}}
\newcommand{\gv}{g_v}
\newcommand{\gjtilde}{g_{\tilde{j}}}
\newcommand{\dvgv}{\Delta_v g_v}
\newcommand{\dTgv}{\Delta_\theta g_v}
\newcommand{\gT}{g_\theta}
\newcommand{\gqtilde}{g_{\tilde{q}}}
\newcommand{\dvgT}{\Delta_v g_\theta}
\newcommand{\dTgT}{\Delta_\theta g_\theta}
\newcommand{\kEbar}{\bar{k}_\textrm{\tiny E}}
\newcommand{\kE}{k_\textrm{\tiny E}}
\newcommand{\kEbarn}{\bar{k}_\textrm{\tiny E} \left( d_n, \Tn \right)}
\newcommand{\kTbar}{\bar{k}_\theta}
\newcommand{\kT}{k_\theta}
\newcommand{\kTbarn}{\bar{k}_\theta \left( d_n, \Tn \right)}
\newcommand{\pibar}{\bar{\Pi}}
\newcommand{\pibarn}{\pibar \left( d_n, \Tn \right)}
\newcommand{\rhoE}{\rho_\textrm{\tiny E}}
\newcommand{\rhoEdot}{\dot{\rho}_\textrm{\tiny E}}
\newcommand{\rhoVbar}{\bar{\rho}_\textrm{\tiny V}}
\newcommand{\rhoVa}{\rho_{\textrm{\tiny V}a}}
\newcommand{\rhoV}{\rho_\textrm{\tiny V}}
\newcommand{\thetadot}{\dot{\theta}}
\newcommand{\Tn}{\theta_n}
\newcommand{\Tnpe}{\theta_{n+1}}
\newcommand{\Tbnpe}{\bar{\theta}_{n+1}}
\newcommand{\dTnpe}{\Delta \theta_{n+1}}
\newcommand{\varv}{\delta v}

\newcommand{\varT}{\delta \theta}

\newcommand{\vdot}{\dot{v}}
\newcommand{\vn}{v_n}
\newcommand{\vnpe}{v_{n+1}}
\newcommand{\vbnpe}{\bar{v}_{n+1}}
\newcommand{\dvnpe}{\Delta v_{n+1}}
\newcommand{\D}{\bm{D}}
\newcommand{\E}{\bm{E}}
\newcommand{\Edot}{\dot{\bm{E}}}
\renewcommand{\j}{\bm{j}}
\newcommand{\jx}{j_x}
\newcommand{\jy}{j_y}
\newcommand{\jz}{j_z}
\newcommand{\jL}{\bm{j}_\textrm{\tiny L}}
\newcommand{\jV}{\bm{j}_\textrm{\tiny V}}
\newcommand{\jS}{\bm{j}_\textrm{\tiny S}}
\newcommand{\n}{\bm{n}}
\newcommand{\q}{\bm{q}}
\newcommand{\qP}{\bm{q}_\textrm{\tiny P}}
\newcommand{\qF}{\bm{q}_\textrm{\tiny F}}
\newcommand{\I}{\mathbf{I}}
\newcommand{\djLdE}{\pd{\jL}{\E}}
\newcommand{\djVdEdot}{\pd{\jV}{\Edot}}
\newcommand{\djSdgradT}{\pd{\jS}{\grad{\Tnpe}}}
\newcommand{\dqFdgradT}{\pd{\qF}{\grad{\Tnpe}}}
\newcommand{\Aelx}{A_{\mathrm{el}x}}
\newcommand{\Aely}{A_{\mathrm{el}y}}
\newcommand{\Aelz}{A_{\mathrm{el}z}}

\newcommand{\Aucx}{A_{\mathrm{uc}x}}
\newcommand{\Aucy}{A_{\mathrm{uc}y}}
\newcommand{\Aucz}{A_{\mathrm{uc}z}}

\newcommand{\Afuntimet}{\mathcal{A}\hspace*{-0.7mm}\left(\bm{x},t\right)}
\renewcommand{\d}{d}
\renewcommand{\ddot}{\dot{d}}

\newcommand{\dn}{d_n}

\newcommand{\EL}{^\textrm{\tiny EL}}
\newcommand{\ME}{^\textrm{\tiny ME}}

\newcommand{\Iinpe}{\left( I_1 \right)_{n+1}}
\newcommand{\Iiinpe}{\left( I_2 \right)_{n+1}}
\newcommand{\Iiiinpe}{\left( I_3 \right)_{n+1}}

\newcommand{\Vdis}{V_\mathrm{dis}}
\newcommand{\Veff}{V_\mathrm{eff}}
\newcommand{\Vref}{V_\mathrm{ref}}
\newcommand{\Vel}{V_\mathrm{el}}

\newcommand{\Vuc}{V_\mathrm{uc}}
\newcommand{\Vco}{V_\mathrm{co}}
\newcommand{\xM}{\bm{x}_\textrm{\tiny M}}
\AtBeginDocument{}

\date{}

\graphicspath{{./02_Figures/}}

\begin{document}

%- Authors
\author{\large {\parbox{\linewidth}{\centering Tim van der Velden$^a$\footnote{Corresponding author: \\ e-mail: tim.van.der.velden@ifam.rwth-aachen.de}\hspace{1.0mm},
                                               Bob Rommes$^b$,
                                               Andreas Klink$^b$,
                                               Stefanie Reese$^a$,
                                               Johanna Waimann$^a$}}\\[0.5cm]
  \hspace*{-0.1cm}
  \normalsize{\em \parbox{\linewidth}{\centering
    \vspace{6mm}
    $^a$Institute of Applied Mechanics, RWTH Aachen University, Mies-van-der-Rohe-Str. 1, D-52074 Aachen, Germany
    
    $^b$Laboratory for Machine Tools and Production Engineering (WZL) of RWTH Aachen University, Campus-Boulevard 30, D-52074 Aachen, Germany}
  }
}

%- Title
\title{\LARGE A novel approach for the efficient modeling of material dissolution in electrochemical machining}

\maketitle

%- Abstract
\small
{\bf Abstract.} {
This work presents a novel approach to efficiently model anodic dissolution in electrochemical machining.
Earlier modeling approaches employ a strict space discretization of the anodic surface that is associated with a remeshing procedure at every time step. Besides that, the presented model is formulated by means of effective material parameters. Thereby, it allows to use a constant mesh for the entire simulation and, thus, decreases the computational costs.
Based on Faraday's law of electrolysis, an effective dissolution level is introduced, which describes the ratio of a dissolved volume and its corresponding reference volume. This inner variable allows the modeling of the complex dissolution process without the necessity of computationally expensive remeshing by controlling the effective material parameters.
Additionally, full coupling of the thermoelectric problem is considered and its linearization and numerical implementation are presented. The model shows good agreement with analytical and experimental validation examples by yielding realistic results. Furthermore, simulations of a pulsed electrochemical machining process yield a process signature of the surface roughness related to the specific accumulated electric charge. The numerical examples confirm the simulation's computational efficiency and accurate modeling qualities.

}

%- Keywords
\vspace*{0.3cm}

\normalsize

%- Content
\section{Introduction}
\label{sec:1}

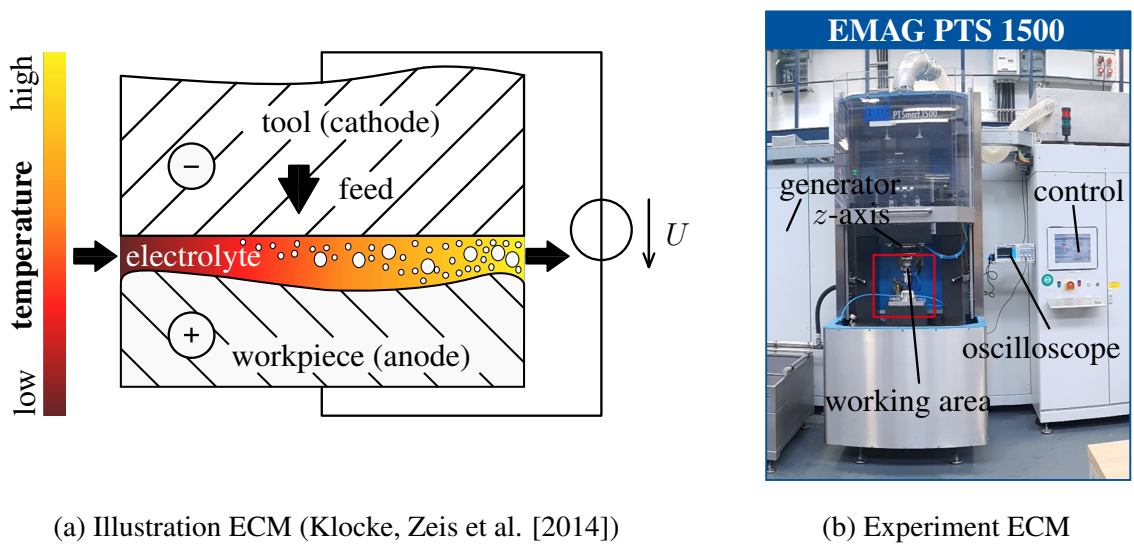
\begin{figure}[htbp] 
  \centering 
  \hspace{-10mm}
  \begin{subfigure}{.66\textwidth} 
      
    \centering 
      
    \begin{tikzpicture} 
      \node[inner sep=0pt] (pic) at (0.00,0) {\includegraphics[height=65mm]
      {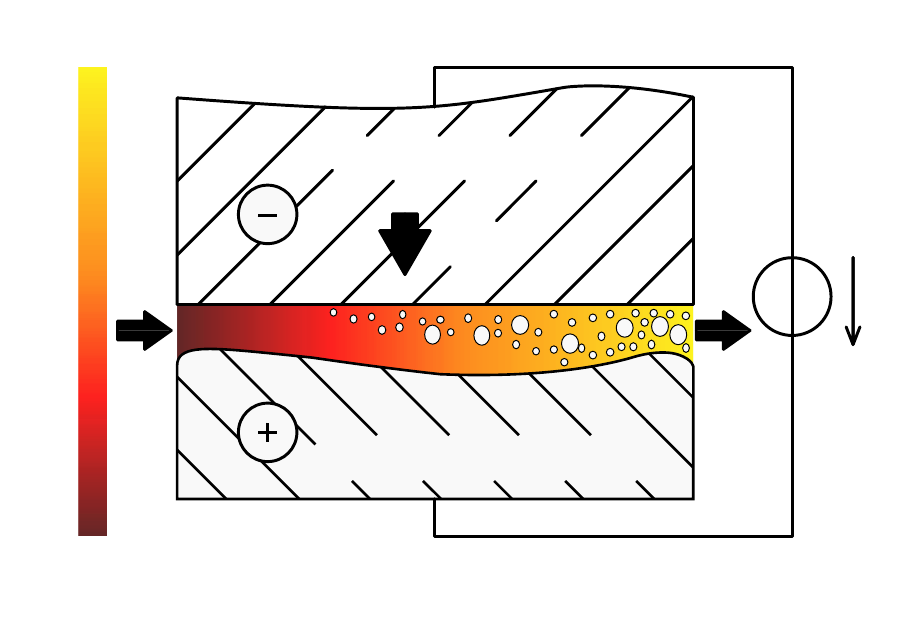}};
      \node[inner sep=0pt] (ca) at ($(pic.center)+( 0.20, 1.60)$)  {tool (cathode)};
      \node[inner sep=0pt] (fe) at ($(pic.center)+( 0.40, 0.75)$)  {feed};
      \node[inner sep=0pt, rotate=90] (te) at ($(pic.center)+(-4.10, 0.00)$)  {\textbf{temperature}};
      \node[inner sep=0pt, rotate=90] (hi) at ($(pic.center)+(-4.10, 2.15)$) {high};
      \node[inner sep=0pt, rotate=90] (lo) at ($(pic.center)+(-4.10,-1.95)$) {low};
      \node[inner sep=0pt] (el) at ($(pic.center)+(-1.87,-0.18)$)  {\textcolor{white}{electrolyte}};
      \node[inner sep=0pt] (an) at ($(pic.center)+( 0.20,-1.50)$)  {workpiece (anode)};
      \node[inner sep=0pt] (U)  at ($(pic.center)+( 4.50, 0.15)$)  {$U$};
    \end{tikzpicture} 
         
%     \vspace{-3mm}
    \caption{Illustration ECM (\citetalias{KlockeZeisEtAl2014} [\citeyear{KlockeZeisEtAl2014}])}
    \label{fig:ECM_Skizze}
         
  \end{subfigure}
  \begin{subfigure}{.33\textwidth} 
      
    \centering 
      
    \begin{tikzpicture} 
      \node[inner sep=0pt] (pic) at (0,0) {\includegraphics[height=65mm]
      {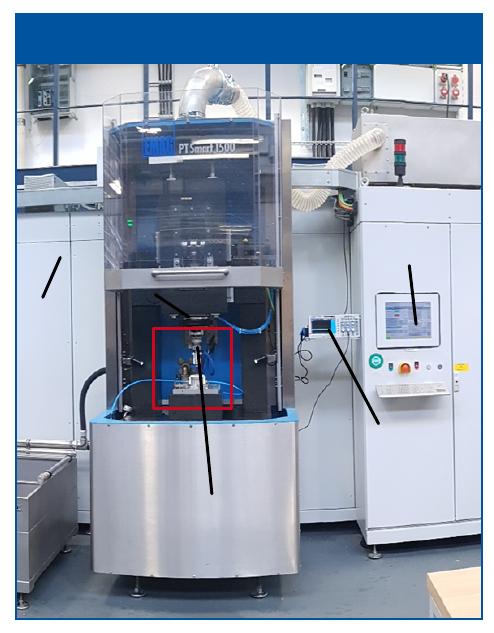}};
      \node[inner sep=0pt] (co)   at ($(pic.center)+( 0.00, 2.85)$)  {\textcolor{white}{\textbf{EMAG PTS 1500}}};
      \node[inner sep=0pt] (co)   at ($(pic.center)+( 1.75, 0.73)$)  {control};
      \node[inner sep=0pt] (os)   at ($(pic.center)+( 1.25,-1.40)$)  {oscilloscope};
      \node[inner sep=0pt] (wa)   at ($(pic.center)+(-0.50,-2.10)$)  {working area};
      \node[inner sep=0pt] (za)   at ($(pic.center)+(-1.25, 0.42)$)  {$z$-axis};
      \node[inner sep=0pt] (ge)   at ($(pic.center)+(-1.40, 0.75)$)  {generator};
    \end{tikzpicture} 
         
%     \vspace{-3mm}
    \caption{Experiment ECM}
    \label{fig:ECM_Experiment}
         
  \end{subfigure} 
  \caption{Illustration and experiment of electrochemical machining.} 
  \label{fig:ECM}     
\end{figure}

In many technical systems, materials with a high mechanical and thermal strength are applied to fulfill the efficiency requirements of individual components. Especially in turbomachinery manufacturing, this poses a challenge for conventional machining processes concerning tool wear and the required geometric tolerances. Hence, processes such as electrochemical machining (ECM), in which the strength of the material does not affect the removal process, are gaining importance (\citetalias{KlockeKlinkEtAl2014} [\citeyear{KlockeKlinkEtAl2014}]). 
In ECM, the material removal is based on the principle of electrolysis, caused by an electric current between the tool (cathode) and the workpiece (anode) (see e.g.~\cite{HamannVielstich2005}). The electric current is enabled by an electrically conductive fluid called electrolyte (see Fig.~\ref{fig:ECM}). This removal mechanism allows an efficient machining of high strength materials such as titanium or nickel-alloys, without the occurrence of stress states within rim zones, due to the lack of mechanical and thermal energy during the process (cf.~\cite{DeBarrOliver1968}, \cite{McGeough1974}, \cite{BergsHarst2020}).

A challenge in ECM is the complex tool development. ECM is an imaging machining process, where the contour of the tool defines the final shape of the workpiece. However, the local material removal rate highly depends on the electric conductivity of the electrolyte, which itself depends on multiple physical phenomena within the gap, such as the two-phase flow and the local temperature (\cite{KlockeKoenig2007}).
Due to multiphysical coupling (cf.~\cite{vanTijumPajak2008}), the resulting geometry after ECM is difficult to predict using deterministic calculations. In the past, ECM tools have been developed using heuristic experimental approaches, which employ a time-consuming iterative methodology. Experimental studies may be found in e.g.~\cite{HopenfeldCole1969}, \cite{CookFooteEtAl1973} and \cite{DattaLandolt1981}.

The precise modeling of machining processes is a challenging topic for modern industries. Furthermore, the accurate prediction of process results enables the reduction of calculation time as well as experimental costs. Initially, \cite{Tipton1964} presents the analytical $\mathrm{cos} \, \theta$ method to compute the equilibrium shape of the work piece. Afterwards, e.g.~\cite{Huembs1975} extends this analytical method to account for unsteady sinking conditions.
Moreover, numerous numerical models for ECM have been developed and, therefore, only a brief overview is given in the following. \cite{Walsch1977} presents the first numerical model for the computation of the gap width in ECM. Additional physical aspects, like the effect of the grain size on the performance of ECM, are considered by e.g.~\cite{RajurkarHewidy1988}. Furthermore, \cite{HardistyMilehamEtAl1993} describe the moving boundary value problem in ECM with a two-dimensional finite element model and, further, \cite{HardistyMileham1999} also extend this model for a parabolic cathode shape.
Transport mechanisms are first considered by \cite{DeconinckVanDamme2012a, DeconinckVanDamme2012b, DeconinckHoogsteen2013}, who, additionally, use the level set method to describe the anodic surface. \cite{Zeis2015} presents a fully coupled multiphysical model of the ECM process that allows for an automated design using iterative simulations. Finally, the dissolution of multiphase materials is modeled by e.g.~\cite{KozakZybura-Skrabalak2016} and \cite{Harst2019}. For a comprehensive overview of the numerical models the reader is kindly referred to e.g.~\cite{HindujaKunieda2013}, \cite{Zeis2015} and \cite{Harst2019}.

Although these approaches have proven to be applicable for the related cases, they are based on complex numerical strategies that require intensive fine-tuning, remeshing and high computational costs. To improve the performance of the simulations, we require more efficient approaches for modeling ECM. Hence, this paper presents a modeling approach for the material dissolution based on the concept of effective physical properties. Moreover, the model avoids remeshing and, thus, allows for the simulation of the entire process with one mesh.
In this paper, the authors utilize a transient, electro-thermally coupled finite element formulation to accurately model the principal impacts in ECM and, further, to account for the interaction between the electric and thermal field. So far, experiments failed to prove the necessity of considering thermoelectric effects. Nevertheless, we consider a fully coupled model to maintain generality and flexibility for possible future applications. Early works related to the solution of thermoelectric problems may be found e.g.~in \cite{Buist1995} who employ the finite element method (FEM) to investigate the steady-state performance of thermoelectric devices and in \cite{LauBuist1997} who study the performance of power generation in thermoelectricity. Other authors, e.g.~\cite{AntonovaLooman2005}, conduct transient investigations of Peltier cooling devices using finite elements. Furthermore, \cite{Perez-AparicioTaylorEtAl2007} present a nonlinear fully coupled finite element formulation for steady-state thermoelectricity. In \cite{PalmaPerez-AparicioEtAl2012}, the formulation is extended for dynamic problems using a hyperbolic heat conduction model. Moreover, coupling of thermal and electrical fields with the mechanical field may be found in e.g.~\cite{Perez-AparicioPalmaEtAl2016b} and, additionally, with the magnetic field in \cite{Perez-AparicioPalmaEtAl2016a}.

\textbf{Outline of the work.\quad} The effective modeling of the anodic dissolution based on Faraday's law of electrolysis is discussed in Section~\ref{sec:dissolution}. Thereafter, in  Section~\ref{sec:etcoupling}, the governing balance equations of thermoelectricity, the constitutive laws and the corresponding weak forms are presented. Moreover, Section~\ref{sec:discretization} serves to define a unit cell and to introduce a time and space discretization based on the backward Euler method and the finite element method, respectively. In Sections~\ref{ssec:Ex1}~-~\ref{ssec:Ex3}, analytical and experimental reference solutions validate the model's performance. Next, in Section~\ref{ssec:Ex4}, the model's predictive capabilities are investigated for the evolution of the surface texture in a pulsed electrochemical machining process. Finally, Section~\ref{sec:conclusion} provides the paper's conclusion.

\textbf{Notational conventions.\quad} Italic characters $a$, $A$ denote scalars and zeroth-order tensors, bold-face italic characters $\bm{b}$, $\bm{B}$ denote vectors and first-order tensors and bold-face roman characters $\mathbf{c}$, $\mathbf{C}$ refer to matrices and second-order tensors. The operators $\div{\bullet}$ and $\grad{\bullet}$ define the divergence and gradient of a quantity with respect to Cartesian coordinates. The transpose of a quantity is defined by $(\bullet)\T$. A dot $\cdot$ defines the single contraction of two tensors. The time derivative of a quantity is defined by $\dot{(\bullet)}$.

\section{Homogenized description of the anodic dissolution}
\label{sec:dissolution}

%----------------------------------------------------------------------------------------------------------------------------------%
The anodic dissolution of an arbitrary metal atom $\mathrm{Me}$ is characterized by the oxidation reaction
\begin{equation}
  \mathrm{Me} \longrightarrow \mathrm{Me}^{z +} + z \, \mathrm{e}^{-}
\end{equation}
where $\mathrm{e}^{-}$ denotes an electron and $z$ the electrochemical valency, which describes the number of electrons which seperate during the chemical process. At the macroscopic level, Faraday's law of electrolysis describes the related dissolved volume and reads for a multi-phase material according to \cite{KlockeKoenig2007}:
\begin{equation}
\label{eq:Vdis}
  \Vdis = \eta \, \sum_a \, \lambda_a \, \frac{M_a}{F \, \rhoVa \, \sum_b \, \nu_b \, z_{ab}} \,\, I \, t
\end{equation}
In Eq.~\eqref{eq:Vdis}, phase $a$ is defined by the volume fraction $\lambda_a$, the molar mass $M_a$ and the volume density $\rhoVa$. The dissolution is a result of different reactions $b$ which take place with a probability described by the factor $\nu_b$ and an individual electrochemical valency $z_{ab}$. Furthermore, the efficiency $\eta$, Faraday's constant $F = 96485$~\si{\ampere\second\per\mol}, the current $I$ and the machining time $t$ are taken into account. Based on the work of \cite{Harst2019}, we introduce the effectively dissolved volume $\Veff$ as an experimentally detected material parameter which considers anodic gas evolution as well as additional chemical reactions. Up to now, focusing on the anodic dissolution, chemical reactions at the cathode are neglected. Thus, $\Veff$ describes the incremental dissolved volume $\mathrm{d}\Vdis$ per incrementally flown electric charge given by $I \mathrm{d}t$. Accordingly, the infinitesimal dissolved volume per time increment reads
\begin{equation}
  \dfrac{\mathrm{d}\Vdis}{\mathrm{d}t} = \Veff \, I.
   \label{eqn:Vdis}
\end{equation}

The objective of this work is the presentation of a new modeling approach for the anodic dissolution, which enables a computation of the entire process without remeshing. Hence, we define a  dissolution level $d \in [0,1]$ as the ratio of the dissolved volume and the corresponding reference volume, which reads per time increment
\begin{equation}
\label{eq:dinc}
\dfrac{\mathrm{d}d}{\mathrm{d}t} = \frac{\mathrm{d}\Vdis}{\mathrm{d}V \, \mathrm{d}t} = \frac{\Veff \, I(\j,d)}{\mathrm{d}V}
\end{equation}
where the scalar electric current $I$ is a function of the electric current density $\j$ and the dissolution level $d$.

In analogy to damage modeling (see e.g.~\cite{BrepolsWulfinghoffEtAl2017, BrepolsWulfinghoffEtAl2020}), where the stiffness of the material degrades when damage evolves, the material parameters of the unit cell in electrochemical machining alter, when electrolyte replaces metal material. Thus, the averaged material parameters $\bar{(\bullet)}$ are defined by the mixture of the metal phases and the electrolyte in dependence of the dissolution level $d$
\begin{equation}
\label{eq:param}
  \bar{(\bullet)} = \left( 1 - \d \right) \, \sum_a \lambda_a \,\, (\bullet)_a \, + \, \d \,\, (\bullet)_\mathrm{\scriptscriptstyle EL}
\end{equation}
where $(\bullet)_a$ denotes the parameters of phase $a$ and $(\bullet)_\mathrm{\scriptscriptstyle EL}$ those of the electrolyte, respectively.
Moreover, the contact of metal to electrolyte is a mandatory requirement for the chemical process. We, thus, define an activation function $\mathcal{A}$:
\begin{equation}
  \Afuntimet = 
  \begin{cases}
    1, & \textrm{contact metal-electrolyte} \\
    0, & \mathrm{else}
  \end{cases}
\end{equation}
The function is active at the position $\boldsymbol{x}$, i.e.~equal to $1$, if at this point the material consists of a metal phase ($d<1$) and has contact with the electrolyte. Due to the process related replacement of the metal by the electrolyte, the activation function is also evolving in time.

\begin{figure}[htbp]
  \centering
  
  \includegraphics{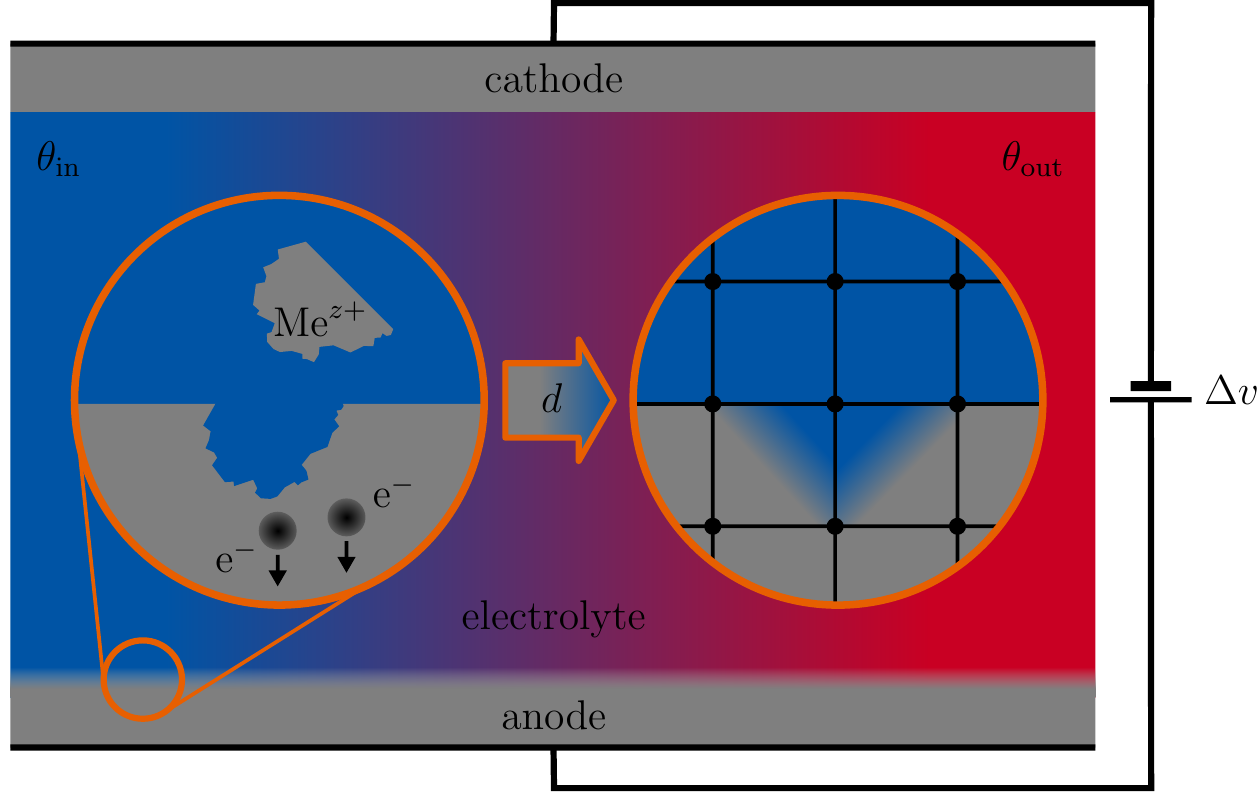}
  
  \caption{Illustration of anodic dissolution during ECM: chemical process (\textit{left}), effective description using the dissolution level $d$ (\textit{right}).}
  \label{fig:sketchECM}
\end{figure}

\section{Electro-thermal coupling}
\label{sec:etcoupling}

In thermoelectricity, the constitutively independent variables are the electric potential $v$ and the absolute temperature $\theta$. In the following, all the material parameters denoted with a bar refer to the effective quantities. As described before, they are a result of the phase-electrolyte mixture and can be calculated using Equation \eqref{eq:param}. The first governing balance equation is the conservation of electric charge (cf.~\cite{Jackson1962}):
\begin{align}
  \rhoEdot + \div{\j} &= 0              \hspace{9.1mm} \mathrm{in} \,\, \Omega    \notag                \\
  v                   &= \widetilde{v}  \hspace{9.0mm} \mathrm{on} \,\, \Gamma_v  \label{eqn:strngfrmv} \\
  \j \cdot \n         &= \widetilde{j}  \hspace{9.2mm} \mathrm{on} \,\, \Gamma_j  \notag
\end{align}
Here, the electric field strength $\E$ reads
\begin{equation}
  \E = - \grad{v},
\end{equation}
and the constitutive law of the electric displacement field $\D$ reads in accordance with the Maxwell equations
\begin{equation}
  \D = \epsnull \epsrbar \, \E,
\end{equation}
where $\epsnull$ and $\epsrbar$ denote the electric constant and the effective relative permittivity, respectively. Moreover, the electric volume charge density $\rhoE$ is defined as
\begin{equation}
  \rhoE = \div{\D}.
\end{equation}
Furthermore, the constitutive law of the electric current density $\j$ consists of three components: the first, related to Ohm's law $\jL$; the second, related to the displacement current $\jV$; the third, related to the Seebeck effect $\jS$. Thus, the electric current density reads
\begin{equation}
  \j \left( v, \vdot, \theta \right) = \jL \left( v \right) + \jV \left( \vdot \right) + \jS \left( \theta \right)
\end{equation}
with the definitions
\begin{align}
  \jL &\coloneq \kEbar \left(- \grad{v} \right) = \kEbar \, \E                              , \\
  \jV &\coloneq \epsnull \epsrbar \left(- \grad{\vdot} \right) = \epsnull \epsrbar \, \Edot , \\
  \jS &\coloneq \kEbar \alphabar \left(- \grad{\theta} \right)
\end{align}
where $\kEbar$ and $\alphabar$ denote the effective quantities for the electric conductivity and for the Seebeck coefficient.

The second governing balance equation is the transient heat conduction equation:
\begin{align}
  \rhoVbar \cTbar \, \thetadot + \div{\q} &= \j \cdot \E + q^\ast \hspace{ 9.0mm} \mathrm{in} \, \Omega        \notag                \\
  \theta                                  &= \widetilde{\theta}   \hspace{24.8mm} \mathrm{on} \, \Gamma_\theta \label{eqn:strngfrmT} \\
  \q \cdot \n                             &= \widetilde{q}        \hspace{24.8mm} \mathrm{on} \, \Gamma_q      \notag
\end{align}
The term $\j \cdot \E$ describes Joule-heating and $q^\ast$ additional heat sources. However, additional heat evolution due to e.g.~chemical reactions is currently neglected. The effective parameters $\rhoVbar$ and $\cTbar$ denote the volume density and the specific heat capacity. The constitutive law of the heat flux $\q$ consists of the part related to the Peltier effect $\qP$ and the part related to Fourier's law $\qF$. Hence, the heat flux reads
\begin{equation}
  \q \left( v, \vdot, \theta \right) = \qP \left( v, \vdot, \theta \right) + \qF \left( \theta \right)
\end{equation}
with the definitions
\begin{align}
  \qP &\coloneq \pibar \, \j                          , \\
  \qF &\coloneq \kTbar \left(- \grad{\theta} \right)
\end{align}
where $\pibar$ and $\kTbar$ denote the Peltier coefficient and the thermal conductivity, both in their efficient representation.
For the non-transient case, an analogous structure of the constitutive equations may be found in e.g.~\cite{Perez-AparicioTaylorEtAl2007}. Moreover, Table~\ref{tab:constants} shows the material parameters' SI units and definitions.
By inserting the constitutive equations into the governing balance equations, multiplying with the arbitrary test functions $\varv$ and $\varT$ and employing partial integration, the weak forms $\gv$ and $\gT$ are obtained (cf.~\cite{Perez-AparicioTaylorEtAl2007}, non-transient):
\begin{equation}
  \gv \coloneq - \intO \left( \, \j + \, \jV \right) \cdot \grad{\varv} \dV + \gjtilde = 0
  \label{eqn:gv}
\end{equation}
\begin{equation}
  \gT \coloneq \intO \left( \rhoVbar \cTbar \thetadot - \j \cdot \E - q^\ast \right) \varT \, \dV - \intO \q \cdot \grad{\varT} \dV + \gqtilde = 0
  \label{eqn:gT}
\end{equation}
The primary unknowns are the electric potential $v$ and the absolute temperature $\theta$. Furthermore, the dissolution level $d$ deals as an internal variable and, thus, influences the effective material parameters within the weak forms \eqref{eqn:gv} and \eqref{eqn:gT}.
The quantities $\gjtilde$ and $\gqtilde$ denote prescribed electric current densities and heat fluxes.

%----------------------------------------------------------------------------------------------------------------------------------%

\begin{table}[htbp]
  \centering
  \caption{Physical constants and material parameters}
  \vspace{-3mm}
  \begin{tabular}{lll}
      \hline
      $\epsnull$          & $[ \si{\ampere\second\per\volt\per\meter} ]$    & Electric constant $\left( 8.854 \times 10^{-12} \right)$  \\
      $F$                 & $[ \si{\ampere\second\per\mol} ]$               & Faraday constant  $\left( 9.648 \times 10^{4}   \right)$  \\
      \hline
      $a$                 & $[ - ]$                                         & Phase $a$                    \\
      $b$                 & $[ - ]$                                         & Reaction $b$                 \\
      \hline
      $\alpha$            & $[ \si{\volt\per\kelvin} ]$                     & Seebeck coefficient          \\
      $\epsr$             & $[ - ]$                                         & Relative permittivity        \\
      $\eta$              & $[ - ]$                                         & Efficiency                   \\
      $\lambda$           & $[ - ]$                                         & Volume fraction              \\
      $\nu$               & $[ - ]$                                         & Probability factor           \\
      $\Pi$               & $[ \si{\volt\per\square\kelvin} ]$              & Peltier coefficient $(\Pi = \alpha \, \theta)$  \\
      $\rhoV$             & $[ \si{\kg\per\cubic\meter} ] $                 & Volume density               \\
      $\cT$               & $[ \si{\joule\per\kg\per\kelvin} ]$             & Specific heat capacity       \\
      $\kE$               & $[ \si{\ampere\per\volt\per\meter} ]$           & Electric conductivity        \\
      $\kT$               & $[ \si{\watt\per\meter\per\kelvin} ]$           & Thermal conductivity         \\
      $M$                 & $[ \si{\kg\per\mol} ]$                          & Molar mass                   \\
      $\Veff$             & $[ \si{\cubic\meter\per\ampere\per\second} ]$   & Effectively dissolved volume \\
      $z$                 & $[ - ]$                                         & Electrochemical valency      \\
      \hline
  \end{tabular}
  \label{tab:constants}
\end{table}

\section{Discretization and finite element implementation}
\label{sec:discretization}

%----------------------------------------------------------------------------------------------------------------------------------%
\subsection{Definition of a unit cell}
\label{ssec:unitcell}

\begin{figure}[htbp]
  \centering
  
  \includegraphics{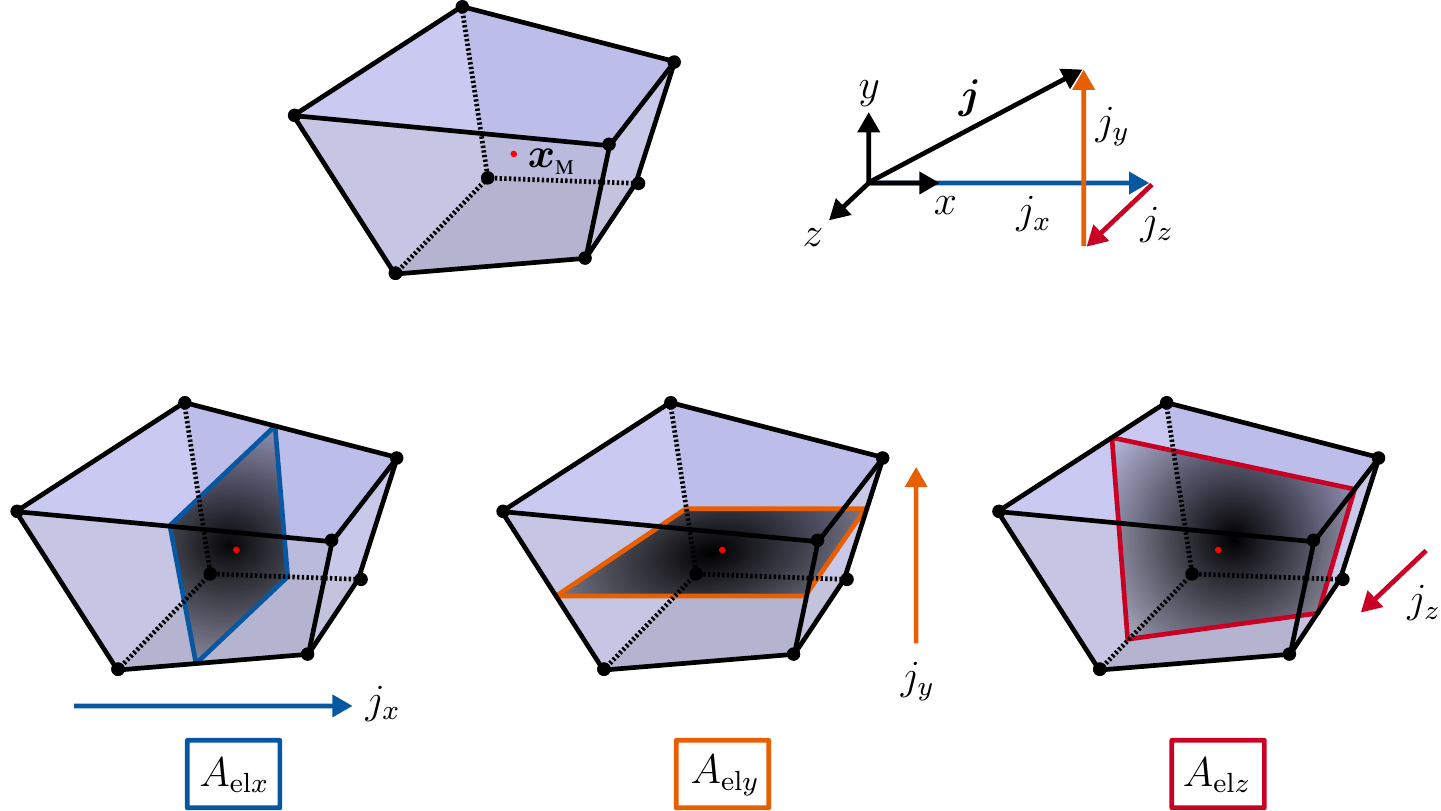}
  
  \caption{Definition of element areas $\Aelx$, $\Aely$ and $\Aelz$.}
  \label{fig:elementareas}
  
\end{figure}

As written in Eq.~\eqref{eq:dinc}, the dissolution level per time $\mathrm{d}d/\mathrm{d}t$ describes the ratio of the related dissolved volume increment $\mathrm{d}\Vdis$ (see Eq.~\eqref{eqn:Vdis}) per volume and time increment. We employ the concept of a unit cell, and, therefore, the dissolution level $d$ defines the relative dissolved volume of a unit cell. Accordingly, the rate of the dissolution level per unit cell volume reads
\begin{equation}
  \dot{d} = \dfrac{1}{\Vuc}\dfrac{\mathrm{d}\Vdis}{\mathrm{d}t} \label{eqn:d_ratio}
\end{equation}
and allows a smoother description of the chemical reaction (see Fig.~\ref{fig:sketchECM}) compared to earlier works. The size of the unit cell may be chosen arbitrarily. However, the unknowns, i.e.~the volume $\Vuc$ and the scalar electric current $I$, must be computed consistently. The vector of the electric current density $\j$ must be transferred to a scalar electric current $I$, because Faraday's law of electrolysis is formulated on a scalar basis (see Eq.~\eqref{eq:Vdis}). Here, in the finite element framework, we define a unit cell at each integration point.
The unit cell's volume $\Vuc$ is deduced from the finite element's volume $\Vel$ and the number of integration points $\ngp$
\begin{equation}
  \Vuc = \frac{\Vel}{\ngp}.
\end{equation}
Moreover, the areas of the unit cell, that enable the transformation of the three-dimensional electric current density vector to a scalar electric current, are computed analogously from the element's areas
\begin{equation}
  \Aucx = \frac{\Aelx}{\ngp}, \quad \Aucy = \frac{\Aely}{\ngp}, \quad \Aucz = \frac{\Aelz}{\ngp}.
\end{equation}
Fig.~\ref{fig:elementareas} defines the element's areas $\Aelx$, $\Aely$ and $\Aelz$ for an arbitrarily shaped element. They are calculated from the intersection points of the element's edges with the $x$-, $y$- and $z$-plane. The position vector $\xM$ in the center of the element and the normals $\bm{n}_x$, $\bm{n}_y$ and $\bm{n}_z$ define these planes.

\begin{figure}[htbp]
  \centering
  
%   \includegraphics{01_Inkscape/02_Unitcell/00_saved/elementareas_partlydissolved.pdf}
  
%   \beginpgfgraphicnamed{elementareas_partlydissolved}
  
  \begin{tikzpicture}
    \node[inner sep=0pt] (pic) at (0,0) {\includegraphics[width=0.9\textwidth]
    {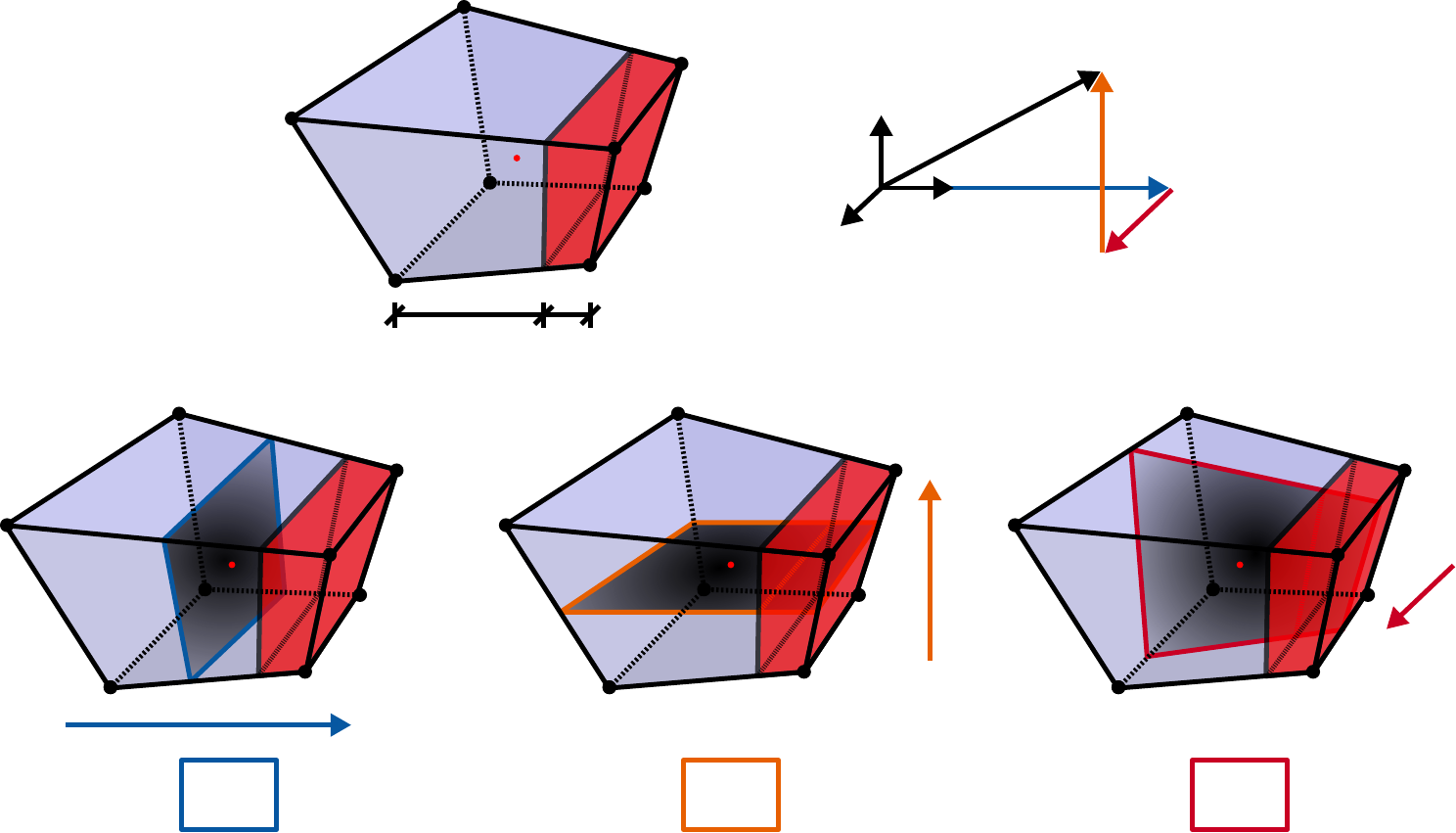}};
    \node[inner sep=0pt] (x)   at ($(pic.north)+( 2.30, -2.10)$)  {$x$};
    \node[inner sep=0pt] (y)   at ($(x.west)   +(-0.65,  1.14)$)  {$y$};
    \node[inner sep=0pt] (z)   at ($(x.west)   +(-1.22, -0.32)$)  {$z$};
    \node[inner sep=0pt] (Ax)  at ($(pic.south)+(-4.925, 0.36)$)  {$\Aelx$};
    \node[inner sep=0pt] (Ay)  at ($(Ax.east)  +( 4.625, 0)$)     {$\Aely$};
    \node[inner sep=0pt] (Az)  at ($(Ay.east)  +( 4.64,  0)$)     {$\Aelz$};
    \node[inner sep=0pt] (xM)  at ($(pic.north)+(-2.12, -2.08)$)  {$\xM$};
    \node[inner sep=0pt] (d1)  at ($(pic.north)+(-2.60, -3.50)$)  {$(1-d)l^e$};
    \node[inner sep=0pt] (d2)  at ($(pic.north)+(-1.53, -3.47)$)  {$d l^e$};
    \node[inner sep=0pt] (j)   at ($(pic.north)+( 2.55, -1.00)$)  {$\j$};
    \node[inner sep=0pt] (jx)  at ($(j.south)  +( 0.65, -0.99)$)  {$\jx$};
    \node[inner sep=0pt] (jy)  at ($(j.south)  +( 1.45, -0.1)$)   {$\jy$};
    \node[inner sep=0pt] (jz)  at ($(j.south)  +( 1.90, -1.1)$)   {$\jz$};
    \node[inner sep=0pt] (jx2) at ($(Ax.east)  +( 1.15,  0.725)$) {$\jx$};
    \node[inner sep=0pt] (jy2) at ($(Ay.east)  +( 1.6,   0.95)$)  {$\jy$};
    \node[inner sep=0pt] (jz2) at ($(Az.east)  +( 1.75,  1.75)$)  {$\jz$};
  \end{tikzpicture}
  
%   \endpgfgraphicnamed
  
  \caption{Reduction of the element areas $\Aelx$, $\Aely$ and $\Aelz$ to take the dissolution level $d$ into account. In this illustration, current $I_1$ is assumed in $x$-direction. The volume marked in red denotes the related dissolved volume.}
  \label{fig:elementareas_partlydissolved}
  
\end{figure}
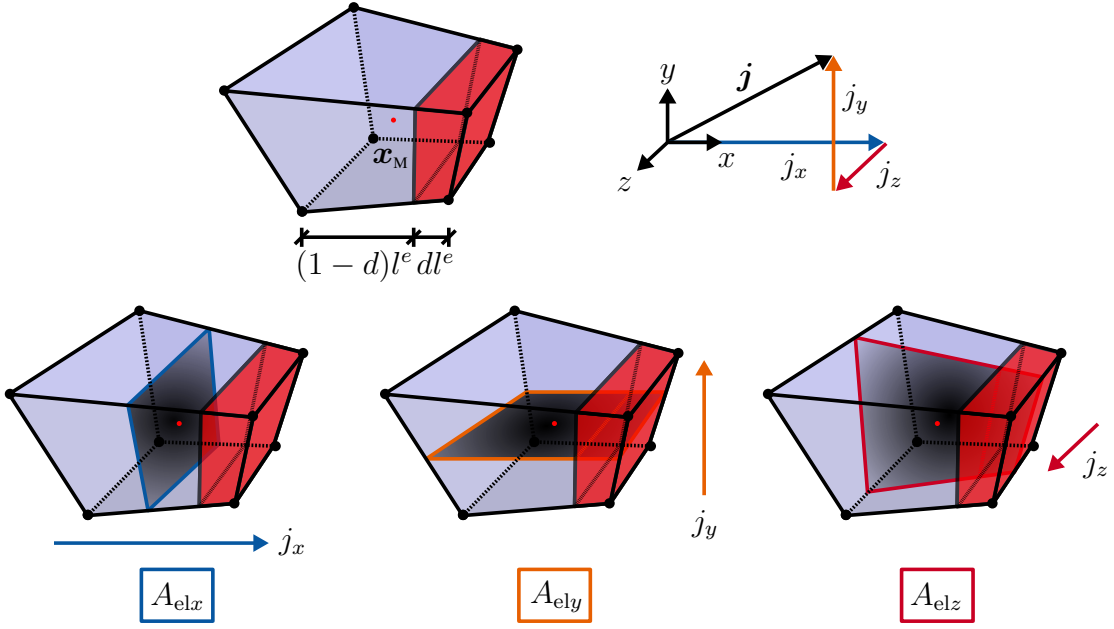

Afterwards, we employ the transfer of the electric current densities and arrange the electric currents in descending order
\begin{equation}
  \begin{Bmatrix}
    \| \Aucx \, \jx \| \\
    \| \Aucy \, \jy \| \\
    \| \Aucz \, \jz \|
  \end{Bmatrix}
  \longmapsto
  \begin{Bmatrix}
    I_1 \\
    I_2 \\
    I_3
  \end{Bmatrix}
  , \qquad I_1 \geq I_2 \geq I_3.
\end{equation}
Finally, the evolution equation of the dissolution level $d$ is expressed in terms of $I_1$, $I_2$ and $I_3$. We assume that current $I_1$ fully contributes to the evolution of the dissolution. The reduction of $I_2$ and $I_3$ by the factor $(1-d)$ considers the dissolution level of the element (see Fig.~\ref{fig:elementareas_partlydissolved}) and, hence, reduces the areas of the element and the unit cell. Therefore, the dissolution rate $\ddot$ reads
\begin{equation}
  \ddot =    \frac{1}{\Vuc} \, \Veff \left[ I_1 + (1-d) \left( I_2 + I_3 \right) \right] \Afuntimet.  \label{eqn:integraldsln}
\end{equation}
The reduction of the electric currents $I_2$ and $I_3$ aims to avoid an overestimation of the dissolution, since the electric current densities, which pass not perpendicular but parallel to the dissolved volume, just partially contribute to the dissolution (see Fig.~\ref{fig:elementareas_partlydissolved}).

%----------------------------------------------------------------------------------------------------------------------------------%
\subsection{Time discretization}
\label{ssec:timediscretization}

The backward Euler method serves to obtain the time discretized expression for the dissolution rate $\ddot$ from Eq.~(\ref{eqn:integraldsln}). It reads, under the assumption that the activation function $\mathcal{A}$ is active,
\begin{equation}
  \frac{\d_{n+1} - d_n}{\dt} = \frac{1}{\Vuc} \, \Veff \left[ \Iinpe + \left( 1 - d_{n+1} \right) \left( \Iiinpe + \Iiiinpe \right)  \right] \label{eqn:dnpeA}
\end{equation}
where $\Vuc$ and $\Veff$ are constant over time. Rewriting Eq.~(\ref{eqn:dnpeA}), an explicit expression for $\d_{n+1}$ is obtained
\begin{equation}
  \d_{n+1} =    \frac{ d_n + \frac{\Veff}{\Vuc} \left( \Iinpe + \Iiinpe + \Iiiinpe \right) \dt }
                     { 1   + \frac{\Veff}{\Vuc} \left(          \Iiinpe + \Iiiinpe \right) \dt } .
\end{equation}
If a value $\d_{n+1} > 1$ is computed, the current value of the dissolution is reset to $\d_{n+1} = 1$.
Analogously, we discretize the time derivatives of the primary variables $v$ and $\theta$ to
\begin{equation}
    \big( \, \vdot     \, \big)_{n+1} = \frac{\vnpe-\vn}{\dt} , \hspace{10mm}
    \big( \, \thetadot \, \big)_{n+1} = \frac{\Tnpe-\Tn}{\dt} . \label{eqn:fdq}
\end{equation}

%----------------------------------------------------------------------------------------------------------------------------------%
\subsection{Finite element discretization}
\label{ssec:fediscretization}

Considering Eq.~(\ref{eqn:fdq}), Appendix~\ref{sec:app_linearization} shows the linearization of the weak forms (Eqs.~(\ref{eqn:gv})-(\ref{eqn:gT})) with respect to $\vnpe$ and $\Tnpe$. Hereby, the effective material parameters' dependence on temperature and dissolution is considered in a staggered approach.

We, thus, assume $\bar{(\bullet)} \coloneq \bar{(\bullet)} \left( \dn, \Tn \right)$, where $\dn$ and $\Tn$ denote the quantities from the previous time step $t_n$. Following the linearization, the problem is spatially discretized employing the finite element method. Therefore, the entire domain is subdivided and approximated by finite elements%
\footnote{Henceforth, the superscript $(\bullet)^e$ denotes the membership of a quantity $(\bullet)$ to the finite element $e$.}
\begin{equation}
  \Omega \approx \bigcup\limits_{e = 1}^{\nel} \, \Omega^e.
\end{equation}
The primary variables and their variations are approximated with standard tri-linear shape functions
\begin{figure}[H]
  \vspace{-8mm}
  \hspace{0.075\textwidth}%
  \begin{minipage}{0.35\textwidth}
    \begin{align}
      v\ofx       &\approx  v^e\ofx      = \Nv\ofx \, \ve,    \notag  \\
      \theta\ofx  &\approx  \theta^e\ofx = \NT\ofx \, \Te,    \notag
    \end{align}
  \end{minipage}%
  \begin{minipage}{0.35\textwidth}
    \begin{align}
      \varv\ofx   &\approx  \varv^e\ofx = \Nv\ofx \, \varve,  \notag  \\
      \varT\ofx   &\approx  \varT^e\ofx = \NT\ofx \, \varTe  \notag
    \end{align}
  \end{minipage}%
  \begin{minipage}{0.2\textwidth}
    \begin{equation}
      \bm{x} \in \Omega^e
      \label{eqn:ShpN}
    \end{equation}
  \end{minipage}%
  \vspace{-5mm}
\end{figure}
where $\Nv$ and $\NT$ (row vectors) contain the shape function values and $\ve$, $\varve$, $\Te$ and $\varTe$ (column vectors) the element's nodal values. The spatial derivatives of these quantities are computed accordingly with the shape function's derivatives that are stored in the matrices $\Bv$ and $\BT$:
\begin{align}
  \grad{v\ofx}      &\approx \grad{v\ofx}^e      = \Bv\ofx \, \ve,  &
  \grad{\varv\ofx}   \approx \grad{\varv\ofx}^e  = \Bv\ofx \, \varve,  \notag \\
  \grad{\theta\ofx} &\approx \grad{\theta\ofx}^e = \BT\ofx \, \Te,  &
  \grad{\varT\ofx}   \approx \grad{\varT\ofx}^e  = \BT\ofx \, \varTe \hphantom{,} \notag \\
                    &                                                    & \hspace{-40mm} \bm{x} \in \Omega^e \hspace{9.3mm}
  \label{eqn:ShpB}                  
\end{align}
Afterwards, the previously derived approximations (Eqs.~(\ref{eqn:ShpN})-(\ref{eqn:ShpB})) are inserted into the linearized weak forms (Eqs.~(\ref{eqn:Lgv})-(\ref{eqn:LgT})):
\begin{align}
  & \sum_{e=1}^{\nel}  \, \varveT    \, \Big\{
                                               \left[ \, \kvv + \cvv \, \right] \, \dve + \left[ \hspace{5.3mm} \kvT \hspace{6.1mm} \right] \, \dTe
                                        \Big\}
  =                                    
  \sum_{e=1}^{\nel}  \, - \, \varveT \, \Big\{
                                               \, \rv \,
                                        \Big\}
  \\
  & \sum_{e=1}^{\nel}  \, \varTeT    \, \Big\{
                                               \left[ \, \kTv + \cTv \, \right] \, \dve + \left[ \, \kTT + \cTT \, \right] \, \dTe
                                        \Big\}
  =                                    
  \sum_{e=1}^{\nel}  \, - \, \varTeT \, \Big\{
                                               \, \rT \,
                                        \Big\}
\end{align}
Here, contributions from prescribed surface electric current densities and prescribed surface heat fluxes are neglected. Appendix~\ref{sec:app_elemvecmat} shows the definition of the matrices $\kvv$, $\kvT$, $\kTv$, $\kTT$, $\cvv$, $\cTv$ and $\cTT$ and of the residual vectors $\rv$ and $\rT$. Finally, the global equation system is assembled by considering the boundary conditions and the arbitrariness of the test functions:
\begin{equation}
  \begin{bmatrix}
    \Kvv + \Cvv & \KvT \\
    \KTv + \CTv & \KTT + \CTT
  \end{bmatrix}
  \begin{Bmatrix}
    \dv \\
    \dT
  \end{Bmatrix}
  =
  -
  \begin{Bmatrix}
    \Rv \\
    \RT
  \end{Bmatrix}
\end{equation} 
The increments $\dv$ and $\dT$ are computed in each global Newton-Raphson iteration and the components of the tangent matrix and the residual vector are assembled according to
\begin{align}
  \Kvv &= \assmbl \left( \kvv \right), \quad \KvT = \assmbl \left( \kvT \right), \quad \KTv = \assmbl \left( \kTv \right), \quad \KTT = \assmbl \left( \kTT \right), \notag \\
  \Cvv &= \assmbl \left( \cvv \right), \quad \CTv = \assmbl \left( \cTv \right), \quad \CTT = \assmbl \left( \cTT \right),  \\
  \Rv  &= \assmbl \left( \rv  \right), \quad \RT  = \assmbl \left( \rT  \right).  \notag
\end{align}

For further information on the finite element method, the reader is kindly referred to the literature of e.g.~\cite{Hughes1987} and \cite{ZienkiewiczTaylorEtAl2005}.

\section{Numerical examples}
\label{sec:examples}

This section presents the application of the previously developed model in numerical examples. First, we validate the model's capabilities to accurately model material dissolution by analytical and experimental reference solutions. Then, we apply the model to exemplarily compute a so-called process signature motivated by the work of the transregional Collaborative Research Center 136 ``Process Signatures'', see \cite{BrinksmeierReeseEtAl2018}.

The temperature-dependent material parameters in the following examples are approximated by cubic polynomials
\begin{equation}
  f(\theta) = c_0 + c_1 \, \theta + c_2 \, \theta^2 + c_3 \, \theta^3
\end{equation}
where the absolute temperature $\theta$ is given in $ [\si{\kelvin}] $ and the coefficients $c_0$, $c_1$, $c_2$ and $c_3$ in Table~\ref{tab:matpar}. The functions stem from the works of \cite{Zeis2015} and \cite{Harst2019} where they have been identified for a steel 42CrMo4 and an electrolyte solution with $20 $~wt.-\si{\percent} NaNO$_3$. Due to the lack of experimental data, we assume the Seebeck coefficients in the range of water and aluminum to $\alpha\EL = 1 \, \si{\micro\volt\per\kelvin}$ and $\alpha\ME = 5 \, \si{\micro\volt\per\kelvin}$, where the superscript $(\bullet)\EL$ indicates the reference to the electrolyte and $(\bullet)\ME$ to steel.
The following examples neglect fluid mechanical effects. Moreover, we model Joule-heating in the electrolyte by defining the in- and outflow temperature of the electrolyte which is obtained from experimental investigations. The contribution from $\j \cdot \E$ is not considered in electrolyte finite elements to avoid an unphysical overheating of the electrolyte, since cooling effects due to flushing are neglected. Future work focuses on the precise modeling of the electrolyte temperature in cooperation with computational fluid dynamical simulations.
\begin{table}[h]
  \centering
  \caption{Coefficients of material parameters}
  \vspace{-3mm}
  \begin{tabular}{llrrrr}
      \hline
      $f(\theta)$        &  unit                                          & $c_0$                  & $c_1$                   & $c_2$                    & $c_3$                          \\
      \hline \hline
      $\cT\EL$           & $[ \si{\joule\per\kg\per\kelvin} ]$            & $ 8.145 \times 10^{ 3}$ & $-3.204 \times 10^{ 1}$ & $ 8.371 \times 10^{-2}$  & $-6.979 \times  10^{-5}$ \\
      $\cT\ME$           & $[ \si{\joule\per\kg\per\kelvin} ]$            & $ 3.554 \times 10^{ 2}$ & $ 2.848 \times 10^{-1}$ & $-5.000 \times 10^{-5}$  & $-$ \\
      $\kE\EL$           & $[ \si{\ampere\per\volt\per\meter} ]$          & $-6.302 \times 10^{ 1}$ & $ 2.530 \times 10^{-1}$ & $-$                      & $-$ \\
      $\kE\ME$           & $[ \si{\ampere\per\volt\per\meter} ]$          & $ 1.131 \times 10^{ 7}$ & $-3.710 \times 10^{ 4}$ & $ 6.020 \times 10^{ 1}$  & $-3.994 \times 10^{-2}$ \\
      $\kT\EL$           & $[ \si{\watt\per\meter\per\kelvin} ]$          & $-8.691 \times 10^{-1}$ & $ 8.949 \times 10^{-3}$ & $-1.584 \times 10^{-5}$  & $ 7.975 \times 10^{-9}$ \\
      $\kT\ME$           & $[ \si{\watt\per\meter\per\kelvin} ]$          & $ 3.651 \times 10^{ 1}$ & $ 4.899 \times 10^{-2}$ & $-1.012 \times 10^{-4}$  & $ 4.654 \times 10^{-8}$ \\
      $\Veff\ME$         & $[ \si{\cubic\meter\per\ampere\per\second} ]$  & $ 3.650 \times 10^{-11}$& $-$                     & $-$                      & $-$ \\
      $\alpha\EL$        & $[ \si{\volt\per\kelvin} ]$                    & $ 1.000 \times 10^{-6}$ & $-$                     & $-$                      & $-$ \\
      $\alpha\ME$        & $[ \si{\volt\per\kelvin} ]$                    & $ 5.000 \times 10^{-6}$ & $-$                     & $-$                      & $-$ \\
      $\epsr\EL$         & $[ - ]$                                        & $ 1.000 \times 10^{ 0}$ & $-$                     & $-$                      & $-$ \\
      $\epsr\ME$         & $[ - ]$                                        & $ 8.000 \times 10^{ 1}$ & $-$                     & $-$                      & $-$ \\
      $\rhoV\EL$         & $[ \si{\kg\per\cubic\meter} ]$                 & $ 8.385 \times 10^{ 2}$ & $ 1.401 \times 10^{ 0}$ & $ 3.011 \times 10^{-3}$  & $ 3.718 \times  10^{-7}$ \\
      $\rhoV\ME$         & $[ \si{\kg\per\cubic\meter} ]$                 & $ 7.849 \times 10^{ 2}$ & $-6.289 \times 10^{-2}$ & $-4.167 \times 10^{-4}$  & $ 1.907 \times  10^{-7}$ \\
      \hline
  \end{tabular}
  \label{tab:matpar}
\end{table}

%----------------------------------------------------------------------------------------------------------------------------------%
\subsection{Stationary dissolution process - analytical validation}
\label{ssec:Ex1}
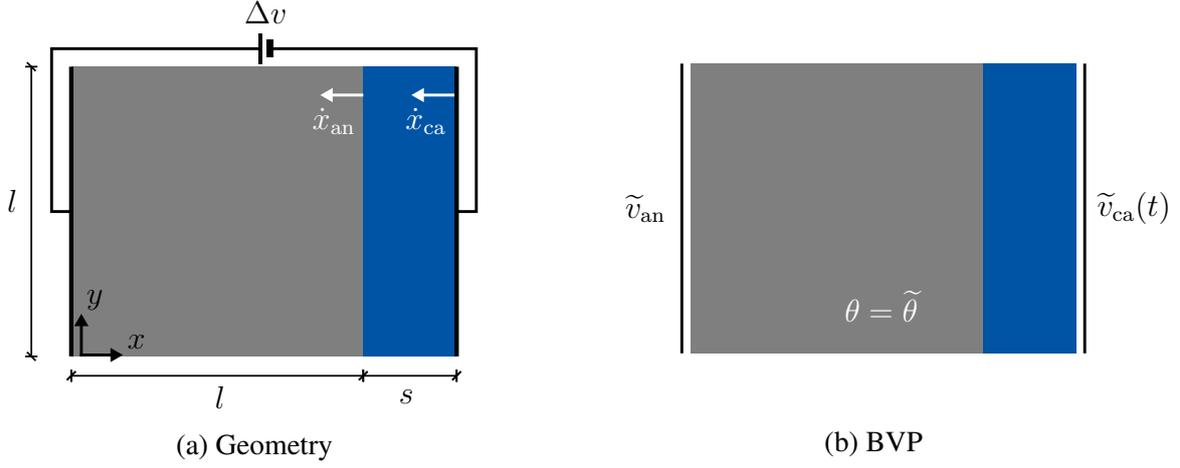
\begin{figure}[htbp] 
     \centering 
     \begin{subfigure}{.45\textwidth} 
         
         \centering 
         
         \begin{tikzpicture} 
           \node[inner sep=0pt] (pic) at (0,0) {\includegraphics[width=\textwidth]
           {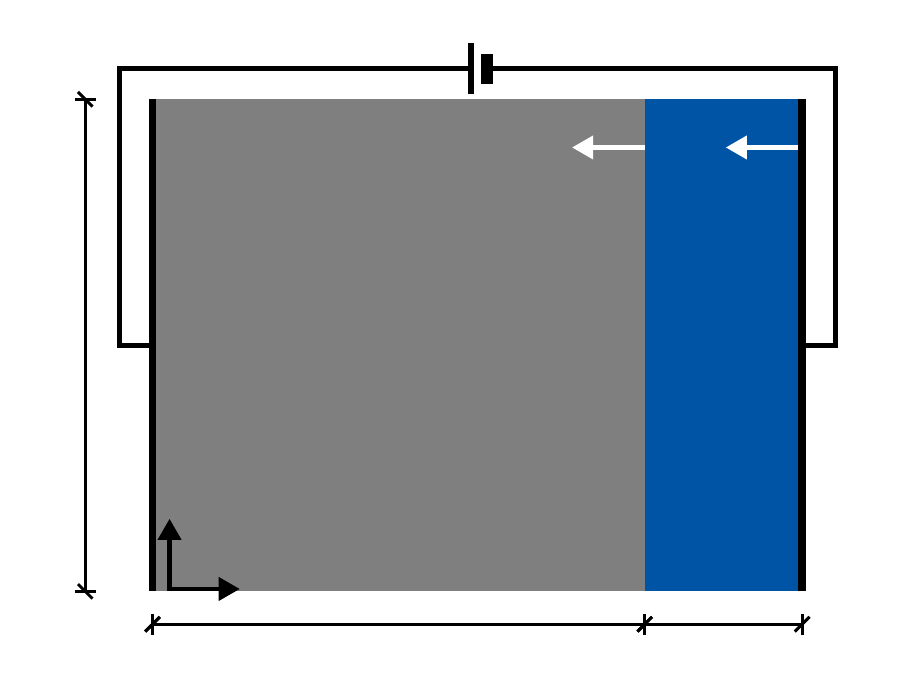}};
           \node[inner sep=0pt] (l1)   at ($(pic.west) +( 0.42, 0.10)$)  {$l$};
           \node[inner sep=0pt] (l2)   at ($(pic.south)+(-0.45, 0.16)$)  {$l$};
           \node[inner sep=0pt] (s)    at ($(pic.south)+( 2.00, 0.16)$)  {$s$};
           \node[inner sep=0pt] (x)    at ($(pic.south)+(-1.55, 0.90)$)  {$x$};
           \node[inner sep=0pt] (y)    at ($(x.north)  +(-0.55, 0.45)$)  {$y$};
           \node[inner sep=0pt] (xda)  at ($(pic.north)+( 1.05,-1.50)$)  {\textcolor{white}{$\dot{x}_\mathrm{an}$}};
           \node[inner sep=0pt] (xdc)  at ($(xda.west) +( 1.50, 0.00)$)  {\textcolor{white}{$\dot{x}_\mathrm{ca}$}};
           \node[inner sep=0pt] (dv)   at ($(pic.north)+( 0.15,-0.07)$)  {$\delv$};
         \end{tikzpicture} 
         
         \caption{Geometry}
         \label{fig:Ex1Geom}
         
     \end{subfigure}
     \qquad
     \begin{subfigure}{.45\textwidth} 

         \centering 
         
         \begin{tikzpicture} 
           \node[inner sep=0pt] (pic) at (0,0) {\includegraphics[width=\textwidth]
           {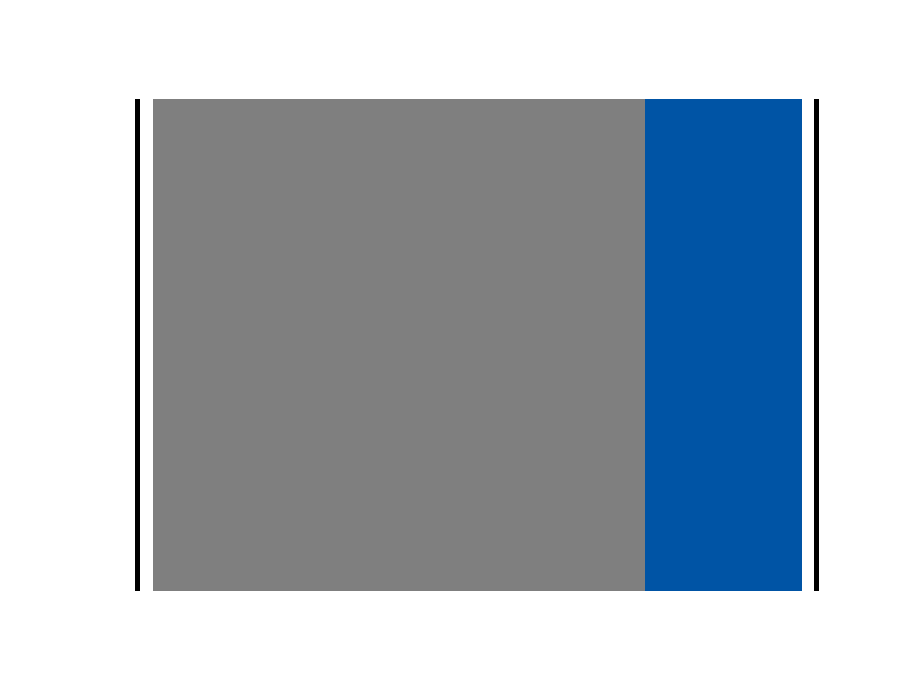}};
           \node[inner sep=0pt] (van)   at ($(pic.north)+(-3.00,-2.70)$)  {$\widetilde{v}_\mathrm{an}$};
           \node[inner sep=0pt] (vca)   at ($(van.west) +( 6.70, 0.00)$)  {$\widetilde{v}_\mathrm{ca}(t)$};
           \node[inner sep=0pt] (van)   at ($(pic.north)+( 0.10,-4.00)$)  {\textcolor{white}{$\theta = \widetilde{\theta}$}};
         \end{tikzpicture} 
         
         \caption{BVP}
         \label{fig:Ex1BVP}
         
     \end{subfigure} 
     
     \caption{Geometry and boundary value problem of the analytical reference solution for the stationary dissolution process.} 
     \label{fig:Ex1}
     
\end{figure}

The first example considers a stationary dissolution process, i.e.~the cathode's feed rate $\dot{x}_\mathrm{ca}$ and the anode's dissolution rate $\dot{x}_\mathrm{an}$ coincide. Fig.~\ref{fig:Ex1Geom} shows the initial geometrical setup with $l = 1 \, \si{\milli\meter}$. A thickness of $g = 0.1 \, \si{\milli\meter}$ is assumed. \cite{KlockeKoenig2007} derive the formula for the gap width $s$ in the stationary dissolution process according to
\begin{equation}
  s = \frac{\kE\EL \left( \delv - \delvpol \right) \Veff}{\dot{x}_\mathrm{ca}}. \label{eqn:gapwidth}
\end{equation}
In this example, we strive to compare the numerical results with a simple, analytical reference solution and, therefore, differing from Table~\ref{tab:matpar}, set the electrolyte's electric conductivity to $\kE\EL = 16 \, \si{\ampere\per\volt\per\meter}$ and the effectively dissolved volume to $\Veff\ME = 10^{-11} \, \si{\cubic\meter\per\ampere\per\second}$. The applied voltage is $\delv = 20 \, \si{\volt}$, a polarization voltage $\delvpol$ is neglected and the feed rate is $\dot{x}_\mathrm{ca} = 10^{-5} \, \si{\meter\per\second}$. With these assumptions, the stationary gap width yields $s = 0.32 \, \si{\milli\meter}$. Additionally, we apply a constant temperature distribution of $\widetilde{\theta} = 323.15 \, \si{\kelvin}$ to avoid inhomogeneous electric current density distributions due to the Seebeck effect and, thereby, ensure consistency with the one-dimensional analytical reference solution.

\begin{figure}[htbp] 
     \centering 
         
     \includegraphics{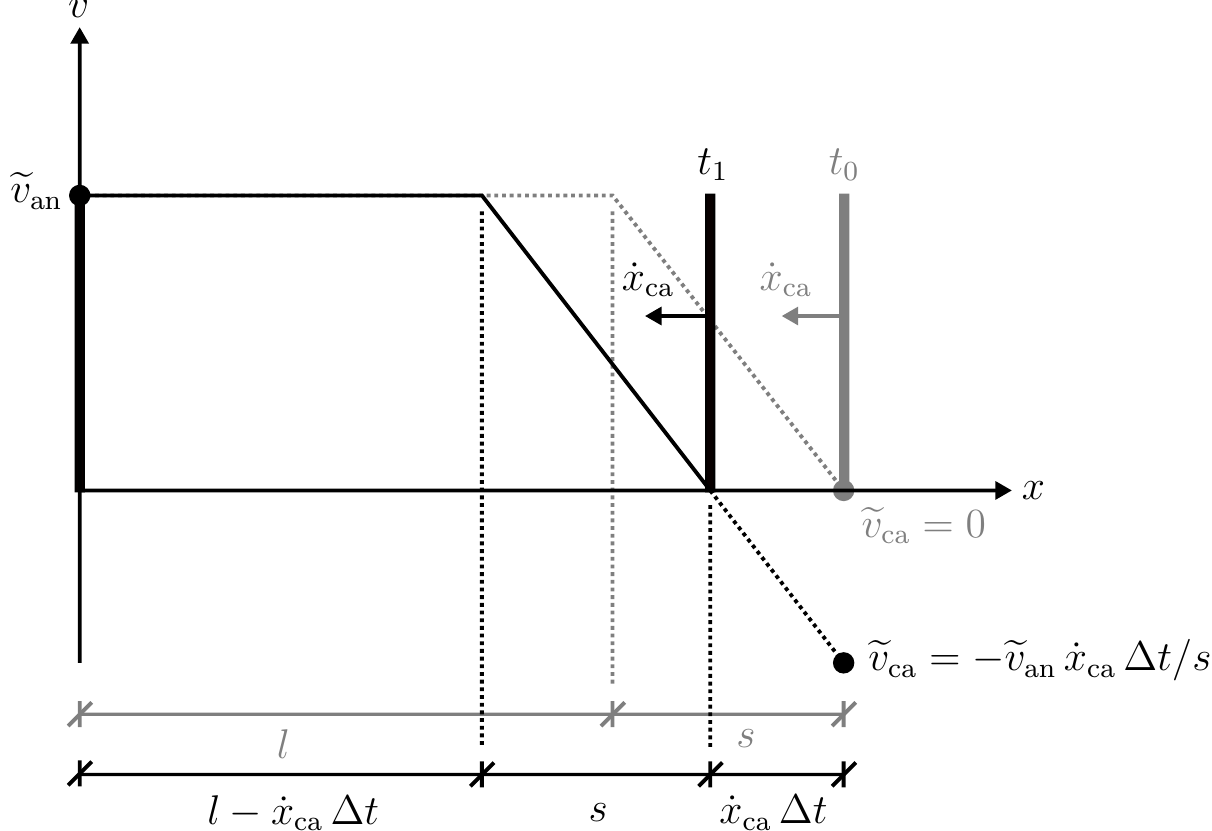}
         
     \caption{Applying the theorem of intersecting lines to prescribe the electric potential distribution at $t_0$ and $t_1$ to ensure that the cathode's position and $v = 0$~\si{\volt} coincide.}
     \label{fig:Ex1vca}
\end{figure}

Fig.~\ref{fig:Ex1BVP} shows the boundary value problem (BVP). A constant electric potential of $\widetilde{v}_\mathrm{an} = 20 \, \si{\volt}$ is applied at $x = 0 \, \si{\milli\meter}$. To model the cathode's feed, the electric potential at $x = 1.32 \, \si{\milli\meter}$ is time varying with an initial value of $\widetilde{v}_\mathrm{ca} = 0 \, \si{\volt}$. Under the assumption of a negligible potential drop in the metal and of a linear potential distribution in the electrolyte, we apply the theorem of intersecting lines. Thus, the function for the electric potential at $x = 1.32 \, \si{\milli\meter}$ reads
\begin{equation}
  \widetilde{v}_\mathrm{ca}(t) = - \frac{ \widetilde{v}_\mathrm{an} \, \dot{x}_\mathrm{ca} }{s} \, t \label{eqn:vcaoft}
\end{equation}
and, thereby, ensures that the theoretical position of the cathode's surface coincides with the position of the electric potential where $v = 0 \, \si{\volt}$ holds (cf.~Fig.~\ref{fig:Ex1vca}).

\begin{figure}[htbp] 
     \centering 
     \begin{subfigure}{.32\textwidth} 
         \centering 
         \includegraphics[width=\textwidth]{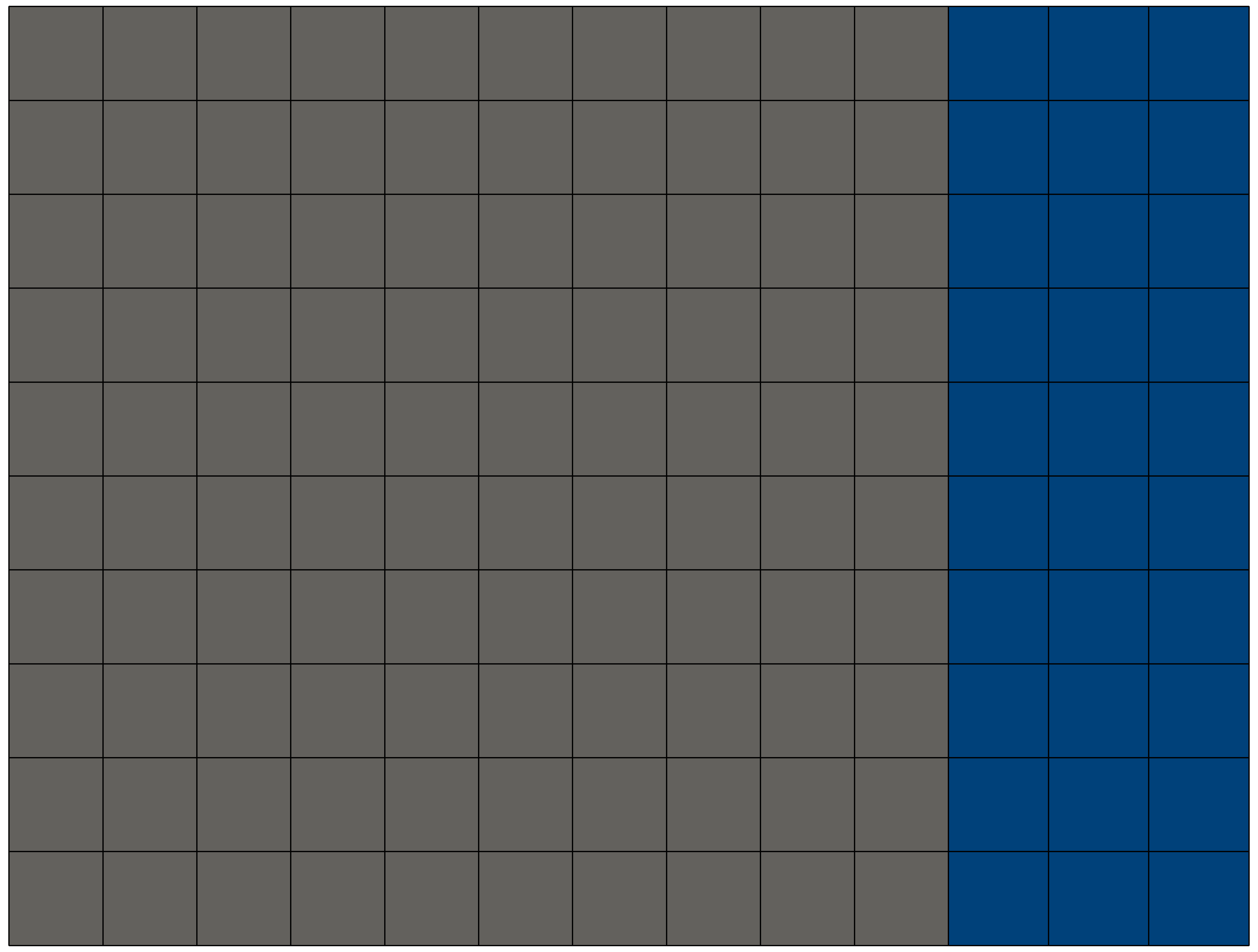}
         \caption{$t = 0$~\si{\second}}
         \label{subfig:Ex1_ed10_t00}
     \end{subfigure}
     \begin{subfigure}{.32\textwidth} 
         \centering 
         \includegraphics[width=\textwidth]{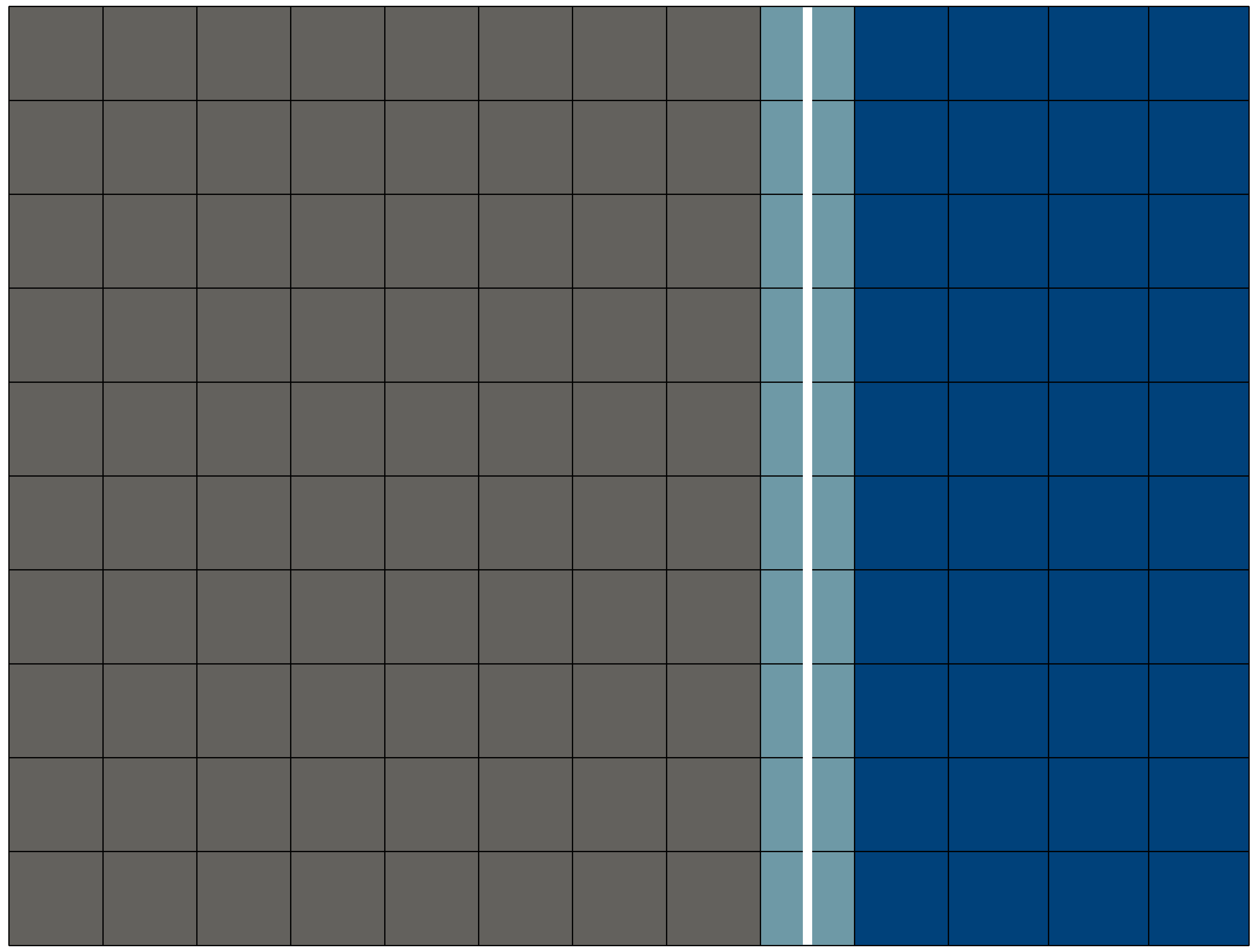}
         \caption{$t = 15$~\si{\second}}
         \label{subfig:Ex1_ed10_t15}
     \end{subfigure}
     \begin{subfigure}{.32\textwidth} 
         \centering 
         \includegraphics[width=\textwidth]{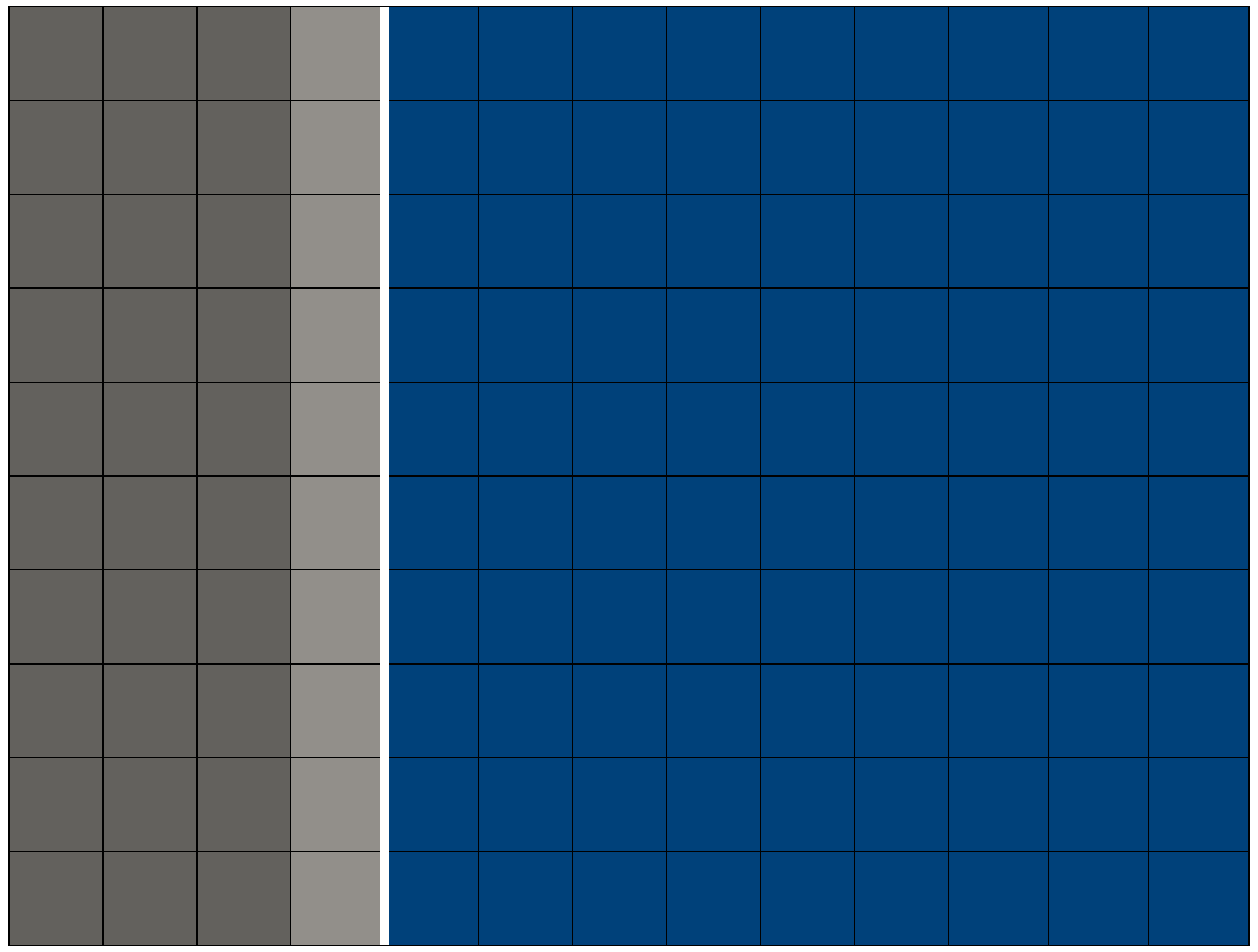}
         \caption{$t = 60$~\si{\second}}
         \label{subfig:Ex1_ed10_t60}
     \end{subfigure}     
     
     \begin{subfigure}{.32\textwidth} 
         \centering 
         \includegraphics[width=\textwidth]{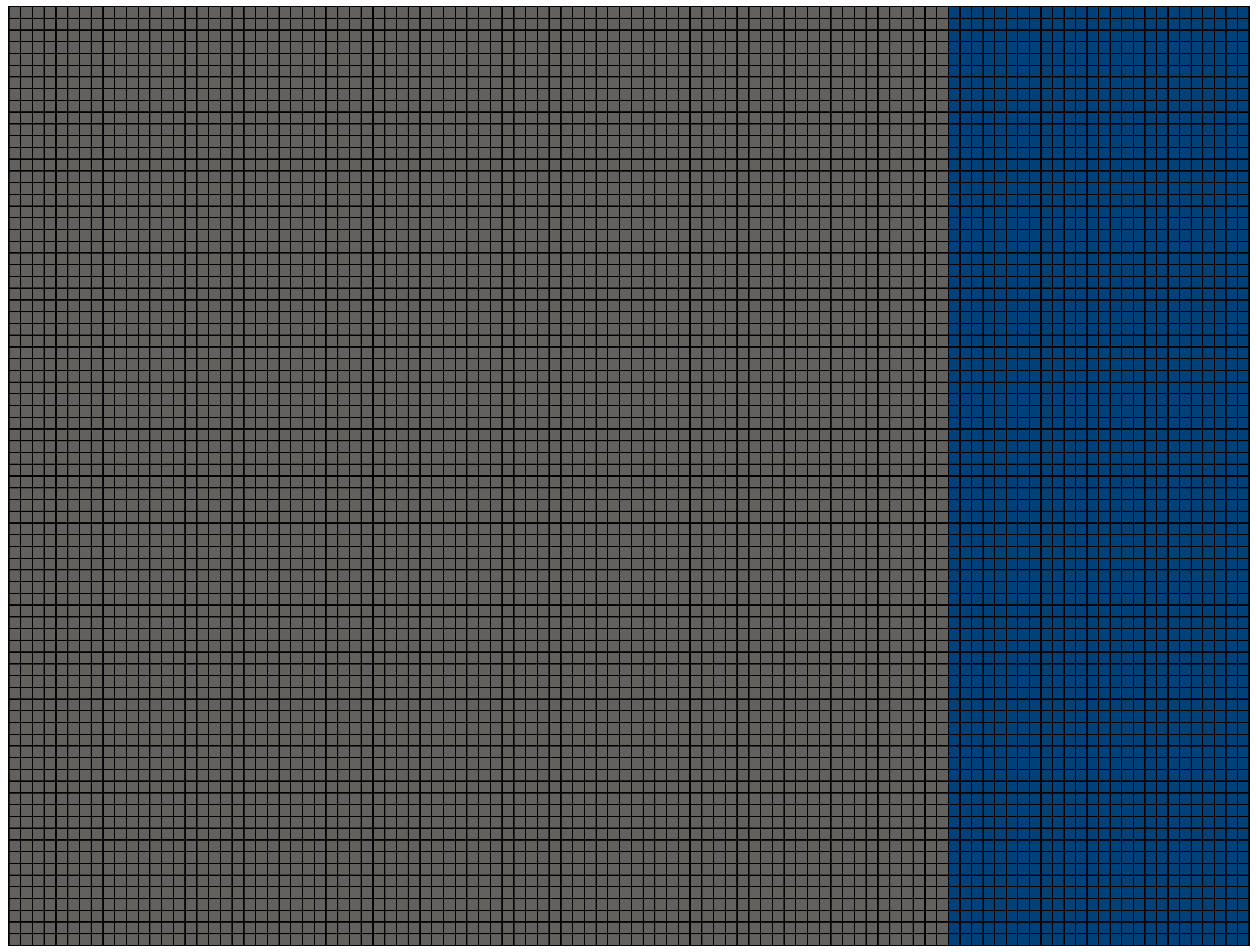}
         \caption{$t = 0$~\si{\second}}
         \label{subfig:Ex1_ed80_t00}
     \end{subfigure}
     \begin{subfigure}{.32\textwidth} 
         \centering 
         \includegraphics[width=\textwidth]{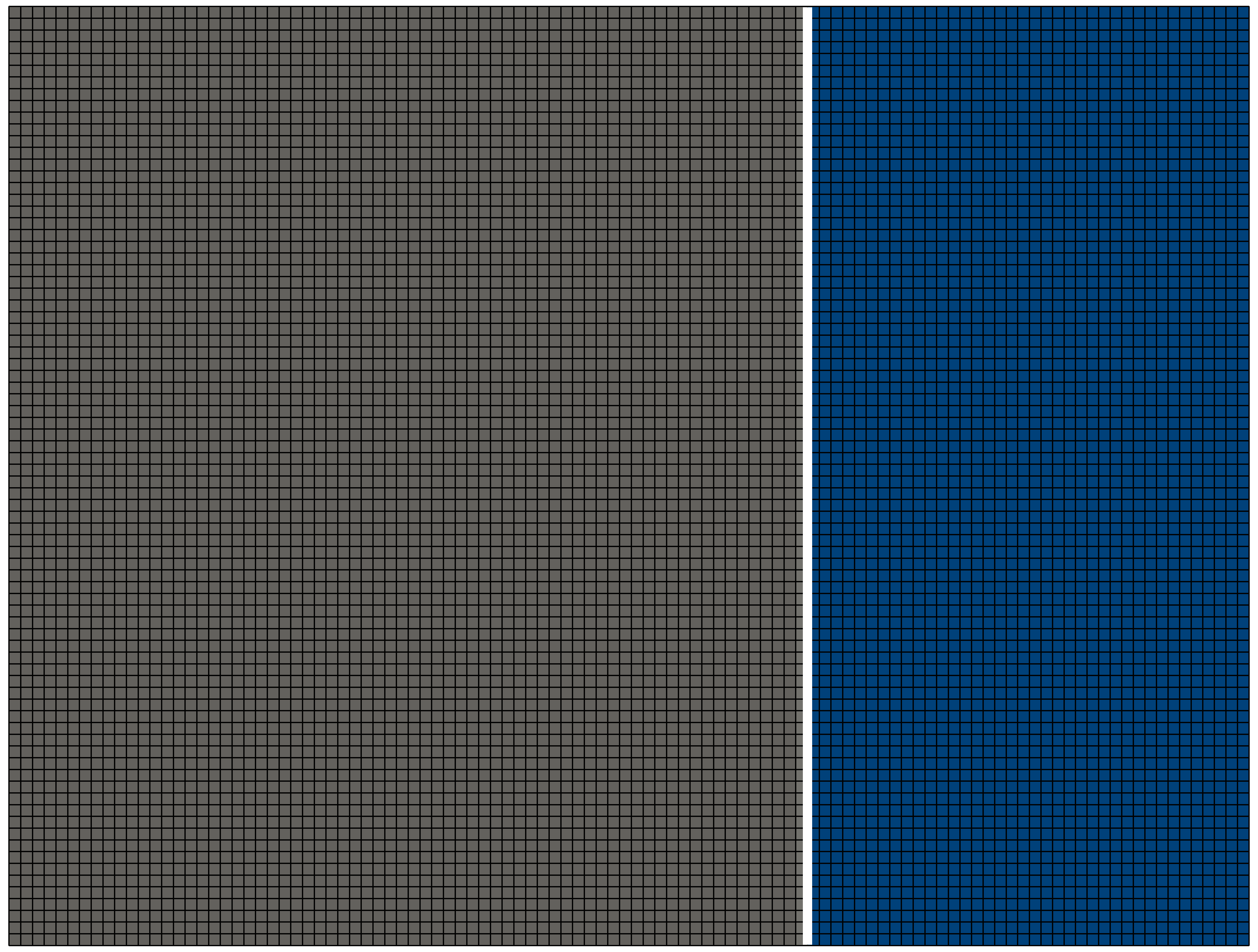}
         \caption{$t = 15$~\si{\second}}
         \label{subfig:Ex1_ed80_t15}
     \end{subfigure}
     \begin{subfigure}{.32\textwidth} 
         \centering 
         \includegraphics[width=\textwidth]{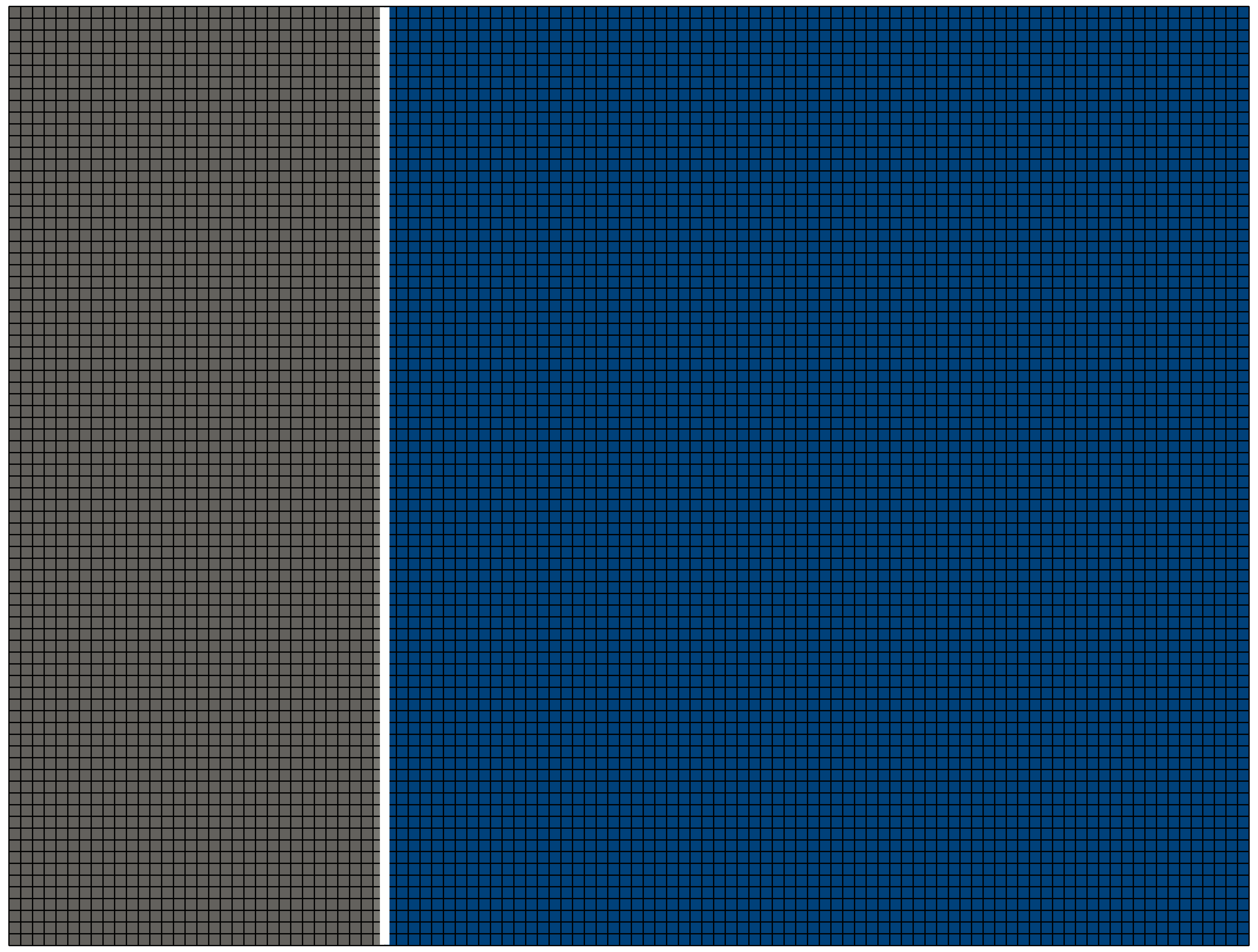}
         \caption{$t = 60$~\si{\second}}
         \label{subfig:Ex1_ed80_t60}
     \end{subfigure}     

     \vspace{3mm}
     \begin{subfigure}{.32\textwidth} 
         \centering 
         \begin{tikzpicture} 
           \node[inner sep=0pt] (pic) at (0,0) {\includegraphics[height=5mm, width=40mm]
           {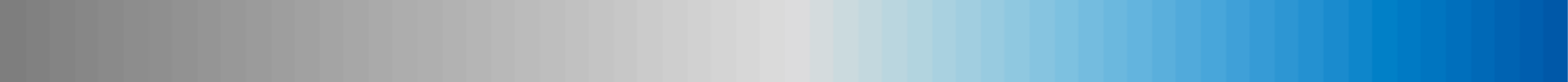}};
           \node[inner sep=0pt] (0)   at ($(pic.south)+(-2.22, 0.26)$)  {$0$};
           \node[inner sep=0pt] (1)   at ($(pic.south)+( 2.22, 0.26)$)  {$1$};
           \node[inner sep=0pt] (d)   at ($(pic.south)+( 0.00, 0.85)$)  {$d \, [-]$};
         \end{tikzpicture} 
     \end{subfigure}
     \caption{Dissolution level $d$ at machining times $0$~\si{\second}, $15$~\si{\second} and $60$~\si{\second} for a coarse (\subref{subfig:Ex1_ed10_t00}-\subref{subfig:Ex1_ed10_t60}) and a fine mesh (\subref{subfig:Ex1_ed80_t00}-\subref{subfig:Ex1_ed80_t60}). The vertical white line indicates the analytical reference solution.}
     \label{fig:Ex1_evolution_d}
\end{figure}

This example serves to investigate different finite element discretizations. For the coarsest mesh, the workpiece consists of $10 \times 10$ elements and for the finest of $80 \times 80$ elements. Fig.~\ref{fig:Ex1_evolution_d} shows the contour plots of the dissolution level $d$ for a machining time of $0$~\si{\second}, $15$~\si{\second} and $60$~\si{\second} for both, the coarsest and finest, meshes with a time increment of $\dt = 0.01~\si{\second}$. A dissolution level of $ d = 0 $ corresponds to pure metal and of $ d = 1 $ to pure electrolyte. The vertical white line visualizes the analytical reference solution after $15$~\si{\second} and $60$~\si{\second}. The results of both meshes show good agreement with the analytical reference solution. However, the coarsest mesh overestimates the dissolution at $t = 60$~\si{\second} by $ + \, 5.4~\si{\percent} $, whereas the finest mesh deviates by negligible $ + \, 0.3~\si{\percent} $.

\begin{figure}[htbp]
\centering

  \includegraphics{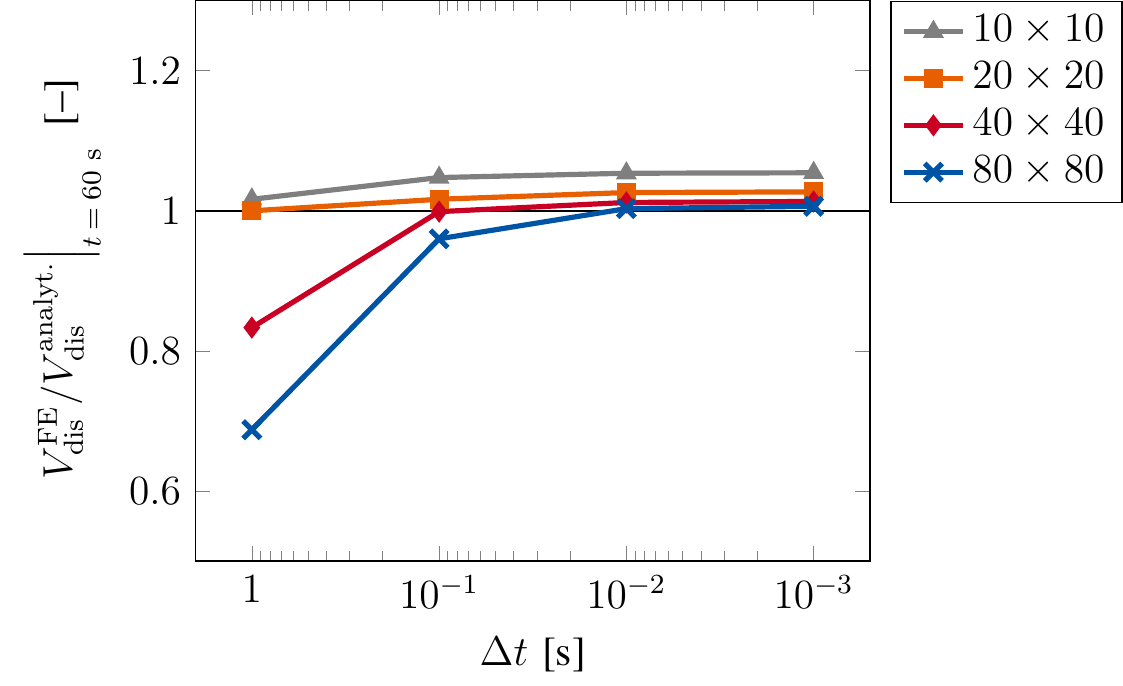}
  
  \caption{Comparison of dissolved volume from simulation $V_\mathrm{dis}^\mathrm{FE}$ and analytical solution $V_\mathrm{dis}^\mathrm{analyt.}$ after a machining time of $60~\si{\second}$.}
  \label{fig:pgf_Ex1_Vdis}

\end{figure}

To investigate the influence of the mesh density and the time step size on the simulation, Fig.~\ref{fig:pgf_Ex1_Vdis} shows the comparison of the numerically and analytically computed dissolved volume $\Vdis$ after $60$~\si{\second}. For the time increment $\dt = 1~\si{\second}$, the coarser meshes ($10 \times 10$, $20 \times 20$) give results close to the analytical solution $( + \, 1.6~\si{\percent} , \,  + \, 0.0~\si{\percent} )$, but slightly differ for smaller time steps, e.g.~$\dt = 0.01~\si{\second}$ $( + \, 5.4~\si{\percent} , \, + \, 2.6~\si{\percent} )$. In contrast, the finer meshes ($40 \times 40$, $80 \times 80$) deviate strongly from the analytical solution for a large time step $\dt = 1~\si{\second}$ $( - \, 16.7~\si{\percent} , \, - \, 30.3~\si{\percent} )$, but converge to the analytical solution for smaller time steps, e.g.~$\dt = 0.01~\si{\second}$ $( + \, 1.2~\si{\percent} , \, + \, 0.3~\si{\percent} )$.

\begin{figure}[htbp]
\centering

  \includegraphics{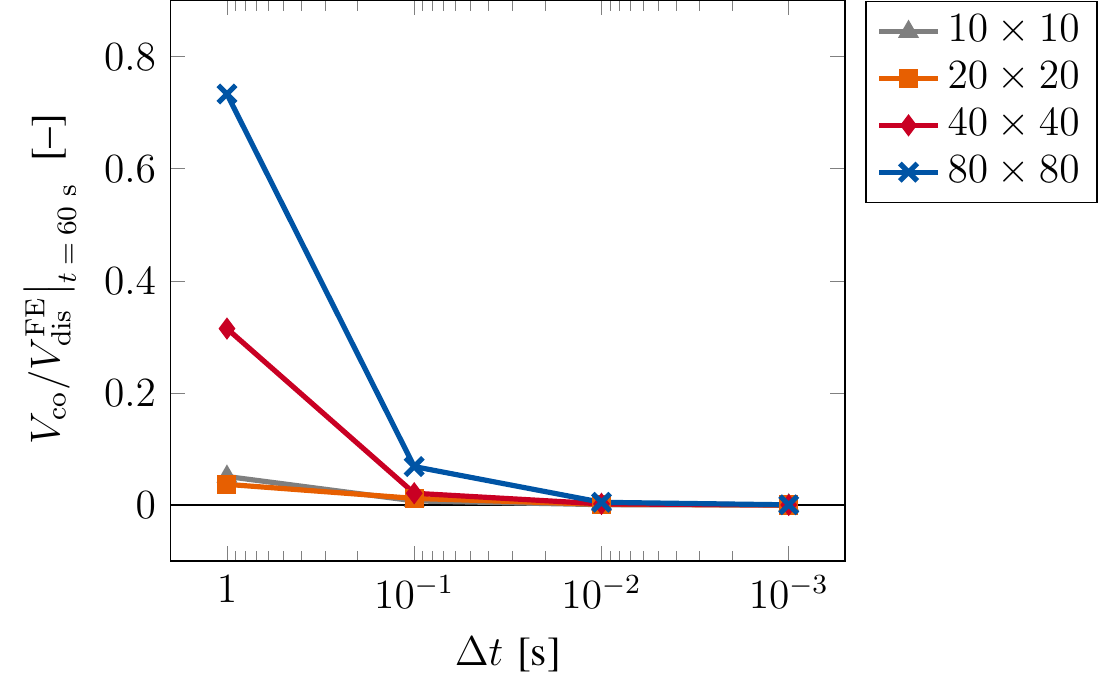}

  \caption{Relative error after a machining time of $60~\si{\second}$: Cut-off volume $V_\mathrm{co}$ over dissolved volume $V_\mathrm{dis}^\mathrm{FE}$ for different time increments and meshes.}
  \label{fig:pgf_Ex1_Vco}
\end{figure}

To analyze this result, the cut-off volume $\Vco$ is introduced that serves as error indicator. It defines the volume that is theoretically dissolved according to Faraday's law, but which is neglected numerically when $d_{n+1}>1$ is reset to $d_{n+1}=1$. Moreover, the cut-off volume is accumulated over all time steps $t$, all elements $e$ and all integration points $i$ as follows
\begin{equation}
  \Vco = \sum_{t=1}^{n_\mathrm{ts}} \sum_{e=1}^{\nel} \sum_{i=1}^{\ngp} \left(\d_{n+1} - 1 \right) \, \Vuc.
\end{equation}
Fig.~\ref{fig:pgf_Ex1_Vco} shows the cut-off volume in relation to the numerically computed dissolved volume after $60$~\si{\second}. The finer meshes, evidently, yield a high relative error $($e.g.~$80 \times 80, \, \dt = 1~\si{\second}: 73.3~\si{\percent} )$ when using a large time increment. However, this error decreases for all meshes when reducing the time step size.

These findings account for the underestimation of the dissolved volume with fine meshes at large time steps in Fig.~\ref{fig:pgf_Ex1_Vdis}. Since we update the activation function $\mathcal{A}$, which enables the elements to dissolve, at the end of every time step, a fine mesh using large time steps inevitably causes an underestimation of the dissolved volume. Therefore, the cut-off volume must always be considered and a relative error of e.g.~$\Vco/\Vdis^\mathrm{FE} \leq 1~\si{\percent}$ should be aimed for.

\begin{figure}[htbp] 
  \centering 
  \begin{subfigure}{.32\textwidth} 
      \centering 
      \includegraphics[width=\textwidth]{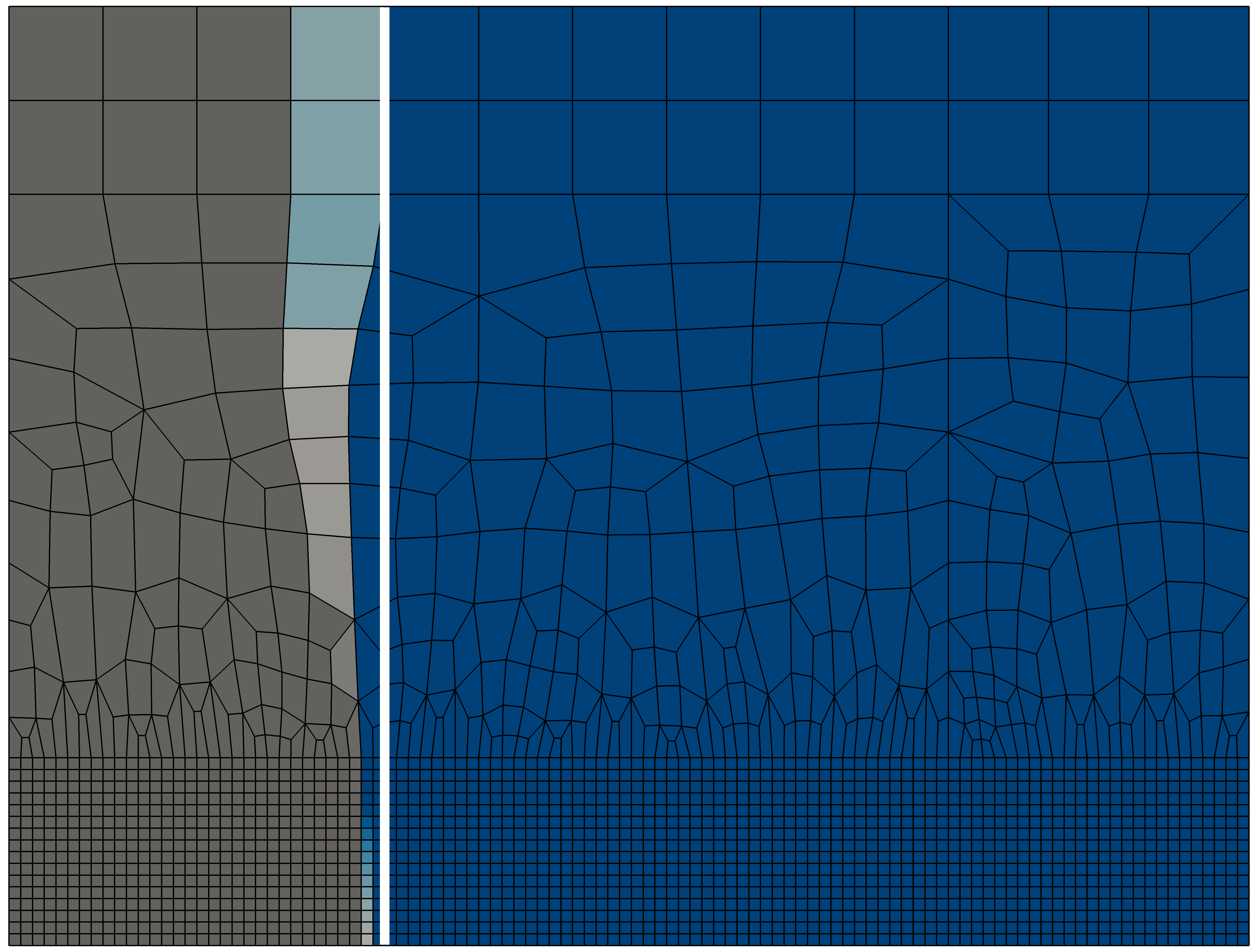}
      \caption{Mesh $10 \rightarrow 80$}
      \label{fig:Ex1_dm1}
  \end{subfigure}
  \begin{subfigure}{.32\textwidth} 
      \centering 
      \includegraphics[width=\textwidth]{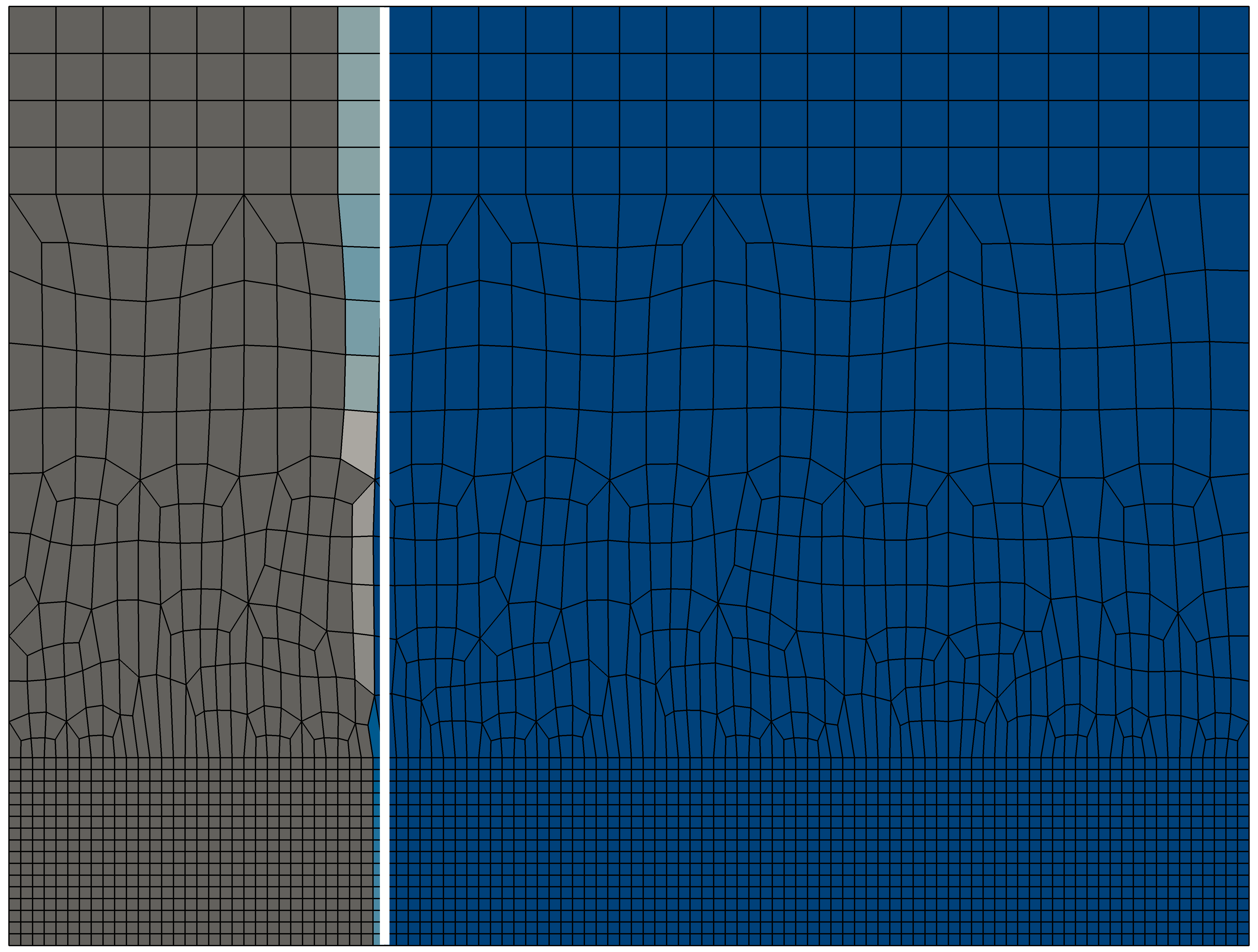}
      \caption{Mesh $20 \rightarrow 80$}
      \label{fig:Ex1_dm2}
  \end{subfigure}
  \begin{subfigure}{.32\textwidth} 
      \centering 
      \includegraphics[width=\textwidth]{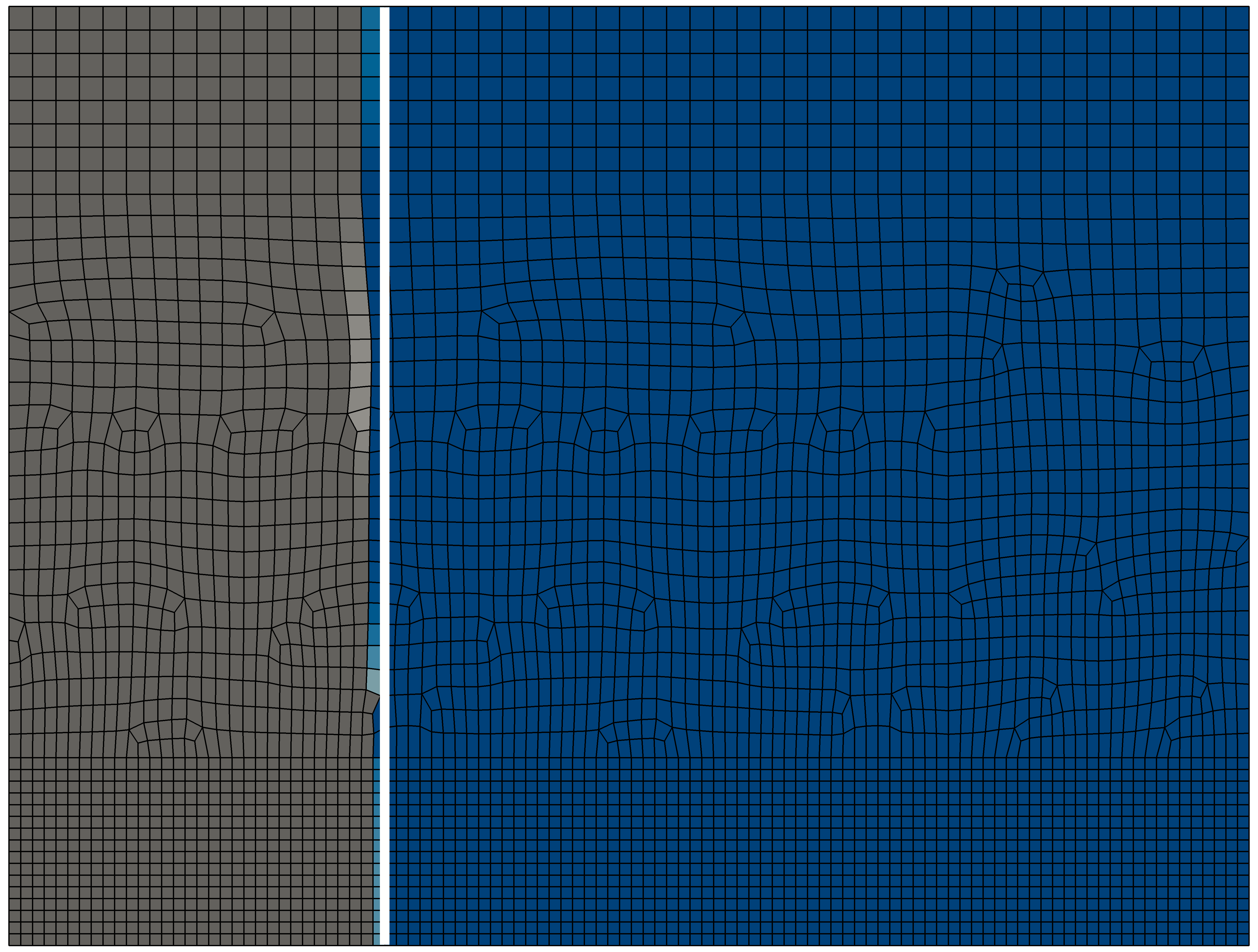}
      \caption{Mesh $40 \rightarrow 80$}
      \label{fig:Ex1_dm3}
  \end{subfigure}

  \vspace{3mm}
  \begin{subfigure}{.32\textwidth} 
      \centering 
      \begin{tikzpicture} 
        \node[inner sep=0pt] (pic) at (0,0) {\includegraphics[height=5mm, width=40mm]
        {03_Contour/01_Examples/00_Color_Maps/02_dsln_hor.pdf}};
        \node[inner sep=0pt] (0)   at ($(pic.south)+(-2.22, 0.26)$)  {$0$};
        \node[inner sep=0pt] (1)   at ($(pic.south)+( 2.22, 0.26)$)  {$1$};
        \node[inner sep=0pt] (d)   at ($(pic.south)+( 0.00, 0.85)$)  {$d \, [-]$};
      \end{tikzpicture} 
  \end{subfigure}
  
  \caption{Dissolution level $d$ for distorted meshes at $t = 60$~\si{\second}. The vertical white line provides the analytical reference solution.}
  \label{fig:Ex1_dm}     
\end{figure}

Moreover, we investigate the influence of distorted meshes. To this end, the simulation utilizes three different meshes, where the bottom edge of the workpiece is discretized with $80$~elements and the top edge with $10$, $20$ and $40$~elements, thus, creating a distortion in the transition area (meshes: $10 \rightarrow 80$, $20 \rightarrow 80$, $40 \rightarrow 80$). Fig.~\ref{fig:Ex1_dm} shows the dissolution level $d$ for the distorted meshes after $60$~\si{\second} with $\dt = 0.01~\si{\second}$. The previously defined criterion $\Vco/\Vdis^\mathrm{FE} \leq 1~\si{\percent}$ holds. The mesh with $10 \rightarrow 80$~elements overestimates the dissolved volume by $ + \, 8.2~\si{\percent} $. However, the meshes with $20 \rightarrow 80$ and $40 \rightarrow 80$~elements converge against the correct solution, overestimating the dissolved volume by just $ + \, 3.6~\si{\percent} $ and $ + \, 3.0~\si{\percent} $. Thus, a dissolving transition zone with extremely different element sizes should be avoided, whereas, moderately distorted meshes yield good results.

In summary, this example proves the model's capability to achieve reasonable results with a coarse mesh at large time steps, and, further, to accurately model material dissolution with fine meshes and time increments.

%----------------------------------------------------------------------------------------------------------------------------------%
\subsection{Planar specimen - experimental validation}
\label{ssec:Ex2}

\begin{figure}[htbp] 
  \centering 
  \begin{subfigure}{.45\textwidth} 
      
    \centering 
      
    \begin{tikzpicture} 
      \node[inner sep=0pt] (pic) at (0,0) {\includegraphics[width=\textwidth]
      {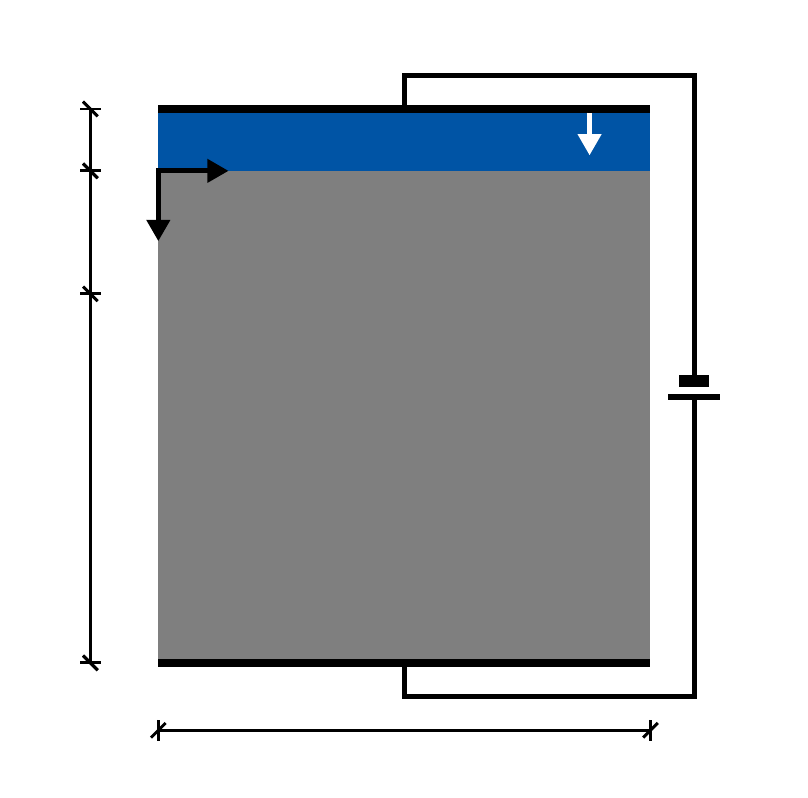}};
      \node[inner sep=0pt] (075l) at ($(pic.west) +( 0.20,-0.60)$)  {$0.75~l$};
      \node[inner sep=0pt] (025l) at ($(075l.north)+(0.00, 1.95)$)  {$0.25~l$};
      \node[inner sep=0pt] (s)    at ($(025l.north)+(0.40, 0.65)$)  {$s$};
      \node[inner sep=0pt] (x)    at ($(pic.north)+(-1.40,-1.70)$)  {$x$};
      \node[inner sep=0pt] (y)    at ($(x.north)  +(-0.55,-0.60)$)  {$y$};
      \node[inner sep=0pt] (xdc)  at ($(pic.east) +(-2.35, 2.25)$)  {\textcolor{white}{$\dot{x}_\mathrm{ca}$}};
      \node[inner sep=0pt] (dv)   at ($(pic.east) +(-0.35, 0.12)$)  {$\delv$};
      \node[inner sep=0pt] (l)    at ($(pic.south)+( 0.00, 0.30)$)  {$l$};
    \end{tikzpicture} 
         
    \vspace{-3mm}
    \caption{Geometry}
    \label{fig:Ex2Geom}
         
  \end{subfigure}
  \quad
  \begin{subfigure}{.45\textwidth} 
    \centering 
    \begin{tikzpicture} 
      \node[inner sep=0pt] (pic) at (0,0) {\includegraphics[width=\textwidth]
      {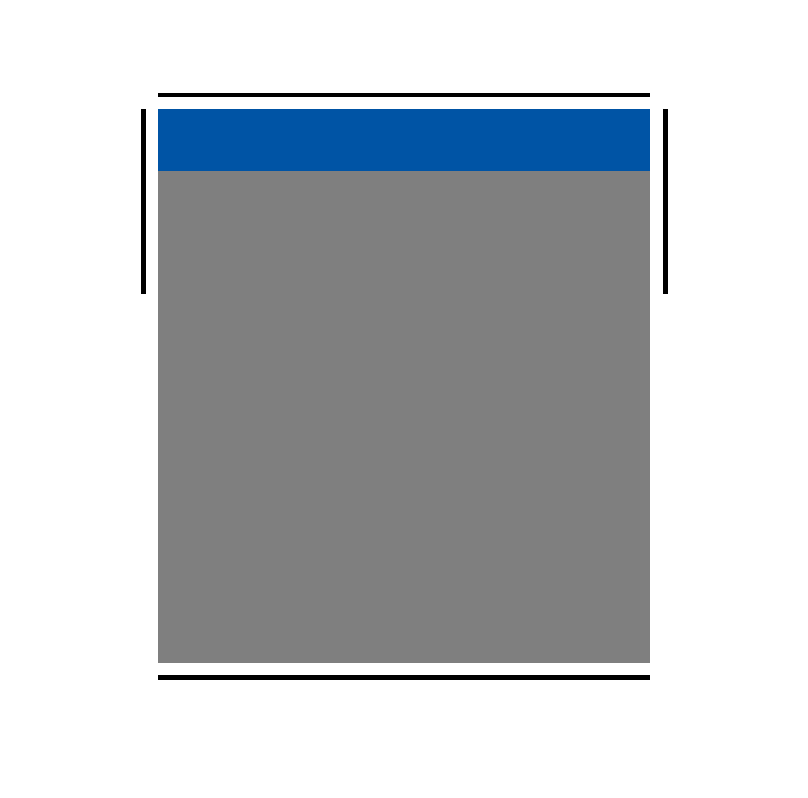}};
      \node[inner sep=0pt] (van)   at ($(pic.south)+( 0.00, 0.70)$)  {$\widetilde{v}_\mathrm{an}$};
      \node[inner sep=0pt] (vca)   at ($(pic.north)+( 0.00,-0.45)$)  {$\widetilde{v}_\mathrm{ca}(t)$};
      \node[inner sep=0pt] (tin)   at ($(pic.west) +( 0.85, 1.75)$)  {$\widetilde{\theta}_\mathrm{in}$};
      \node[inner sep=0pt] (tout)  at ($(tin.east) +( 5.40, 0.00)$)  {$\widetilde{\theta}_\mathrm{out}$};
    \end{tikzpicture} 

    \vspace{-3mm}
    \caption{BVP}
    \label{fig:Ex2BVP}
  \end{subfigure} 
  \caption{Geometry and boundary value problem of planar specimen (cf.~\cite{BergsRommesEtAl2019}).} 
  \label{fig:Ex2}
     
\end{figure}
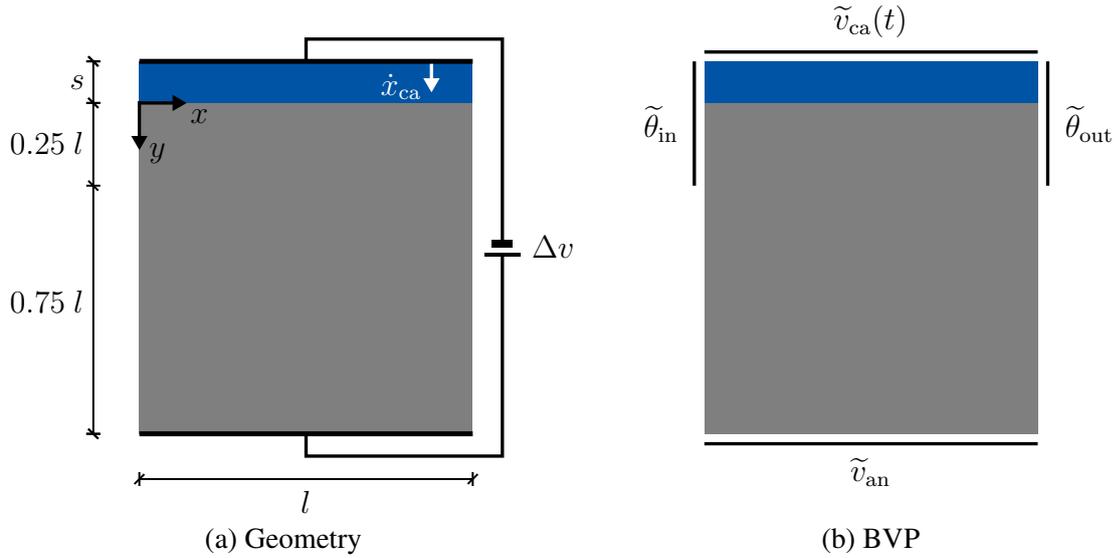

The second example focuses on validating the model's performance by means of the  investigations of \cite{BergsRommesEtAl2019}, who electrochemically machine a planar specimen with $l = 7 \, \si{\milli\meter}$ (Fig.~\ref{fig:Ex2Geom}). The experiment yields an electrolyte's inflow temperature of $\widetilde{\theta}_\mathrm{in} = 298.15 \, \si{\kelvin}$. Since the outflow temperature is not determined in this experiment, we assume $\widetilde{\theta}_\mathrm{out} = 308.15 \, \si{\kelvin}$ due to Joule heating in the electrolyte in accordance with \cite{Zeis2015}. We prescribe the inflow temperature at $x = 0 \, \si{\milli\meter}$ and $y \in [ \, -s, \, 0.25 \, l \, ]$ and the outflow temperature at $x = l$ and $y \in [ \, -s, \, 0.25 \, l \, ]$ (Fig.~\ref{fig:Ex2BVP}). For simplicity, we prescribe the temperature at $y \in [ \, -s, \, 0.25 \, l \, ]$ during the entire simulation to account for the movement of the electrolyte's position. Motivated by \cite{Zeis2015} and \cite{Harst2019}, we alter the experimental voltage of $\delv = 15~\si{\volt}$ in the simulation by a global reduction of $3~\si{\volt}$ to take a polarization voltage $\delvpol$ at anode and cathode into account. The experimental feed rate is $\dot{x}_\mathrm{ca} = 1~\si{\milli\meter\per\minute}$. For the determination of the working gap width, we compute the electrolyte's electric conductivity to $\kE\EL = ( ( \widetilde{\theta}_\mathrm{in} + \widetilde{\theta}_\mathrm{out} ) / 2 ) = 13.7~\si{\ampere\per\volt\per\meter}$ and, using Eq.~\eqref{eqn:gapwidth}, obtain $s = 0.36~\si{\milli\meter}$. We apply the same procedure as in Section~\ref{ssec:Ex1} to model the cathode's feed and compute $\widetilde{v}_\mathrm{ca}(t)$ using Eq.~(\ref{eqn:vcaoft}). Moreover, the thickness reads $g = 0.1~\si{\milli\meter}$ and the time increment is $\dt = 0.01~\si{\second}$.

\begin{figure}[htbp] 
     \centering 
     \begin{subfigure}{.32\textwidth} 
         \centering 
         \includegraphics[width=\textwidth]{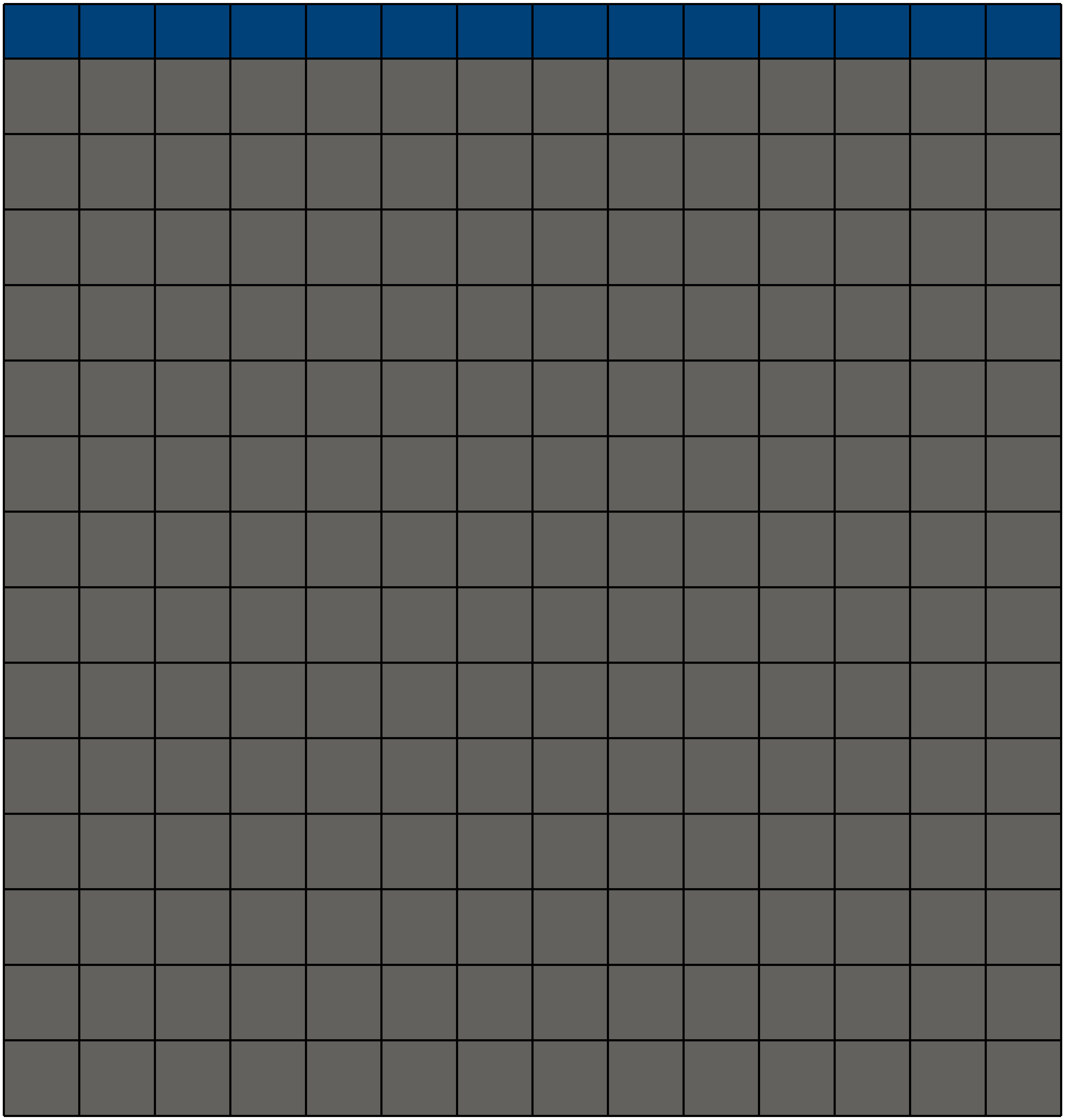}
         \caption{Mesh~1 (210 elements)}
         \label{fig:Ex2_mesh1}
     \end{subfigure}
     \hspace{12mm}
     \begin{subfigure}{.32\textwidth} 
         \centering 
         \includegraphics[width=\textwidth]{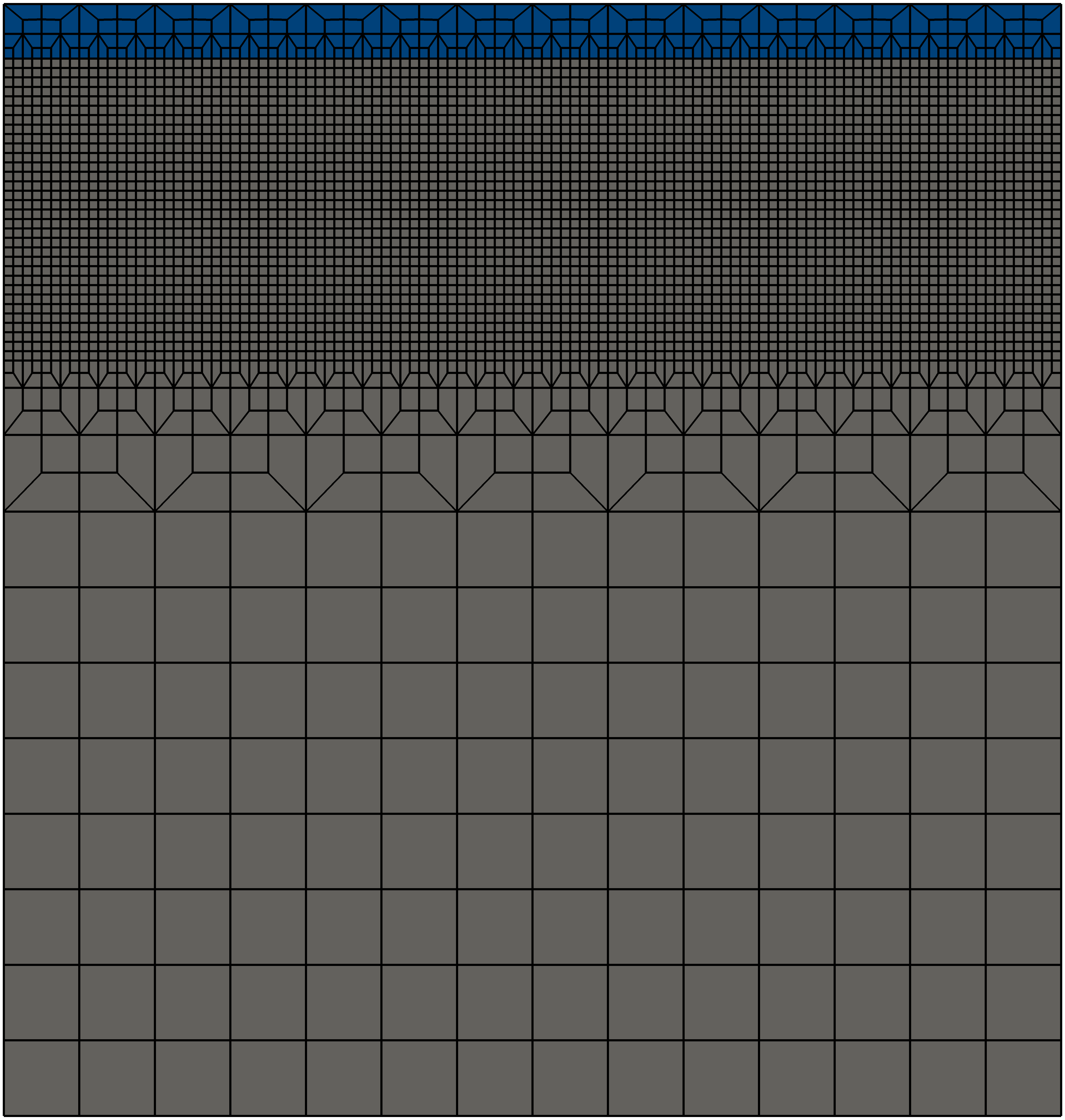}
         \caption{Mesh~4 (4242 elements)}
         \label{fig:Ex2_mesh4}
     \end{subfigure}
     \caption{Coarsest and finest mesh for planar specimen with mesh refinement at the workpiece's surface.}     
     \label{fig:Ex2_meshes}
\end{figure}

The simulation utilizes four different meshes starting with a coarse, structured mesh (Fig.~\ref{fig:Ex2_mesh1}) followed by a successive mesh refinement in the area of interest (Fig.~\ref{fig:Ex2_mesh4}).

\begin{figure}[htbp]
  \centering
  \includegraphics{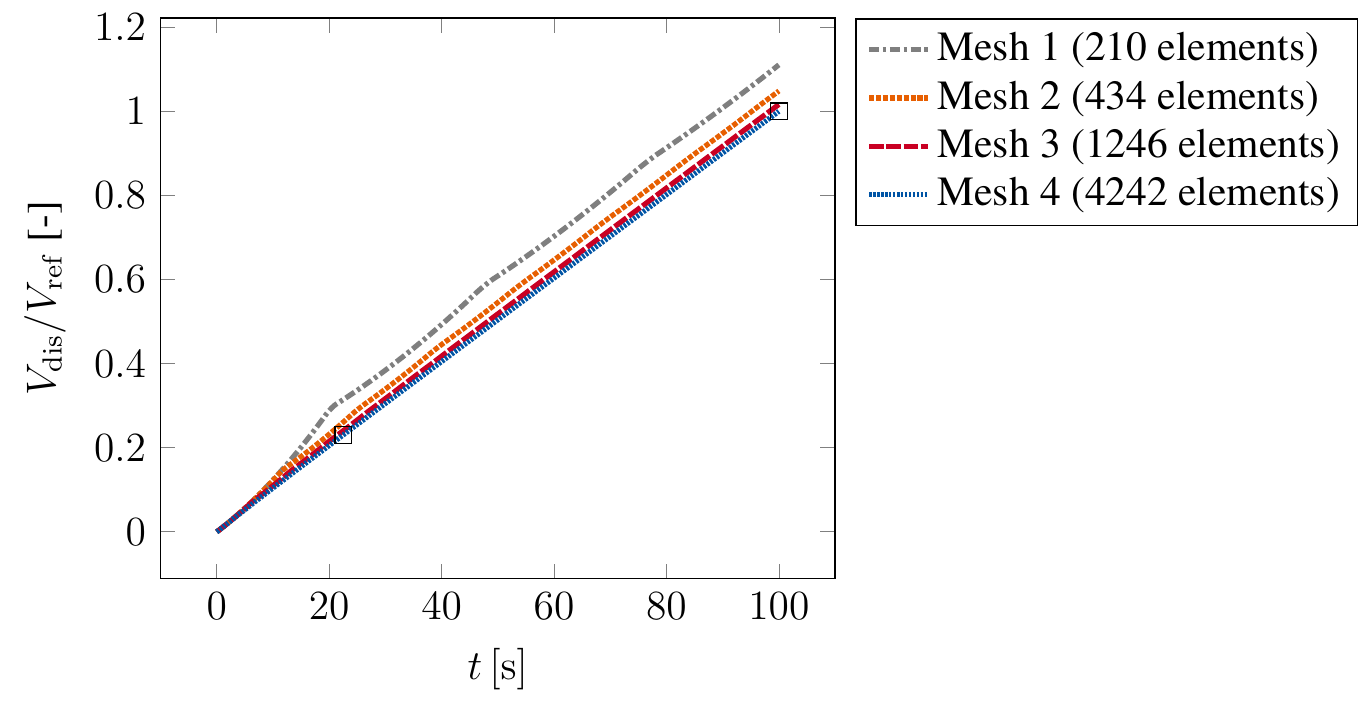}
  \caption{Evolution of dissolved volume $\Vdis$ over machining time $t$ of planar specimen for different meshes.}
  \label{fig:Ex2_Convergence}
\end{figure}

\begin{figure}[htbp] 
     \centering 
     \begin{subfigure}{.32\textwidth} 
         \centering
         \begin{tikzpicture} 
           \node[inner sep=0pt] (pic) at (0,0) {\includegraphics[ width=\textwidth ]
           {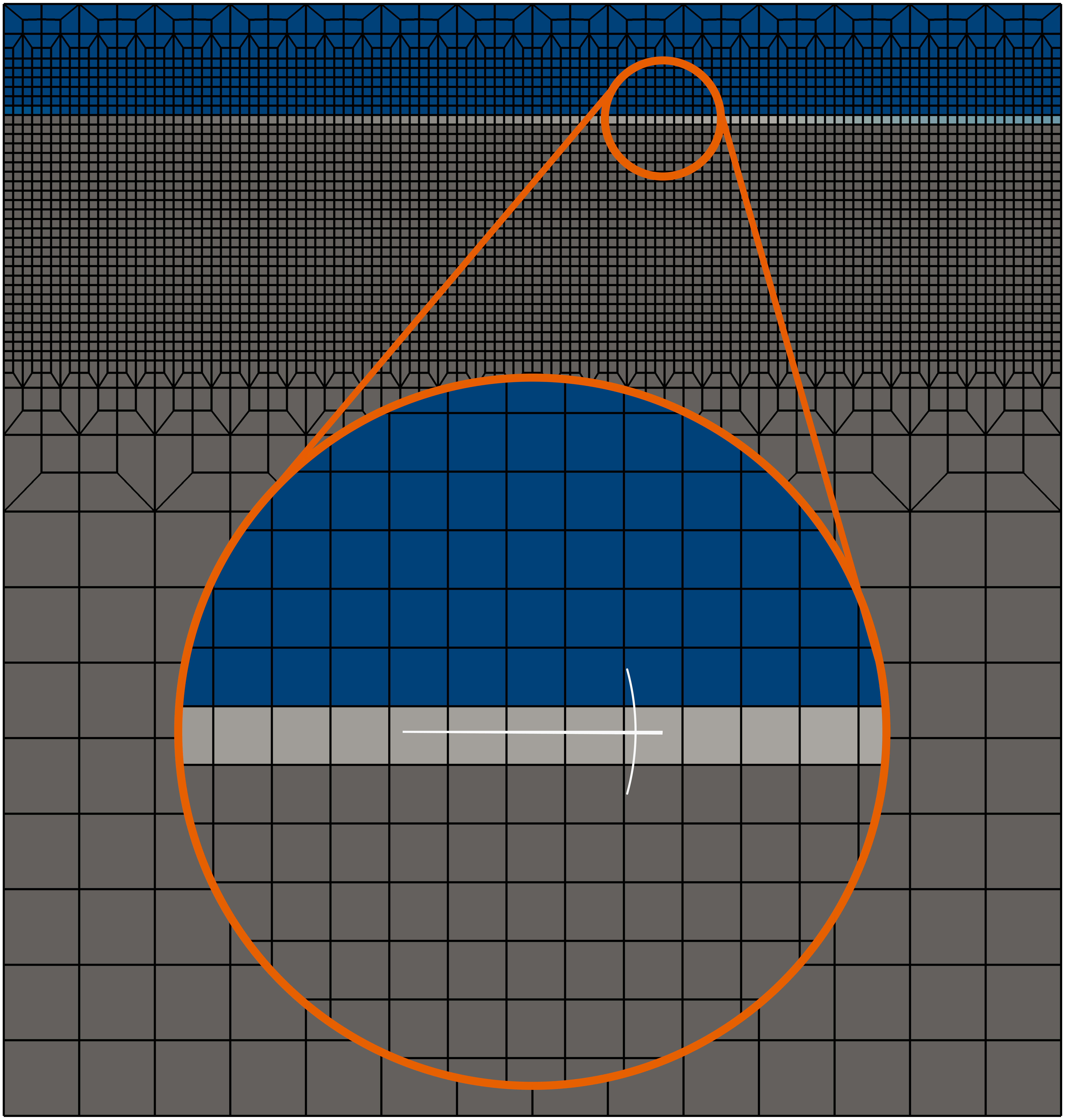}};
           \node[inner sep=0pt] (be) at ( 0.50,-1.50)  {\textcolor{white}{$\beta_1$}};
         \end{tikzpicture}
         \caption{Small machining depth}
         \label{fig:Ex2_dlmd}
     \end{subfigure}
     \hspace{4mm}
     \begin{subfigure}{.32\textwidth} 
         \centering
         \begin{tikzpicture} 
           \node[inner sep=0pt] (pic) at (0,0) {\includegraphics[ width=\textwidth ]
           {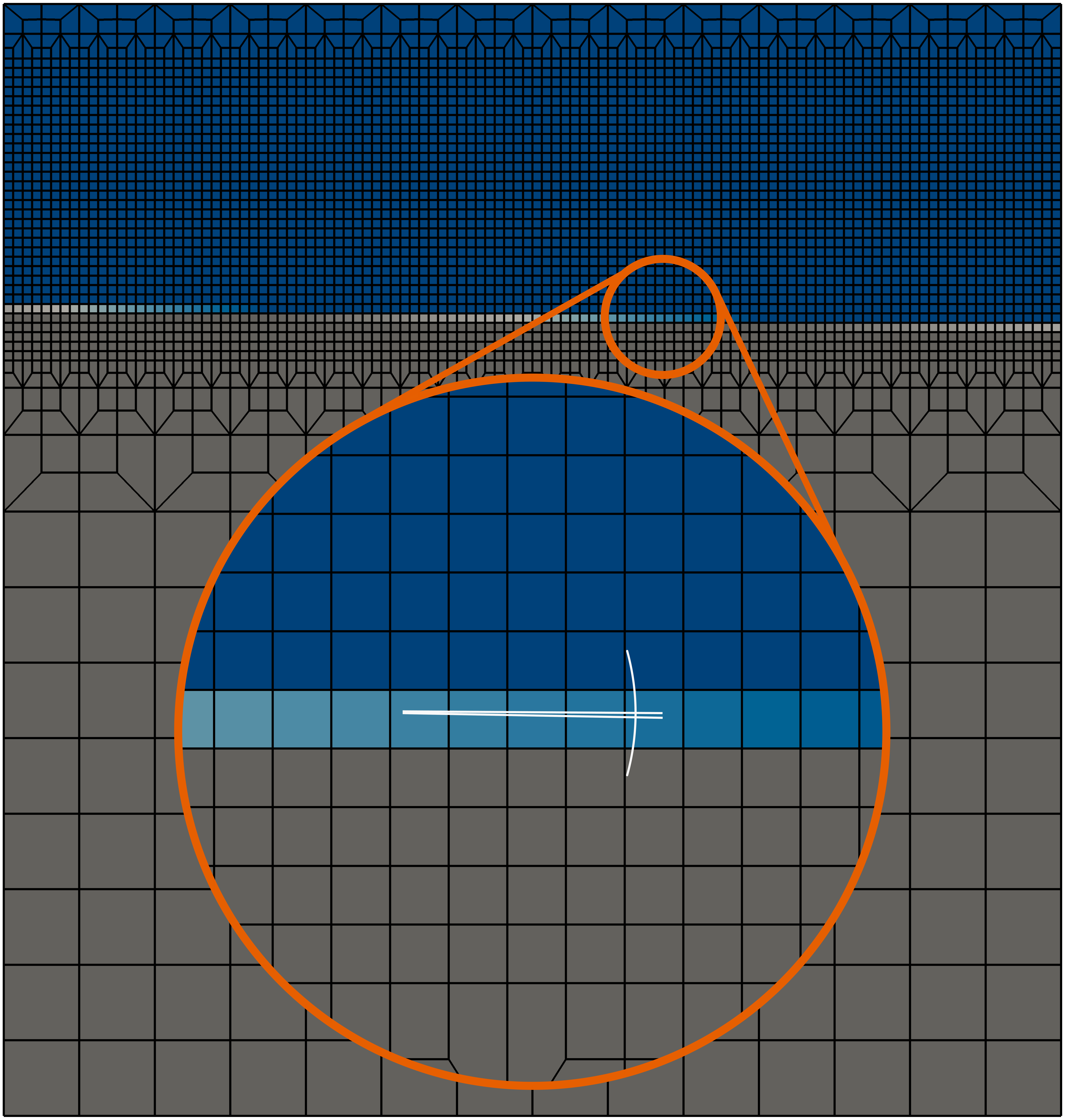}};
           \node[inner sep=0pt] (be) at ( 0.50,-1.50)  {\textcolor{white}{$\beta_2$}};
         \end{tikzpicture}         
         \caption{Large machining depth}
         \label{fig:Ex2_dhmd}
     \end{subfigure}
     \hspace{7mm}
     \begin{subfigure}{.10\textwidth} 
         \centering 
         \begin{tikzpicture} 
           \node[inner sep=0pt] (pic) at (0,0) {\includegraphics[height=40mm, width=5mm]
           {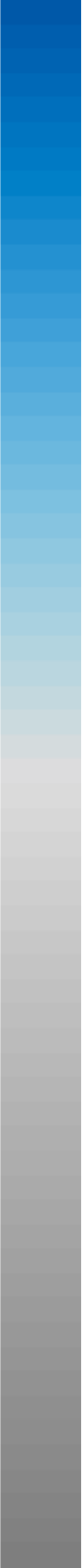}};
           \node[inner sep=0pt] (0)   at ($(pic.south)+( 0.50, 0.05)$)  {$0$};
           \node[inner sep=0pt] (1)   at ($(pic.south)+( 0.50, 4.00)$)  {$1$};
           \node[inner sep=0pt] (d)   at ($(pic.south)+( 0.00, 4.60)$)  {$d \, [-]$};
         \end{tikzpicture} \\
         \vspace{6mm}
     \end{subfigure}     
     
     \caption{Dissolution level $d$ of planar specimen at different machining depths. The inclination $\beta$ of the machined surface increases with the machining depth $\left( \beta_2 \approx 1.04\si{\degree} > \beta_1 \approx 0.38\si{\degree} \right)$.}
     \label{fig:Ex2_dsln}
\end{figure}
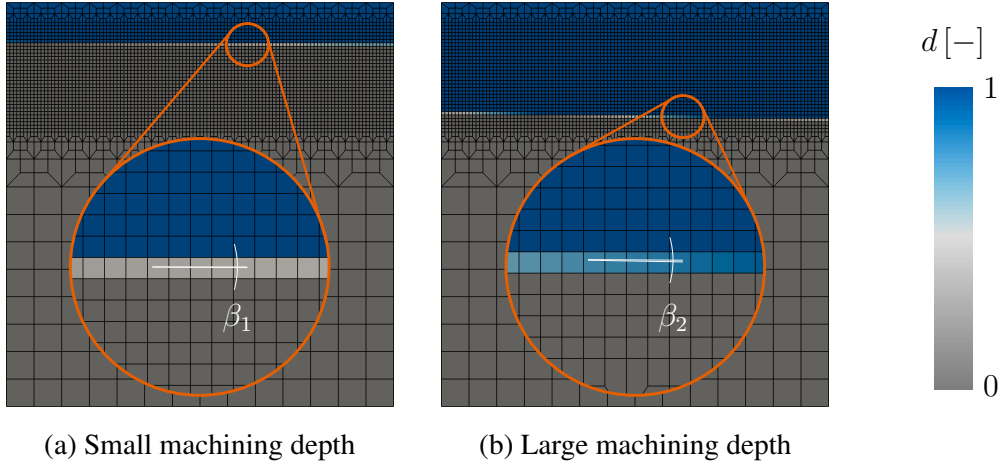

Fig.~\ref{fig:Ex2_Convergence} shows the evolution of the dissolved volume $\Vdis$ of Mesh~1~-~4 normalized to $ \Vref = 120.3~\si{\cubic\milli\meter} $, the dissolved volume of Mesh~4 at $t = 100~\si{\second}$. Only marginal differences between Mesh~3 and Mesh~4 are visible and, therefore, convergence is assumed. The black boxes indicate the snapshots of the dissolution level $d$ for Mesh~4 in Fig.~\ref{fig:Ex2_dsln}. The error indicator is $\Vco/\Vdis^\mathrm{FE} = 0.6~\si{\percent}$ for the finest mesh.

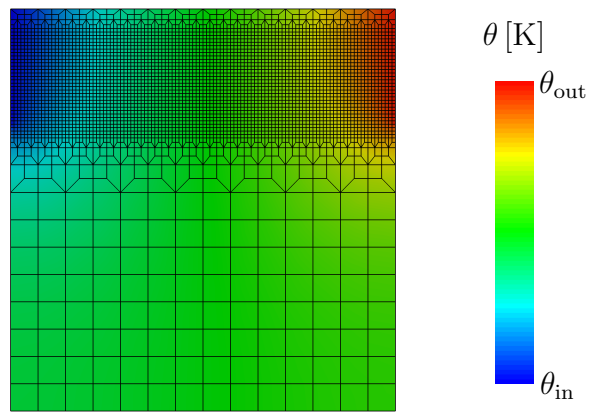
\begin{figure}[htbp] 
  \centering 
  \hspace{23mm}
  \begin{subfigure}{.32\textwidth} 
      \centering 
      \includegraphics[width=\textwidth]{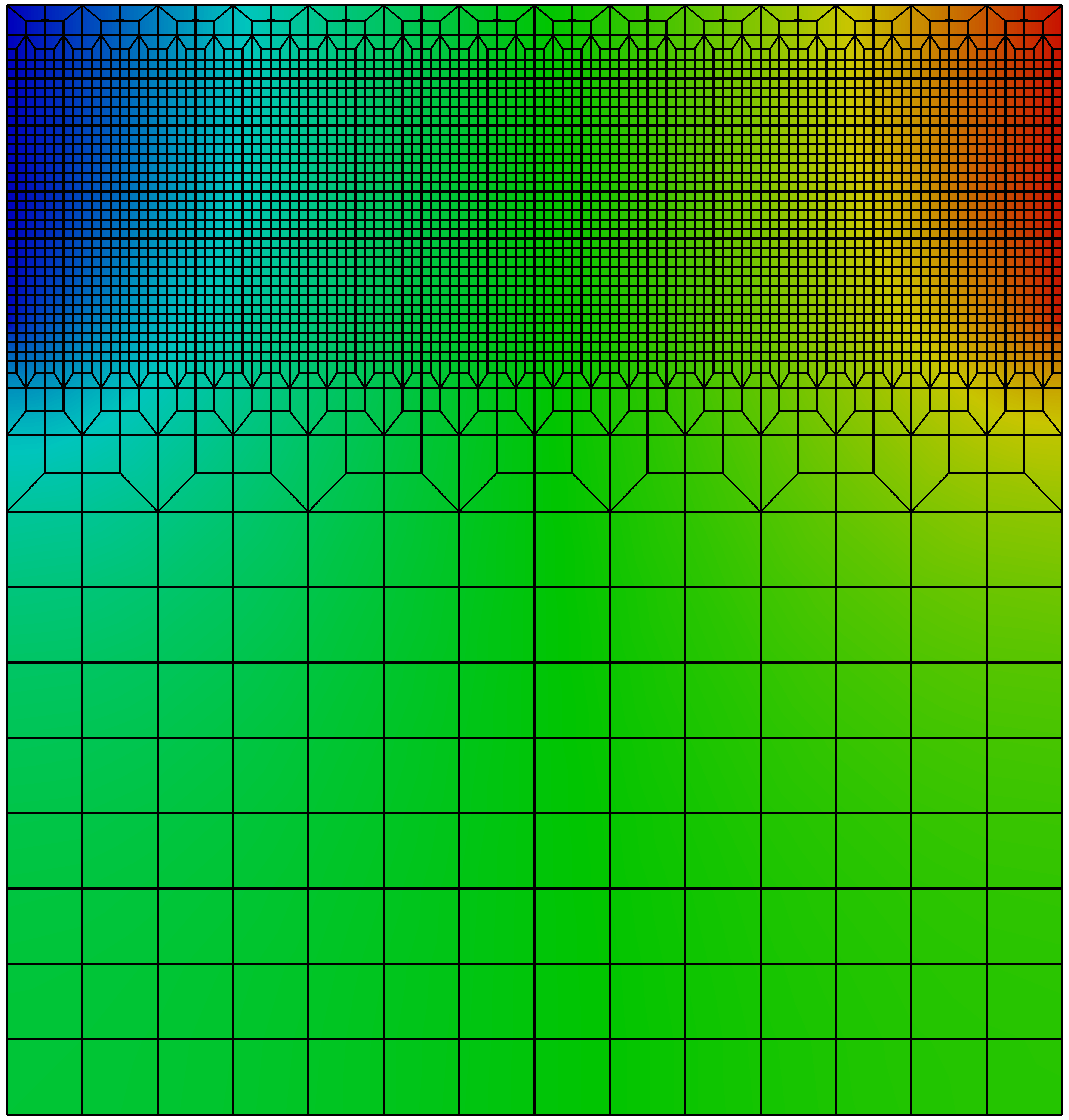}
  \end{subfigure}
  \hspace{8mm}
  \begin{subfigure}{.10\textwidth} 
      \centering 
      \begin{tikzpicture} 
        \node[inner sep=0pt] (pic) at (0,0) {\includegraphics[height=40mm, width=5mm]
        {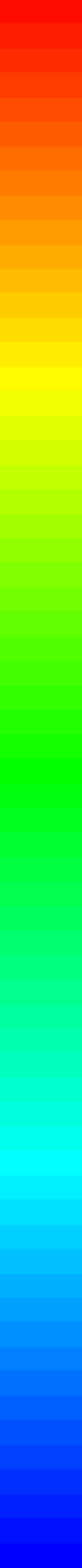}};
        \node[inner sep=0pt] (tin) at ($(pic.south)+( 0.57,-0.02)$)  {$\theta_\mathrm{in}$};
        \node[inner sep=0pt] (tout)at ($(pic.south)+( 0.65, 3.97)$)  {$\theta_\mathrm{out}$};
        \node[inner sep=0pt] (T)   at ($(pic.south)+( 0.00, 4.60)$)  {$\theta \, [\si{\kelvin}]$};
      \end{tikzpicture} \\
      \vspace{0mm}
  \end{subfigure}  
  
  \caption{Distribution of temperature $\theta$ for prescribed in- and outflow temperature.}     
  \label{fig:Ex2_Theta}
  
\end{figure}

Furthermore, Fig.~\ref{fig:Ex2_Theta} shows the temperature distribution which results from the assumed boundary conditions. It increases along the working gap and, thereby, models Joule heating in the electrolyte. In experiments, the inflow-temperature usually matches the workpiece's temperature, which differs in the simulation due to the simplified boundary conditions.

As reported in e.g.~\cite{Zeis2015}, the electrolyte's electric conductivity increases with temperature. Therefore, also the electric current density increases along the working gap. Hence, the simulation yields an increased removal of material and widening%
\interfootnotelinepenalty=10000 % prevents split of footnote
\footnote{The established method for modeling the cathode's feed from Section~\ref{ssec:Ex1} was derived for plane parallel electrode surfaces. Here, this assumption is violated. However, due to close correlation with the experimental results, this appears tolerable.} of the working gap.
\interfootnotelinepenalty=100
For a small machining depth, the anode's surface is slightly inclined (Fig.~\ref{fig:Ex2_dlmd}), but for a large machining depth, a pronounced inclination and working gap widening occurs (Fig.~\ref{fig:Ex2_dhmd}). These results show satisfactory agreement with the investigations of \cite{BergsRommesEtAl2019}. However, \cite{BergsRommesEtAl2019} observe a slightly higher inclination of the machined surface than $\beta_2 \approx 1.04\si{\degree}$ for a large machining depth. Due to the lack of detailed information about the temperature distribution in the machining gap, the numerical results may vary.

%----------------------------------------------------------------------------------------------------------------------------------%
\subsection{Curved specimen with elevation - experimental validation}
\label{ssec:Ex3}

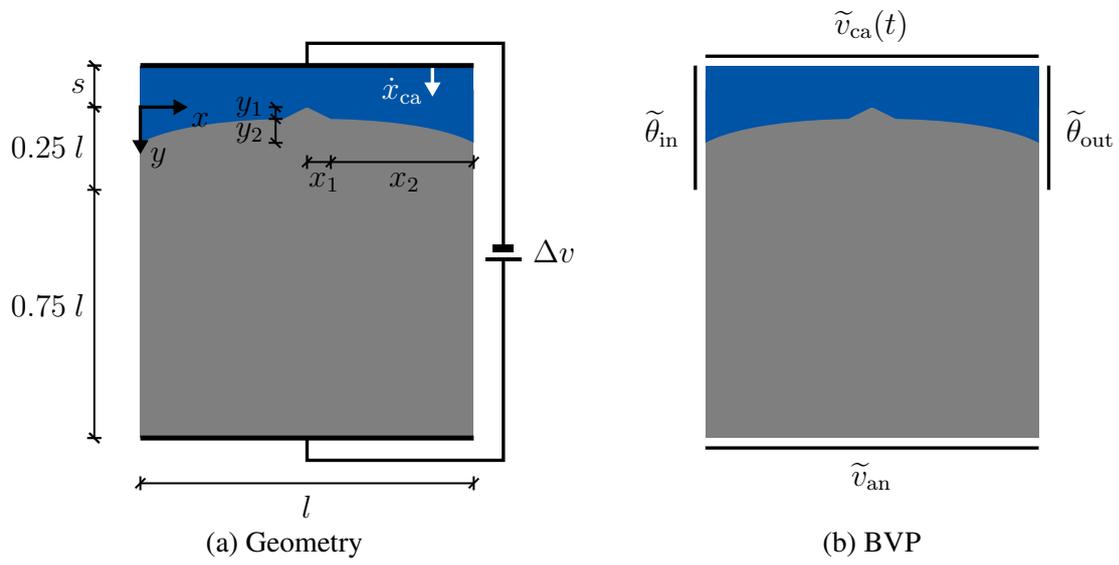
\begin{figure}[htbp] 
     \centering 
     \begin{subfigure}{.45\textwidth} 
         
         \centering 
         
         \begin{tikzpicture} 
           \node[inner sep=0pt] (pic) at (0,0) {\includegraphics[width=\textwidth]
           {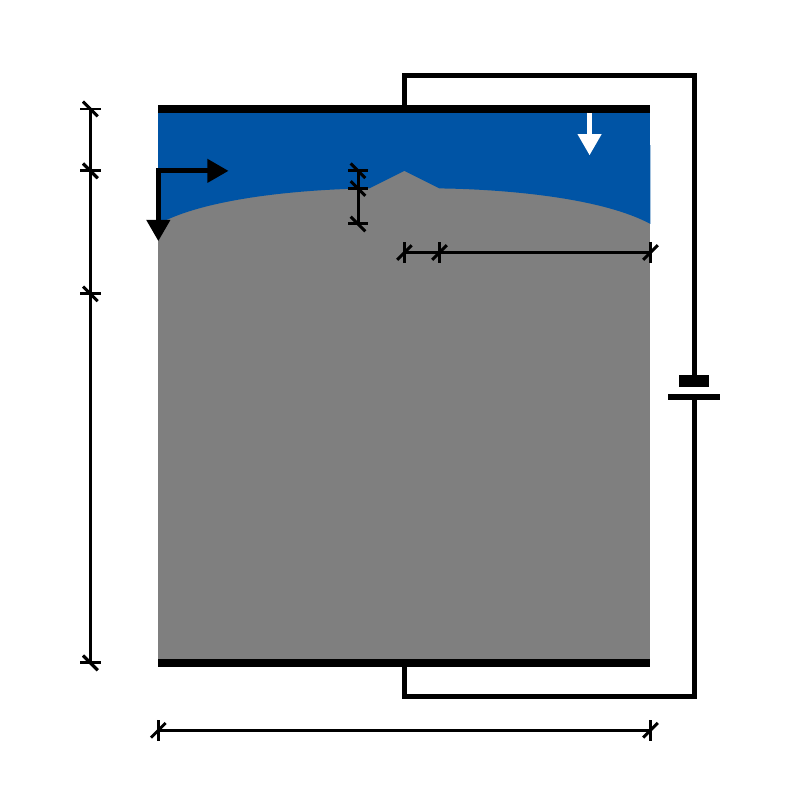}};
           \node[inner sep=0pt] (075l) at ($(pic.west) +( 0.20,-0.60)$)  {$0.75~l$};
           \node[inner sep=0pt] (025l) at ($(075l.north)+(0.00, 1.95)$)  {$0.25~l$};
           \node[inner sep=0pt] (s)    at ($(025l.north)+(0.40, 0.65)$)  {$s$};
           \node[inner sep=0pt] (x)    at ($(pic.north)+(-1.40,-1.70)$)  {$x$};
           \node[inner sep=0pt] (y)    at ($(x.north)  +(-0.55,-0.60)$)  {$y$};
           \node[inner sep=0pt] (x1)   at ($(pic.north)+( 0.23,-2.52)$)  {$x_1$};
           \node[inner sep=0pt] (x2)   at ($(x1.east)  +( 0.85, 0.00)$)  {$x_2$};
           \node[inner sep=0pt] (y1)   at ($(pic.north)+(-0.75,-1.53)$)  {$y_1$};
           \node[inner sep=0pt] (y2)   at ($(y1.south) +( 0.00,-0.20)$)  {$y_2$};
           \node[inner sep=0pt] (xdc)  at ($(pic.east) +(-2.35, 2.25)$)  {\textcolor{white}{$\dot{x}_\mathrm{ca}$}};
           \node[inner sep=0pt] (dv)   at ($(pic.east) +(-0.35, 0.12)$)  {$\delv$};
           \node[inner sep=0pt] (l)    at ($(pic.south)+( 0.00, 0.30)$)  {$l$};
         \end{tikzpicture} 
         
         \vspace{-3mm}
         \caption{Geometry}
         \label{fig:Ex3Geom}
         
     \end{subfigure}
\quad
     \begin{subfigure}{.45\textwidth} 

         \centering 
         
         \begin{tikzpicture} 
           \node[inner sep=0pt] (pic) at (0,0) {\includegraphics[width=\textwidth]
           {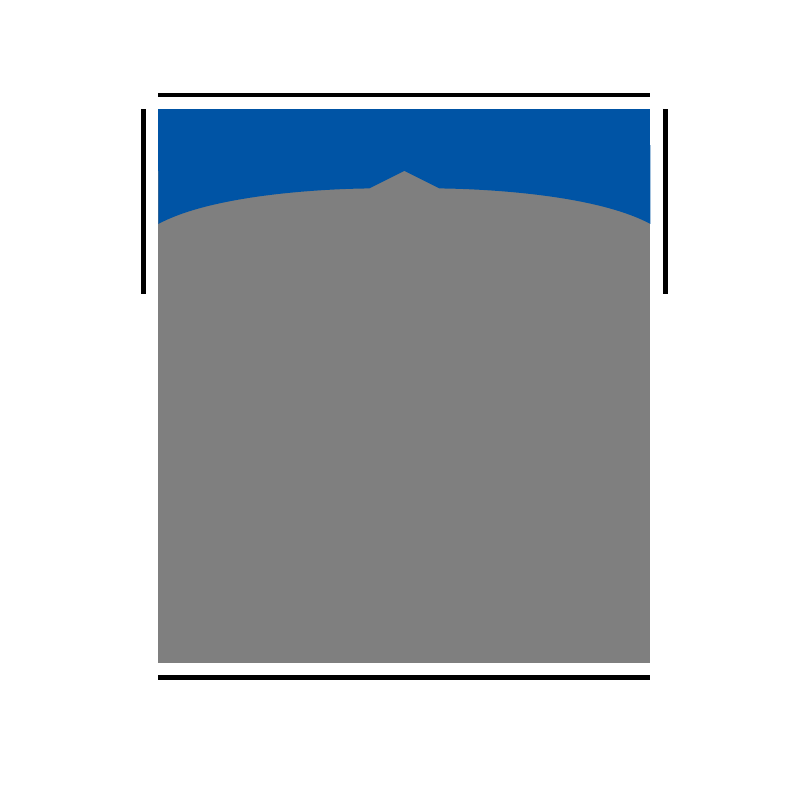}};
           \node[inner sep=0pt] (van)   at ($(pic.south)+( 0.00, 0.70)$)  {$\widetilde{v}_\mathrm{an}$};
           \node[inner sep=0pt] (vca)   at ($(pic.north)+( 0.00,-0.45)$)  {$\widetilde{v}_\mathrm{ca}(t)$};
           \node[inner sep=0pt] (tin)   at ($(pic.west) +( 0.85, 1.75)$)  {$\widetilde{\theta}_\mathrm{in}$};
           \node[inner sep=0pt] (tout)  at ($(tin.east) +( 5.40, 0.00)$)  {$\widetilde{\theta}_\mathrm{out}$};
         \end{tikzpicture} 
         
         \vspace{-3mm}
         \caption{BVP}
         \label{fig:Ex3BVP}
         
     \end{subfigure} 
     
     \caption{Geometry and boundary value problem of curved specimen with elevation (cf.~\cite{BergsRommesEtAl2019}).} 
     \label{fig:Ex3}
     
\end{figure}

This example investigates a specimen with a non-planar surface, which is curved and possesses an elevation. Thus, inhomogeneous material removal at the beginning of the ECM-process can be examined. \cite{BergsRommesEtAl2019} also study this geometry and the parameters read $x_1 = 0.5~\si{\milli\meter}$, $x_2 = 3~\si{\milli\meter}$, $y_1 = 0.25~\si{\milli\meter}$, $y_2 = 0.5~\si{\milli\meter}$ and $l = 7~\si{\milli\meter}$ (Fig.~\ref{fig:Ex3Geom}). A parabola defines the curvature with the vertex of the parabola at position $( 1.5~\si{\milli\meter} \, | \, 0.3~\si{\milli\meter} )$. The gap width $s = 0.36~\si{\milli\meter}$ is adopted from Section~\ref{ssec:Ex2}. The remaining process parameters and boundary conditions%
\footnote{Similar to Section~\ref{ssec:Ex2}, the cathode's feed is approximated, since the anode's surface is not planar. Nevertheless, high consistency with experimental results is obtained.} are analogous to Section~\ref{ssec:Ex2} (Fig.~\ref{fig:Ex3BVP}).

\begin{figure}[htbp] 
     \centering 

     \includegraphics[width=0.32\textwidth]{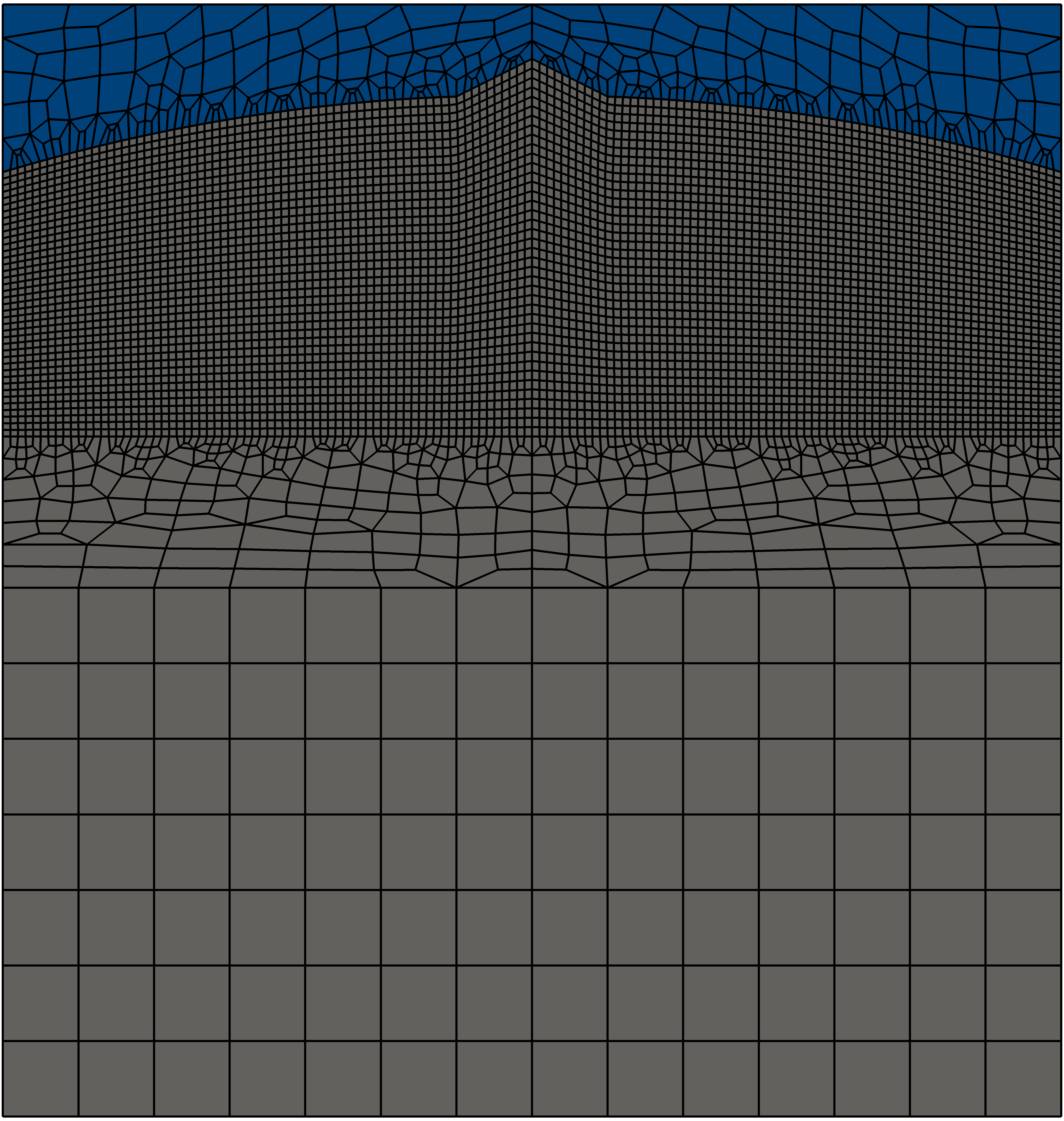}

     \caption{Mesh (6448~elements) of curved specimen with elevation.}     
     \label{fig:Ex3_mesh}
     
\end{figure} 

Further, Fig.~\ref{fig:Ex3_mesh} shows the mesh which is employed in the simulation. The mesh possesses strong mesh refinement at the upper part of the workpiece where material dissolution occurs. The simulation yields $\Vco/\Vdis^\mathrm{FE} = 0.7~\si{\percent}$ at the end of the simulation and, therefore, mesh density and time step size fit.

\begin{figure}[htbp] 
  \centering 
  \hspace{12mm}
  \begin{minipage}{0.68\textwidth}
    \begin{subfigure}{.47\textwidth} 
      \centering 
      \includegraphics[width=\textwidth]{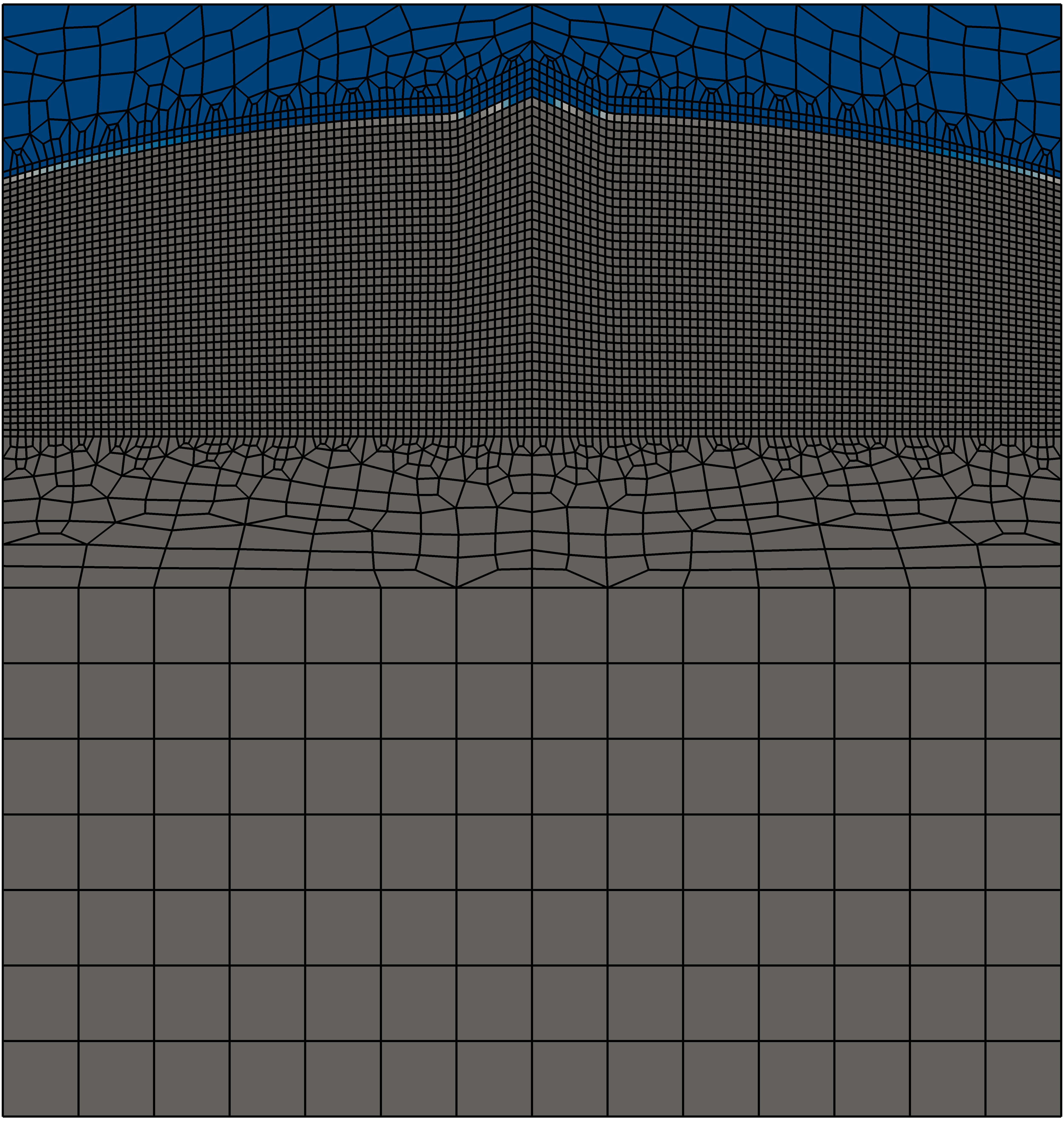}
      \caption{$t = 10$~\si{\second}}
      \label{fig:Ex3_t_10s}
    \end{subfigure}
    \hspace{4mm}
    \begin{subfigure}{.47\textwidth} 
      \centering 
      \includegraphics[width=\textwidth]{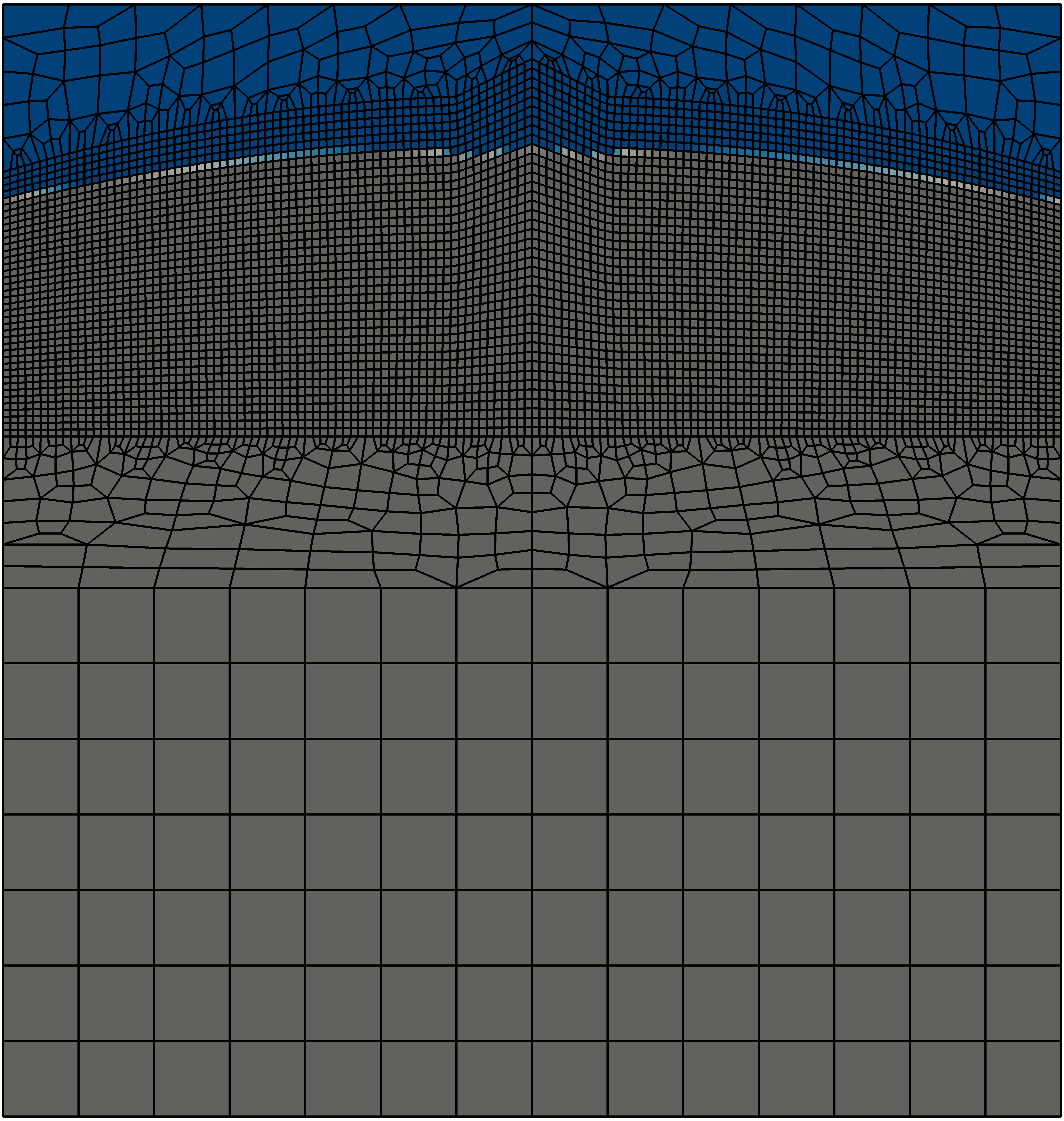}
      \caption{$t = 25$~\si{\second}}
      \label{fig:Ex3_t_25s}
    \end{subfigure} \\
    \vspace{2mm}
    
    \begin{subfigure}{.47\textwidth} 
        \centering 
        \includegraphics[width=\textwidth]{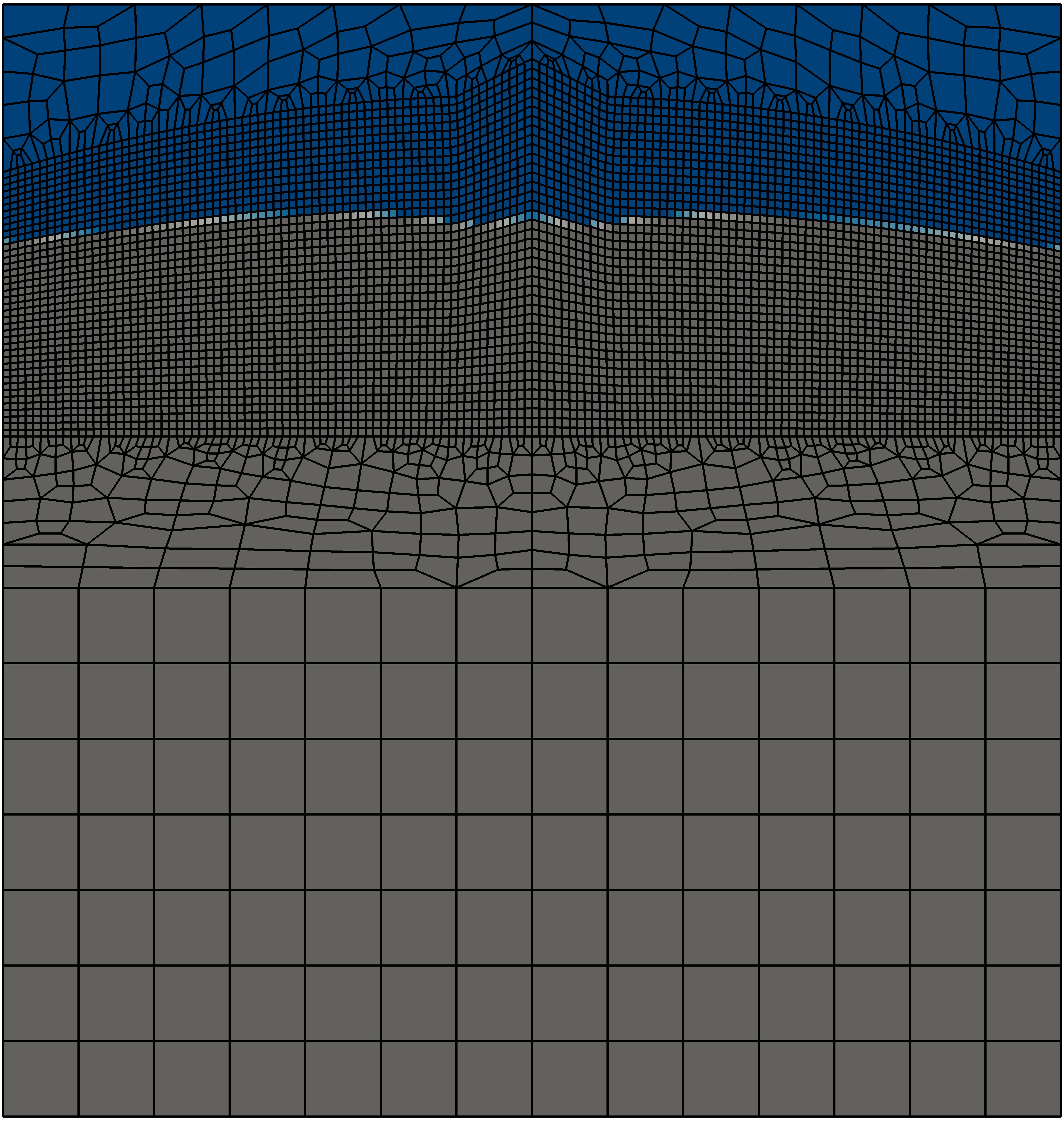}
        \caption{$t = 50$~\si{\second}}
        \label{fig:Ex3_t_50s}
    \end{subfigure}
    \hspace{4mm}
    \begin{subfigure}{.47\textwidth} 
        \centering 
        \includegraphics[width=\textwidth]{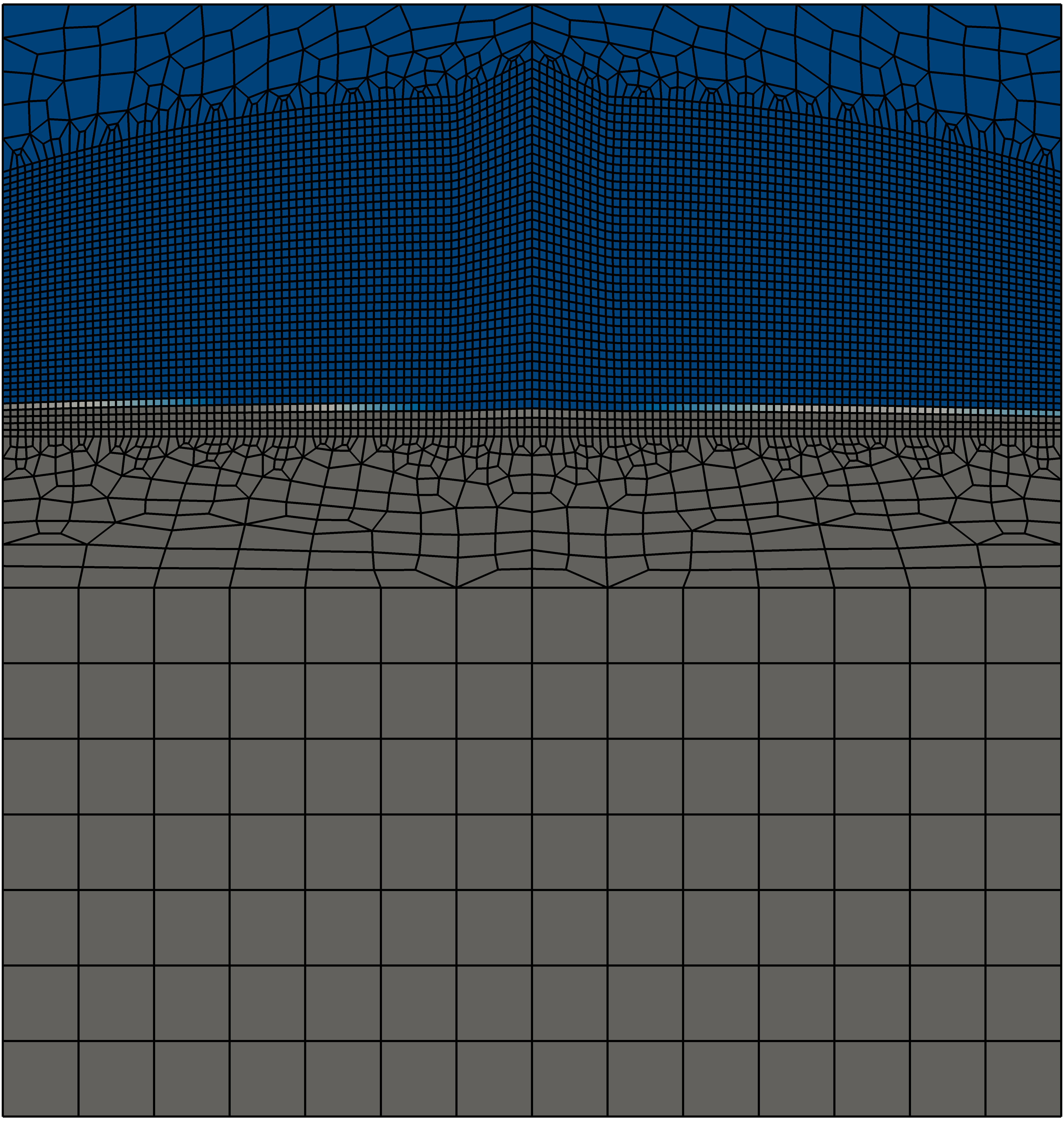}
        \caption{$t = 125$~\si{\second}}
        \label{fig:Ex3_t_125s}
    \end{subfigure}
  \end{minipage}
  \hspace{7mm}
  \begin{minipage}{0.1\textwidth}
    \begin{subfigure}{.10\textwidth} 
      \centering 
      \begin{tikzpicture} 
        \node[inner sep=0pt] (pic) at (0,0) {\includegraphics[height=40mm, width=5mm]
        {03_Contour/01_Examples/00_Color_Maps/01_dsln_ver.pdf}};
        \node[inner sep=0pt] (0)   at ($(pic.south)+( 0.50, 0.05)$)  {$0$};
        \node[inner sep=0pt] (1)   at ($(pic.south)+( 0.50, 4.00)$)  {$1$};
        \node[inner sep=0pt] (d)   at ($(pic.south)+( 0.00, 4.60)$)  {$d \, [-]$};
      \end{tikzpicture} \\
      \vspace{12mm}
    \end{subfigure}
  \end{minipage}
 
  \caption{Dissolution level $d$ at different machining times of curved specimen with elevation. The material removal focuses on the elevation before a level surface emerges.}
  \label{fig:Ex3_dsln}
\end{figure}

Fig.~\ref{fig:Ex3_dsln} shows the dissolution level $d$ after $10~\si{\second}$, $25~\si{\second}$, $50~\si{\second}$ and $125~\si{\second}$ machining time. In the initial stages of the process, increased material removal occurs at the tip of the specimen (Figs.~\ref{fig:Ex3_t_10s}~and~\ref{fig:Ex3_t_25s}). At this position, the electric field lines increase in density and, thus, lead to a higher electric current density and removal of the elevation. As the process continues, the specimen's curvature starts to smooth out (Fig.~\ref{fig:Ex3_t_50s}) until an approximately planar surface evolves (Fig.~\ref{fig:Ex3_t_125s}). To remove the curvature, we require a machining depth larger than two and a half times of the initial imperfection of the specimen. These results conform with the findings of \cite{BergsRommesEtAl2019}.

The example validates the model's capability to exactly simulate material dissolution in the electrochemical machining process.

%----------------------------------------------------------------------------------------------------------------------------------%
\subsection{Pulsed electrochemical machining}
\label{ssec:Ex4}

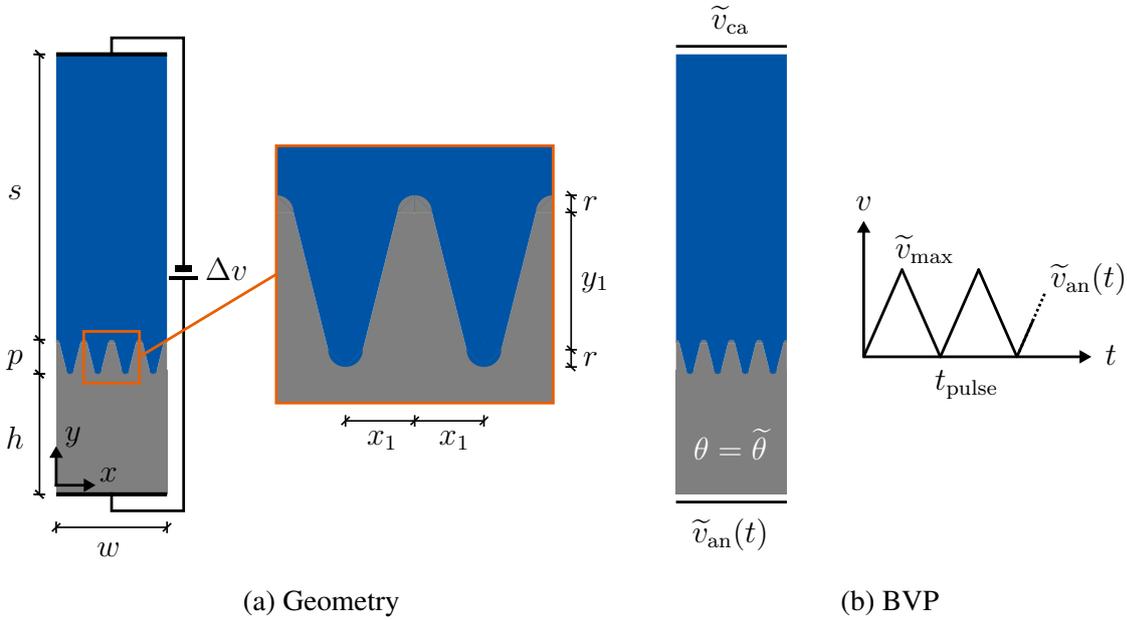
\begin{figure}[htbp] 
     \centering 
     \begin{subfigure}{.5344\textwidth} 
         
         \centering 
         
         \begin{tikzpicture} 
           \node[inner sep=0pt] (pic) at (0,0) {\includegraphics[width=\textwidth]
           {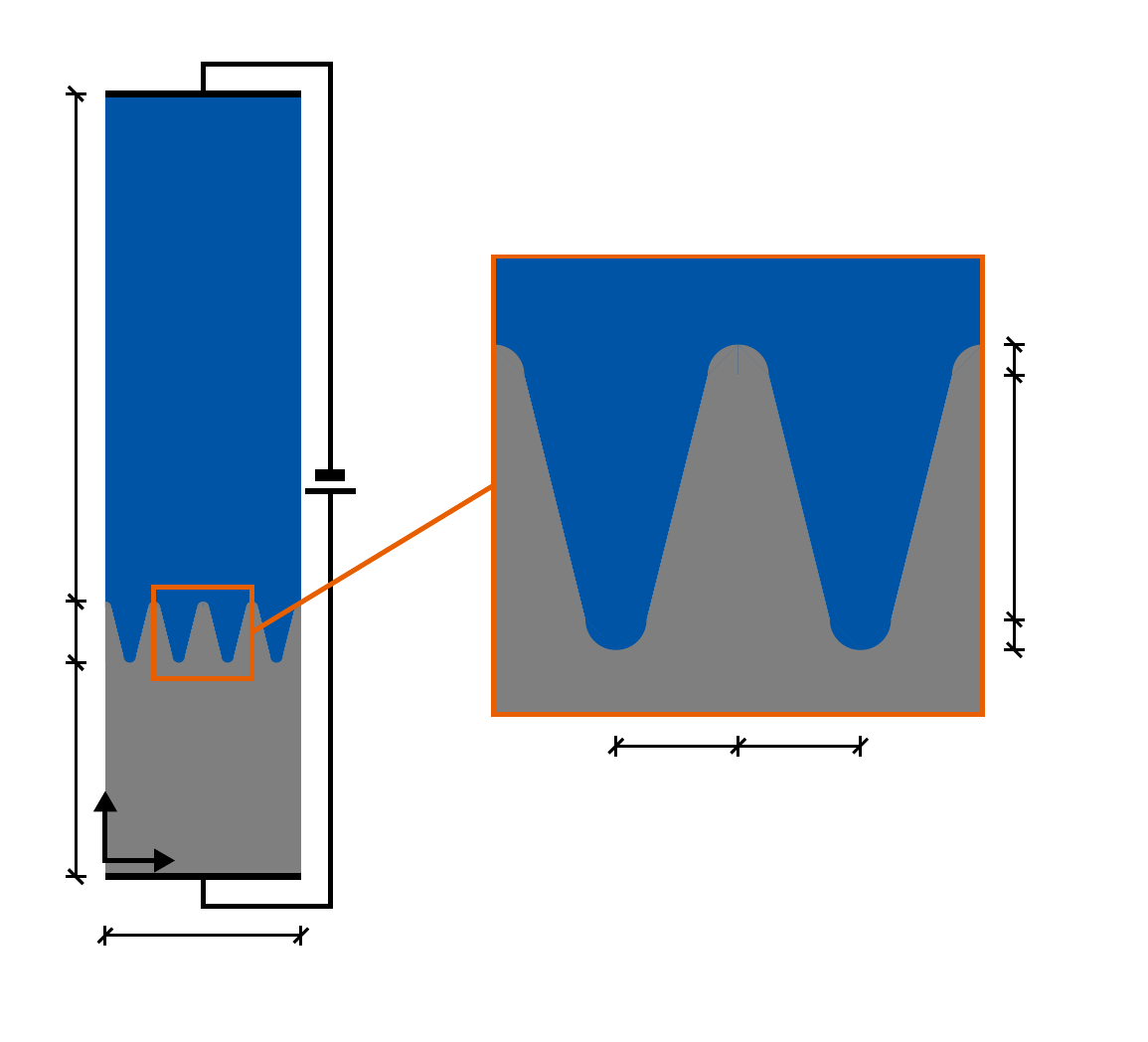}};
           \node[inner sep=0pt] (s)    at ($(pic.west) +( 0.25, 1.35)$)  {$s$};
           \node[inner sep=0pt] (p)    at ($(s.south)  +( 0.00,-2.12)$)  {$p$};
           \node[inner sep=0pt] (h)    at ($(s.south)  +( 0.00,-3.15)$)  {$h$};
           \node[inner sep=0pt] (x)    at ($(pic.south)+(-2.80, 1.50)$)  {$x$};
           \node[inner sep=0pt] (y)    at ($(x.north)  +(-0.47, 0.40)$)  {$y$};
           \node[inner sep=0pt] (w)    at ($(pic.south)+(-2.80, 0.50)$)  {$w$};
           \node[inner sep=0pt] (x1a)  at ($(pic.south)+( 0.80, 1.95)$)  {$x_1$};
           \node[inner sep=0pt] (x1b)  at ($(pic.south)+( 1.75, 1.95)$)  {$x_1$};
           \node[inner sep=0pt] (y1)   at ($(pic.south)+( 3.60, 4.05)$)  {$y_1$};
           \node[inner sep=0pt] (ra)   at ($(y1.north) +(-0.06, 0.90)$)  {$r$};
           \node[inner sep=0pt] (rb)   at ($(y1.south) +(-0.06,-0.88)$)  {$r$};
           \node[inner sep=0pt] (dv)   at ($(pic.north)+(-1.25,-3.55)$)  {$\delv$};
         \end{tikzpicture} 
         
         \vspace{-3mm}
         \caption{Geometry}
         \label{fig:Ex4Geom}
         
     \end{subfigure}%
     \hspace{-5mm}
     \begin{subfigure}{.45\textwidth} 

         \centering 
         
         \begin{tikzpicture} 
           \node[inner sep=0pt] (pic) at (0,0) {\includegraphics[width=\textwidth]
           {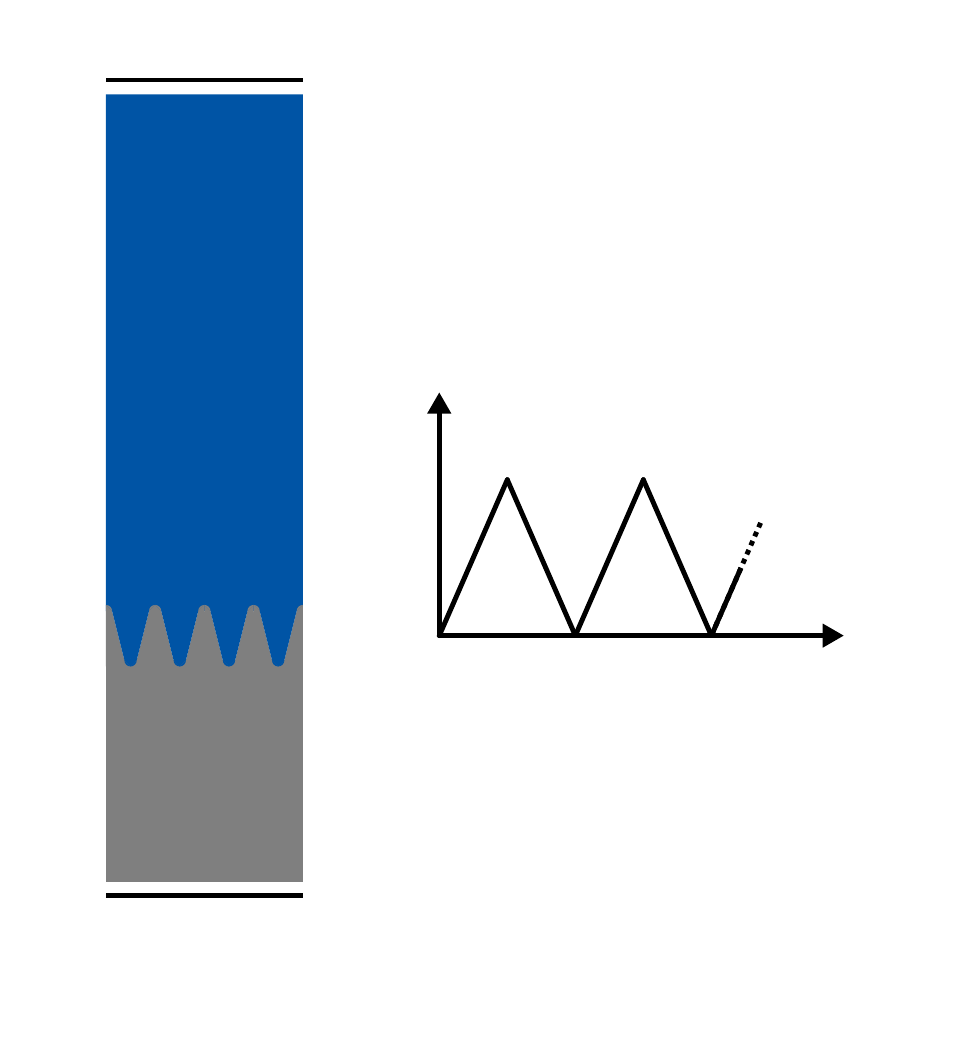}};
           \node[inner sep=0pt] (van)  at ($(pic.south)+(-2.10, 0.70)$)  {$\widetilde{v}_\mathrm{an}(t)$};
           \node[inner sep=0pt] (vca)  at ($(pic.north)+(-2.10,-0.25)$)  {$\widetilde{v}_\mathrm{ca}$};
           \node[inner sep=0pt] (t)    at ($(pic.south)+(-2.10, 1.90)$)  {\textcolor{white}{$\theta = \widetilde{\theta}$}};
           \node[inner sep=0pt] (t)    at ($(pic.east) +(-0.70,-0.80)$)  {$t$};
           \node[inner sep=0pt] (tp)   at ($(pic.east) +(-2.60,-1.20)$)  {$t_\mathrm{pulse}$};
           \node[inner sep=0pt] (v)    at ($(pic.east) +(-3.95, 1.20)$)  {$v$};
           \node[inner sep=0pt] (van2) at ($(pic.east) +(-0.99, 0.20)$)  {$\widetilde{v}_\mathrm{an}(t)$};
           \node[inner sep=0pt] (vmax) at ($(pic.east) +(-3.15, 0.60)$)  {$\widetilde{v}_\mathrm{max}$};
         \end{tikzpicture} 
         
         \vspace{-3mm}
         \caption{BVP}
         \label{fig:Ex4BVP}
         
     \end{subfigure} 
     
     \caption{Geometry and boundary value problem of exemplary PECM setup.} 
     \label{fig:Ex4}
     
\end{figure}

Finally, we apply the model to predict the evolution of the surface roughness in a electrochemical machining application with electrical pulses (PECM). Fig.~\ref{fig:Ex4Geom} shows the exemplary setup of the PECM-process. The working gap width measures%
\footnote{Experimental investigations neglect distances < 1~\si{\micro\meter}. Here, we utilize multiple decimal places to generate a smooth surface profile and mesh.}
$s = 51.875~\si{\micro\meter}$ and the width of the analyzed surface $w = 20~\si{\micro\meter}$. We consider an idealized roughness profile with an initial peak value of $p = 6.25~\si{\micro\meter}$. The remaining parameters read $h = 21.875~\si{\micro\meter}$, $r = 0.625~\si{\micro\meter}$, $x_1 = 2.5~\si{\micro\meter}$, $y_1 = 5~\si{\micro\meter}$ and thickness $g = 1~\si{\micro\meter}$. Due to the short flow length, we neglect a temperature gradient and set the temperature to $\widetilde{\theta} = 298.15~\si{\kelvin}$. The time increment is $\dt = 0.01~\si{\milli\second}$. We neglect the polarization voltage, define the electric potential at the cathode to $\widetilde{v}_\mathrm{ca} = 0~\si{\volt}$ and apply a sawtooth cyclic loading pattern at the anode with $\widetilde{v}_\mathrm{max} = 20~\si{\volt}$ and $t_\mathrm{pulse} = 4~\si{\milli\second}$ for the electric potential (Fig.~\ref{fig:Ex4BVP}). This example assumes stationary electrode's positions and, thus, precise boundary conditions for the electric potential.

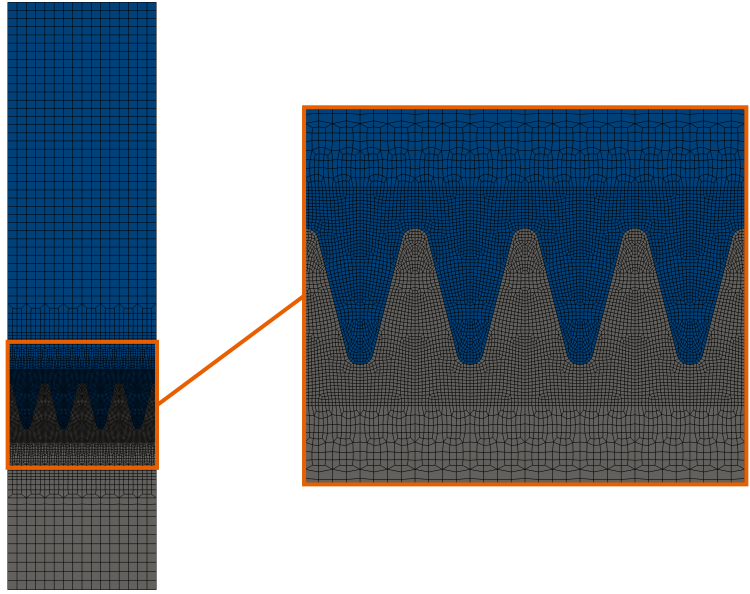
\begin{figure}[htbp] 
         
  \centering 
         
  \begin{tikzpicture} 
    \node[inner sep=0pt] (mesh) at (0,0) {\includegraphics[height=80mm]
    {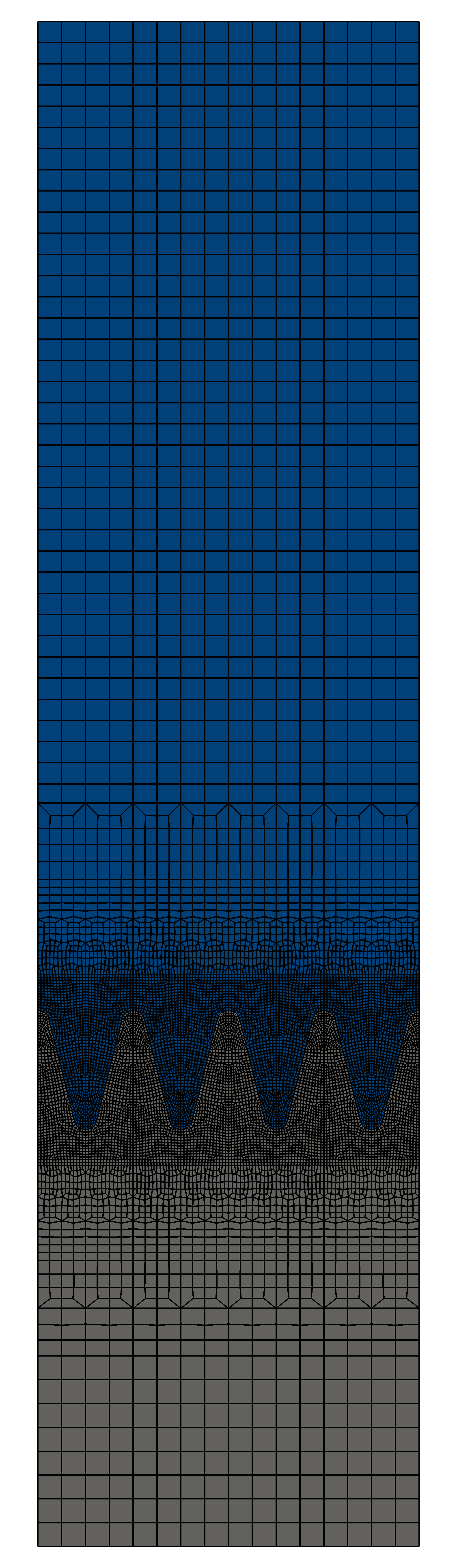}};
    \node[inner sep=0pt] (zoom) at ($(mesh.west) +( 7.00, 0.00)$) {\includegraphics[height=50mm]
    {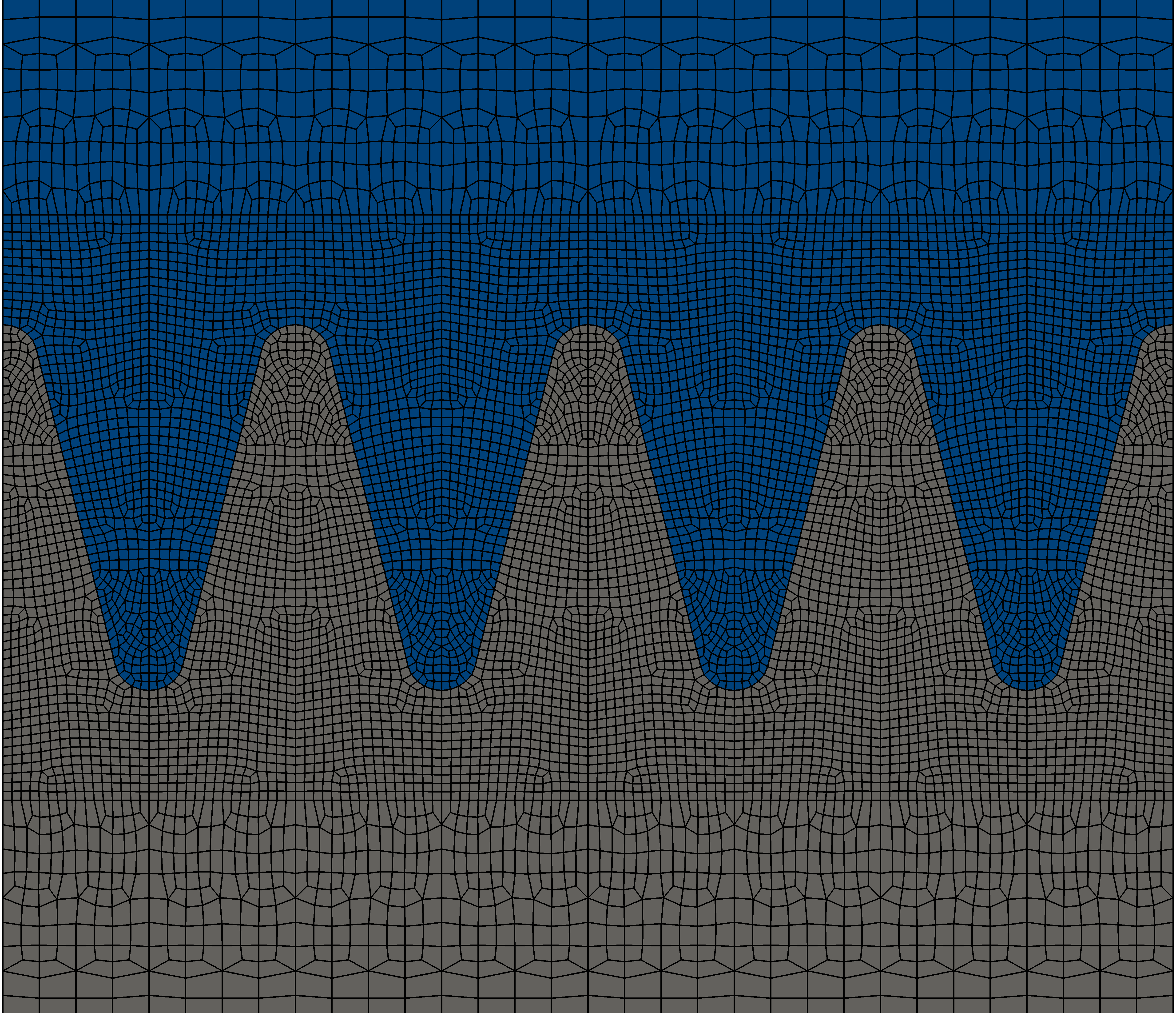}};
    \draw[line width = 1.5pt, draw = sfb3] ($(mesh.west) +( 0.19,-0.61)$)  rectangle ($(mesh.east) +(-0.19,-2.28)$);
    \draw[line width = 1.5pt, draw = sfb3] ($(zoom.west) +( 0.00, 2.50)$)  rectangle ($(zoom.east) +( 0.00,-2.50)$);
    \draw[line width = 1.5pt, draw = sfb3] ($(mesh.east) +(-0.19,-1.45)$)  --        ($(zoom.west) +( 0.00, 0.00)$);
  \end{tikzpicture} 
         
  \vspace{-3mm}
  
  \caption{Mesh (11264 elements) for PECM with refinement at the anode's surface.} 
  \label{fig:Ex4_mesh}
     
\end{figure}

Furthermore, Fig.~\ref{fig:Ex4_mesh} shows the mesh which exhibits strong refinement at the workpiece's surface. The simulation yields $\Vco/\Vdis^\mathrm{FE} = 1.5~\si{\percent}$ which is deemed reasonable, since this is an example of principle.

Next, we introduce two measures for the surface roughness according to \cite{DINENISO25178-2-2012}: First, the maximum height $Rz$ that defines the distance from the maximum peak height to the minimum pit depth
\begin{equation}
  Rz = \| y_\mathrm{max} - y_\mathrm{min} \|.
\end{equation}
Second, the arithmetical mean height $Ra$ which is computed according to
\begin{equation}
  Ra = \frac{1}{L_x} \, \int_0^{L_x} \| y(x) - \bar{y} \| \, \mathrm{d}x,
  \hspace{8mm}
  \bar{y} = \frac{1}{L_x} \, \int_0^{L_x} y(x) \, \mathrm{d}x.
\end{equation}
The center of each element, where the activation function is active, serves to define the surface's roughness profile. With these discrete values, we compute the roughness values $Rz$ and $Ra$ at every time step.

\begin{figure}[htbp]
  \centering
  
  \includegraphics{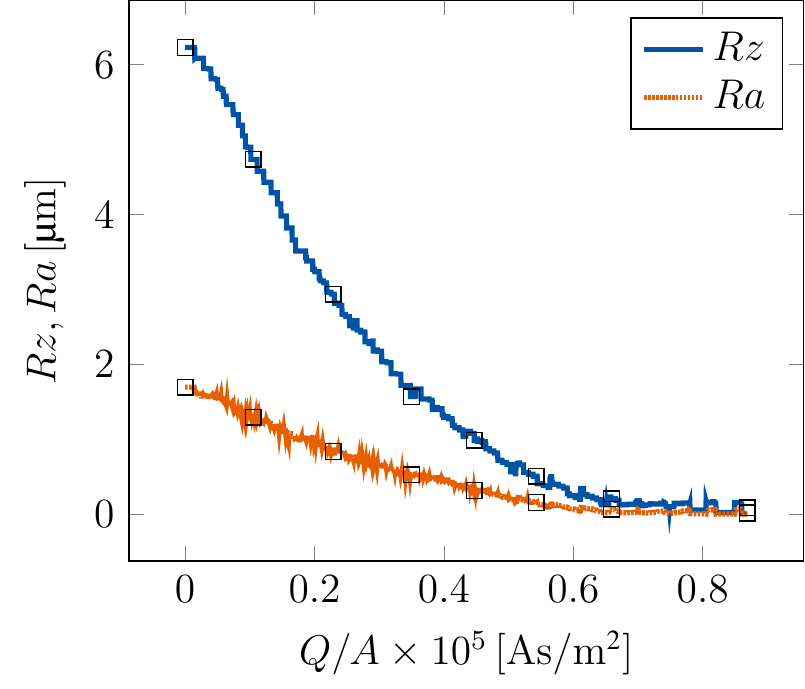}

  \caption{Process signature of surface roughness.} 
  \label{fig:Ex4_PSK1}
\end{figure}

\begin{figure}[htbp] 
  \centering 
  \hspace{22mm}
  \begin{minipage}{0.68\textwidth}
    \centering
    \begin{subfigure}{.47\textwidth} 
      \centering 
      \includegraphics[width=\textwidth]{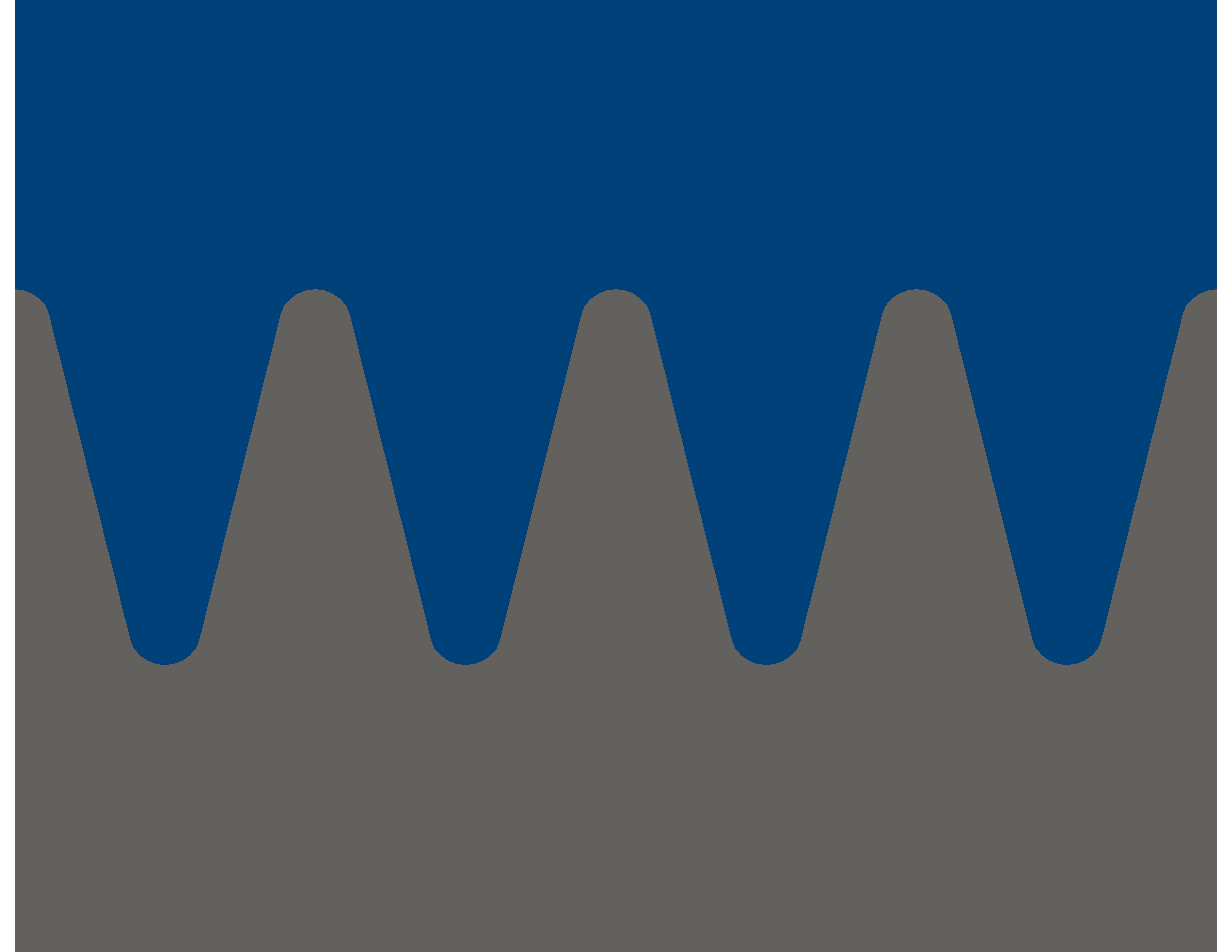}
      \caption{$0~\si{\milli\second}$}
      \label{fig:Ex4_t_0}
    \end{subfigure}
    \hspace{4mm}
    \begin{subfigure}{.47\textwidth} 
      \centering 
      \includegraphics[width=\textwidth]{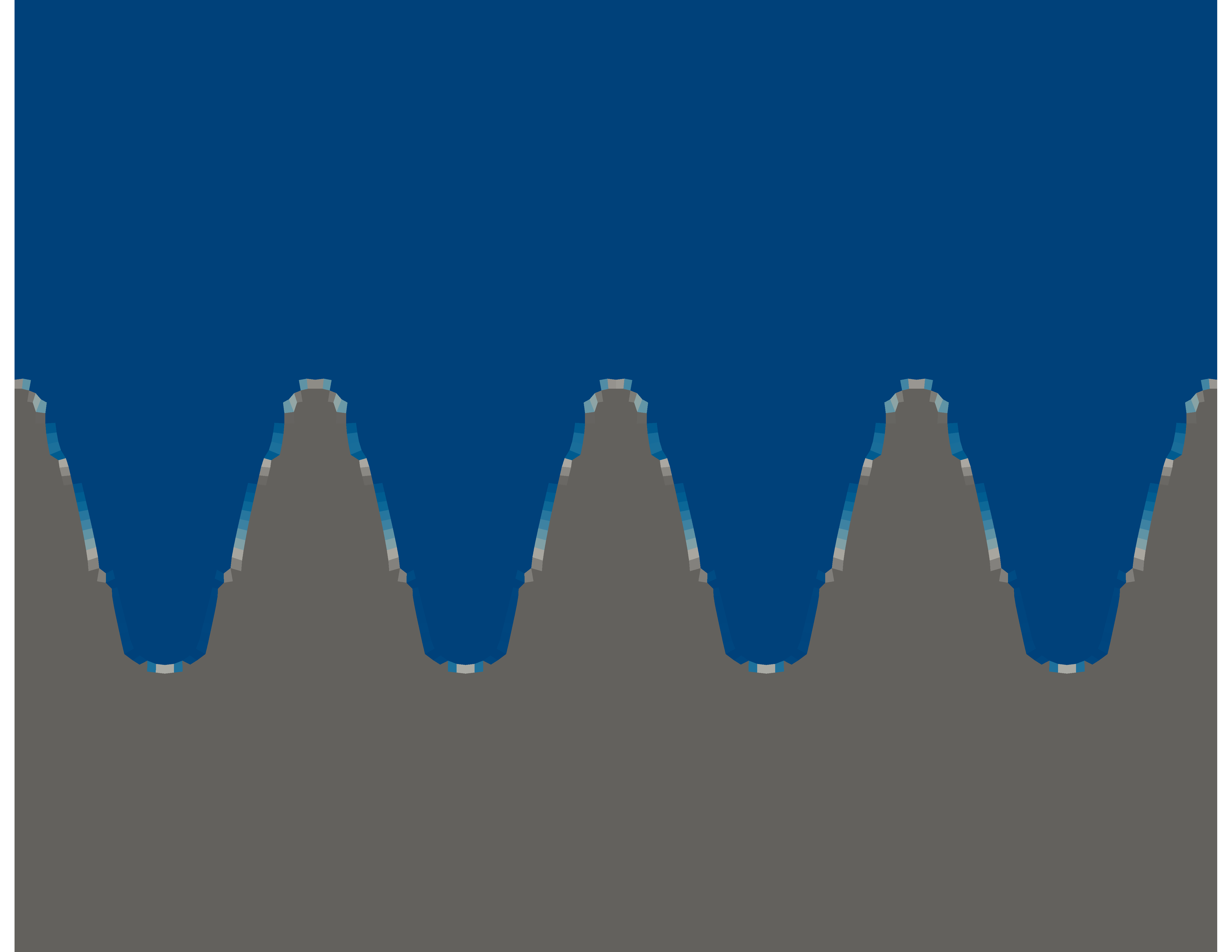}
      \caption{$5~\si{\milli\second}$}
      \label{fig:Ex4_t_50}
    \end{subfigure} \\
    \vspace{2mm}
    
    \begin{subfigure}{.47\textwidth} 
      \centering 
      \includegraphics[width=\textwidth]{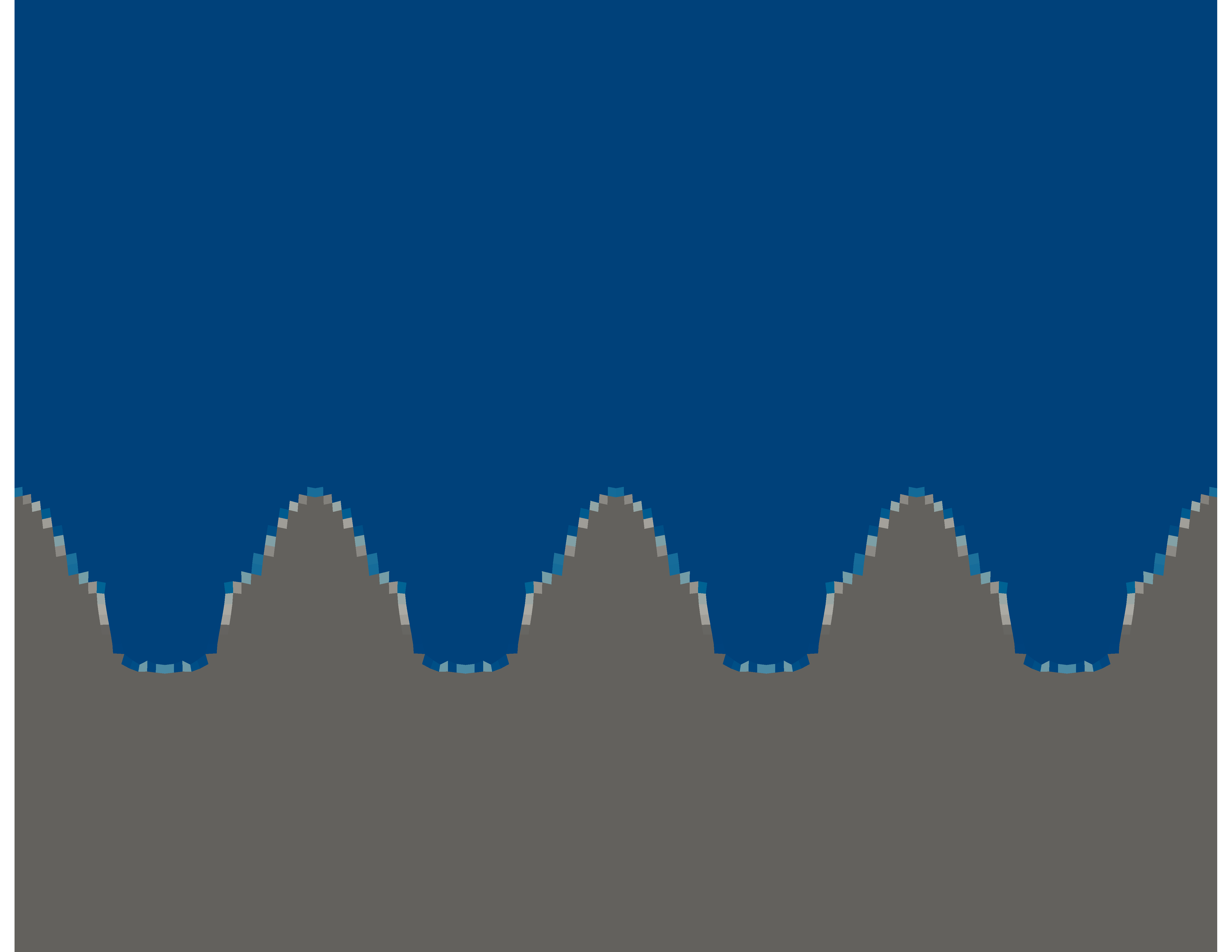}
      \caption{$10~\si{\milli\second}$}
      \label{fig:Ex4_t_100}
    \end{subfigure}
    \hspace{4mm}
    \begin{subfigure}{.47\textwidth} 
      \centering 
      \includegraphics[width=\textwidth]{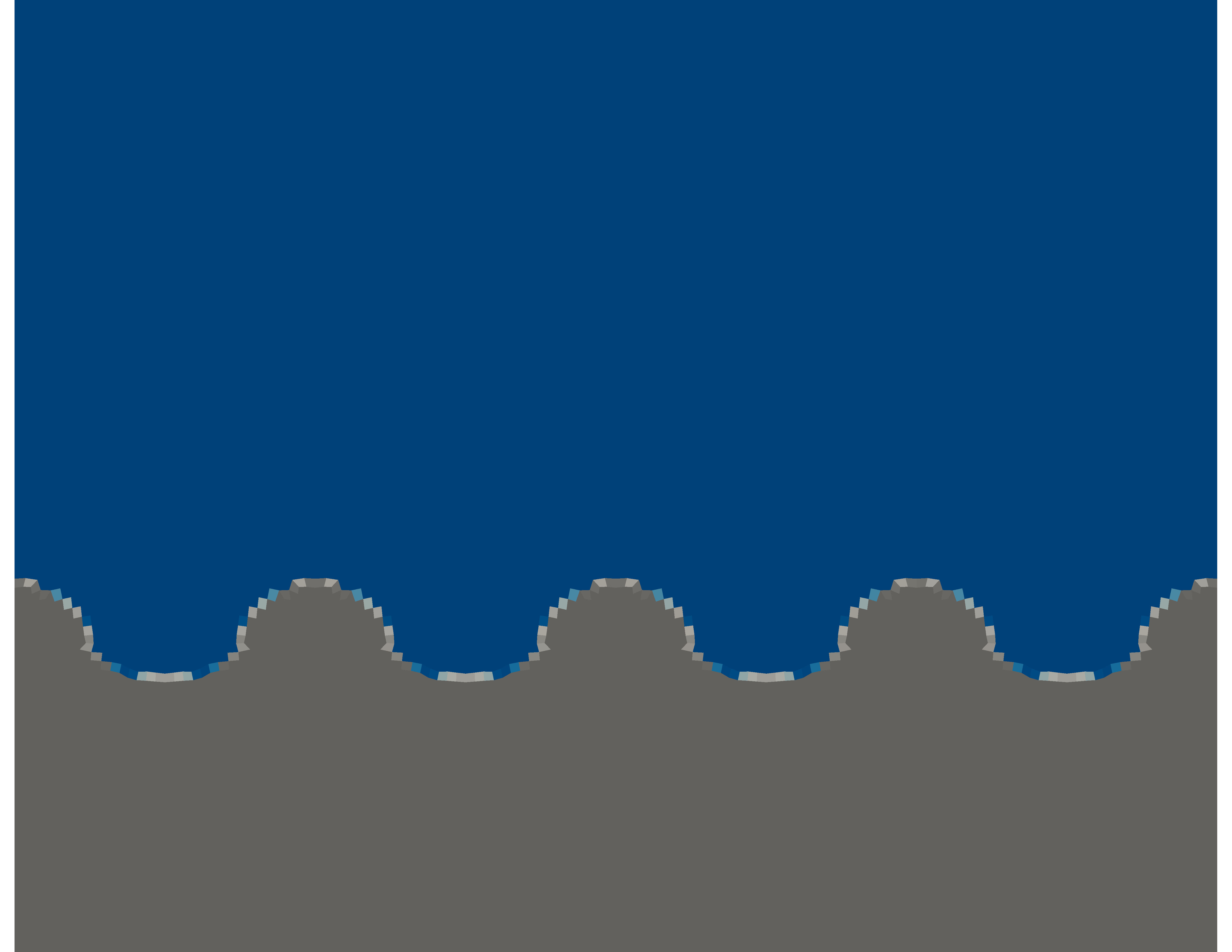}
      \caption{$15~\si{\milli\second}$}
      \label{fig:Ex4_t_150}
    \end{subfigure} \\
    \vspace{2mm}
    
    \begin{subfigure}{.47\textwidth} 
      \centering 
      \includegraphics[width=\textwidth]{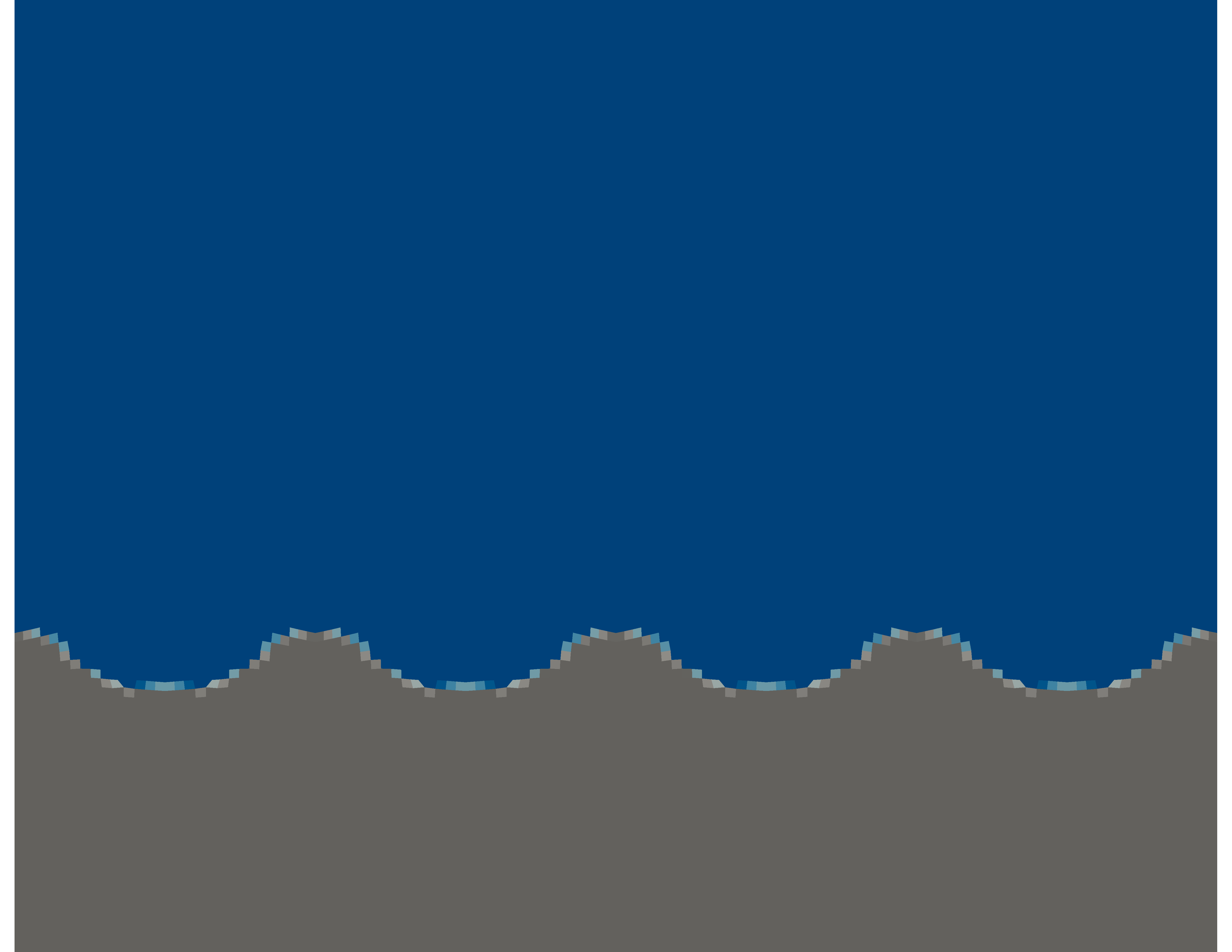}
      \caption{$20~\si{\milli\second}$}
      \label{fig:Ex4_t_200}
    \end{subfigure}
    \hspace{4mm}
    \begin{subfigure}{.47\textwidth} 
      \centering 
      \includegraphics[width=\textwidth]{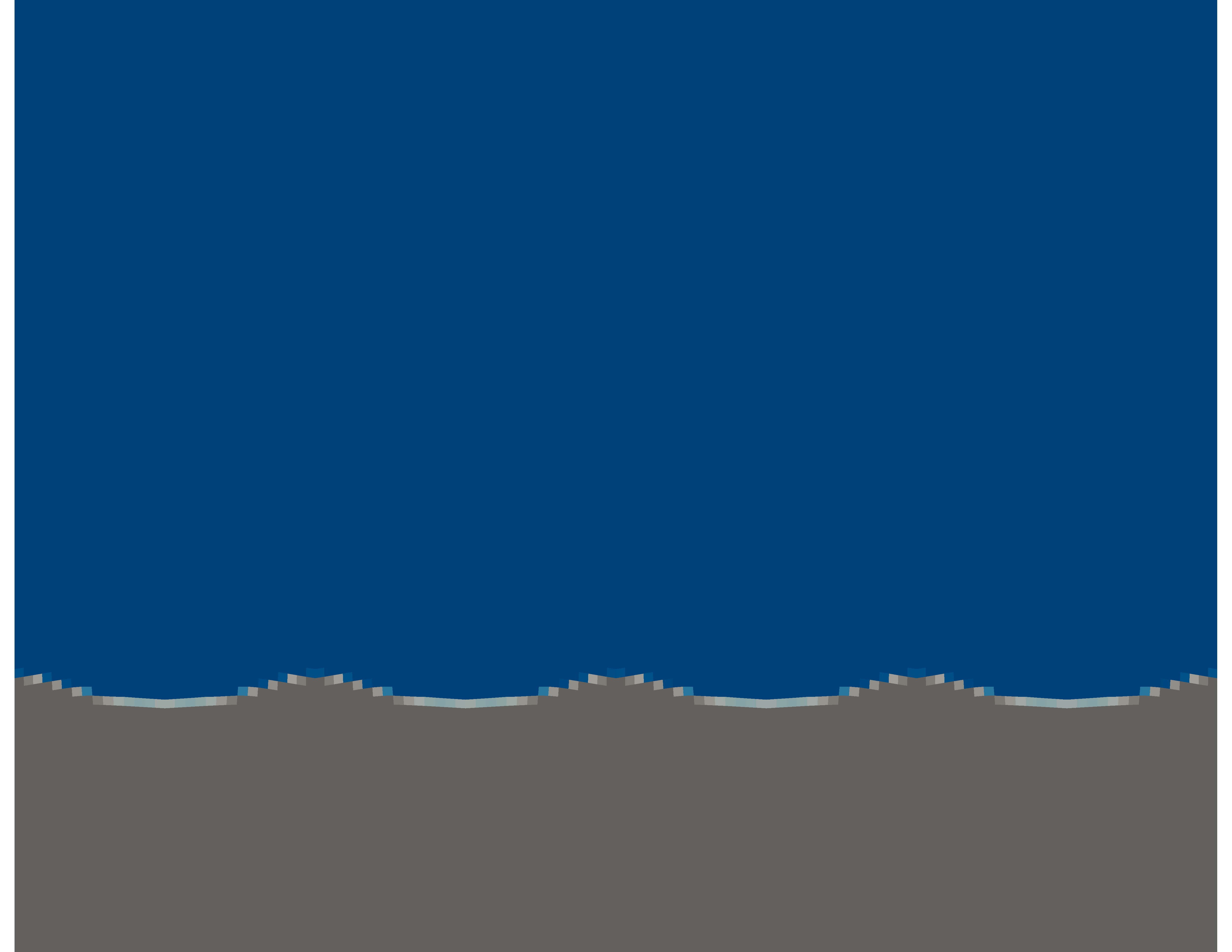}
      \caption{$25~\si{\milli\second}$}
      \label{fig:Ex4_t_250}
    \end{subfigure} \\
    \vspace{2mm}    
    
    \begin{subfigure}{.47\textwidth} 
      \centering 
      \includegraphics[width=\textwidth]{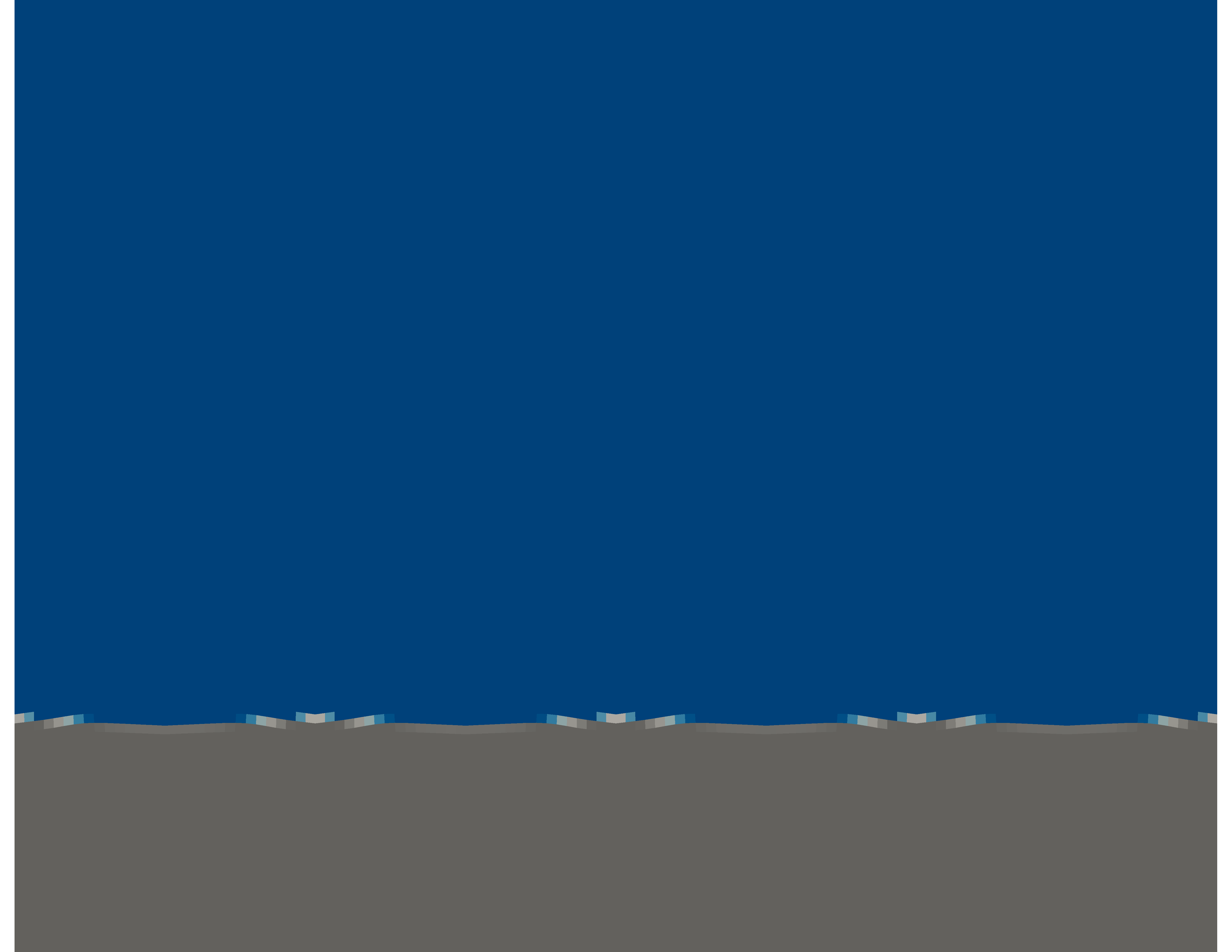}
      \caption{$30~\si{\milli\second}$}
      \label{fig:Ex4_t_300}
    \end{subfigure}
    \hspace{4mm}
    \begin{subfigure}{.47\textwidth} 
      \centering 
      \includegraphics[width=\textwidth]{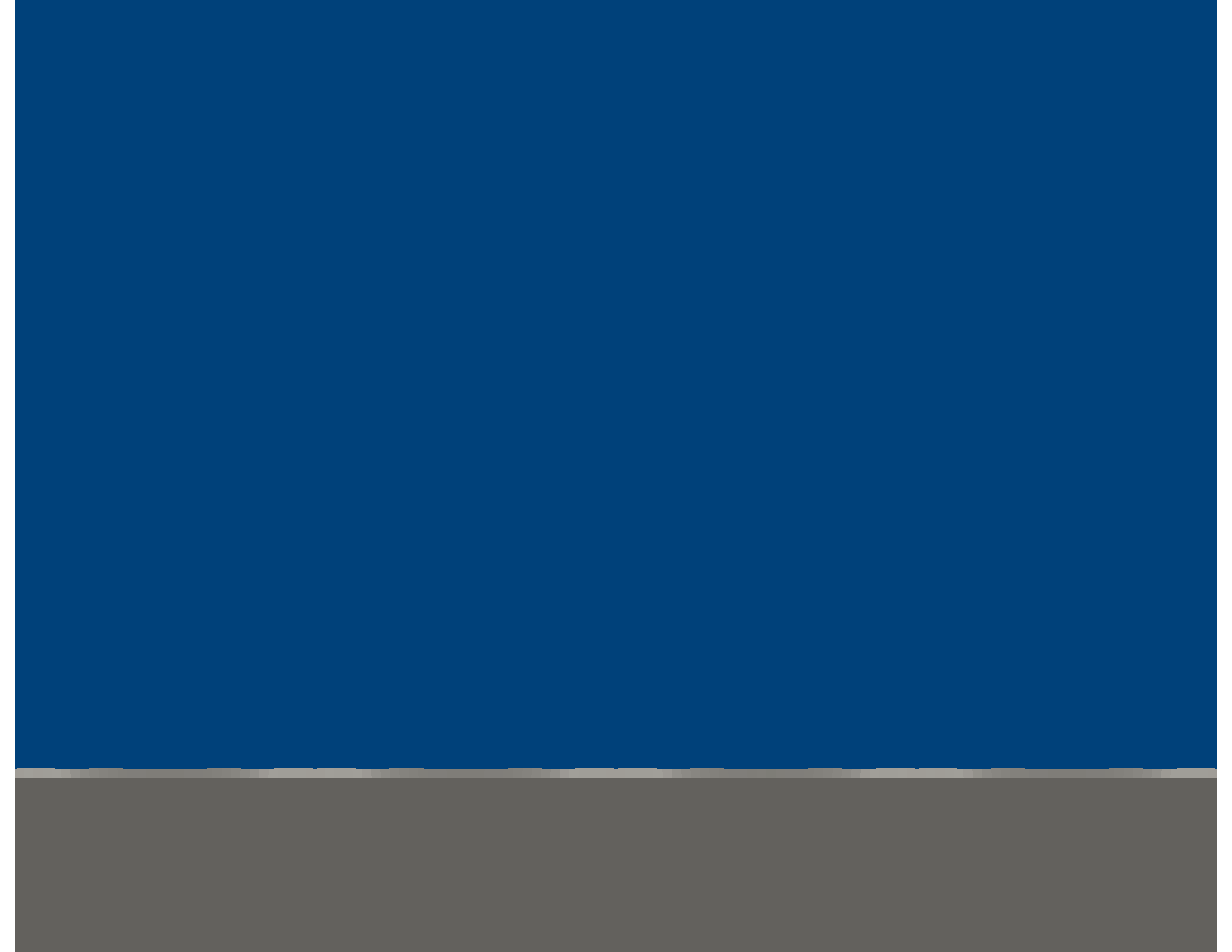}
      \caption{$40~\si{\milli\second}$}
      \label{fig:Ex4_t_400}
    \end{subfigure}
  \end{minipage}
  \hspace{13mm}
  \begin{minipage}{0.05\textwidth}
    \begin{subfigure}{.10\textwidth} 
      \centering 
      \begin{tikzpicture} 
        \node[inner sep=0pt] (pic) at (0,0) {\includegraphics[height=40mm, width=5mm]
        {03_Contour/01_Examples/00_Color_Maps/01_dsln_ver.pdf}};
        \node[inner sep=0pt] (0)   at ($(pic.south)+( 0.50, 0.05)$)  {$0$};
        \node[inner sep=0pt] (1)   at ($(pic.south)+( 0.50, 4.00)$)  {$1$};
        \node[inner sep=0pt] (d)   at ($(pic.south)+( 0.00, 4.60)$)  {$d \, [-]$};
      \end{tikzpicture} \\
      \vspace{12mm}
    \end{subfigure}
  \end{minipage}
 
  \caption{Dissolution level $d$ and surface profile for different machining times in PECM.}     
  \label{fig:Ex4_dsln}
\end{figure}

``A Process Signature is based on the correlation between the internal material loads in manufacturing processes (e.g., stress, strain, temperature) and the resulting material modifications`` (\cite{BrinksmeierReeseEtAl2018}). Here, the material load $Q/A$ is the accumulated electric charge, which passes in vertical direction, divided by the specimen's cross section. The material modification is the evolution of the surface roughness. Other authors, e.g.~\cite{Harst2019} employ the electric field strength $\E$ as material load in ECM. Fig.~\ref{fig:Ex4_PSK1} shows the corresponding process signature of both roughness measures $Rz$ and $Ra$ for an exemplary%
\interfootnotelinepenalty=10000 % prevents split of footnote
\footnote{This example investigates only one surface profile to prove the functionality of the procedure and the model. To obtain a generally valid process signature, further investigations with different surface profiles and experimental validation are required.}
\interfootnotelinepenalty=100
initial surface roughness. Starting with the initial values%
\interfootnotelinepenalty=10000 % prevents split of footnote
\footnote{The initial value of $Rz = 6.23~\si{\micro\meter}$ deviates from $p = 6.25~\si{\micro\meter}$, because we utilize the centers of the activated elements for the computation of the roughness. For a sufficiently fine mesh, we consider this procedure acceptable.}
\interfootnotelinepenalty=100
$Rz = 6.23~\si{\micro\meter}$ and $Ra = 1.70~\si{\micro\meter}$, both roughness measures decrease hyperbolically to zero. The black boxes indicate the snapshots given in Fig.~\ref{fig:Ex4_dsln}.

Fig.~\ref{fig:Ex4_dsln} shows the surface profile for different machining times. First, the material dissolves at the tip of the spikes (Figs.~\ref{fig:Ex4_t_50} - \ref{fig:Ex4_t_100}). Then, the bodies of the spikes dissolve (Figs.~\ref{fig:Ex4_t_150} - \ref{fig:Ex4_t_300}) until a level surface evolves (Fig.~\ref{fig:Ex4_t_400}).

This example proves the applicability of the model to compute process signatures that focus on the surface roughness in PECM and, additionally, to simulate a process with multiple electrical loads.

\section{Conclusion}
\label{sec:conclusion}
This paper presented an innovative method to efficiently model anodic dissolution in ECM, which circumvents the need for computationally expensive remeshing. At first, we define the dissolution level and the corresponding effective material parameters at integration point level. Next, we discuss the coupled problem of thermoelectricity and the numerical implementation in detail. Thereafter, numerical investigations validate the model's performance and accuracy by analytical and experimental reference solutions. In particular, the influence of the finite element mesh density and the time step size is investigated. The model shows -- even in the case of rather coarse meshes -- a highly satisfactory predictability. Moreover, the comparison with experiments confirms the realistic results obtained by means of numerical simulations. Finally, the model enables the computation of a process signature of the surface roughness with multiple electrical loads. The process signature's corresponding material modification is the evolution of the maximum and the arithmetical mean height. Additionally, the specific accumulated electric charge defines the corresponding material load.
Future work includes the modeling of multiphase materials with different polarization voltages, the exact description of the moving boundary value problem and the incorporation of fluid mechanical effects.

%----------------------------------------------------------------------------------------------------------------------------------%
\subsection*{Acknowledgements}

Funding granted by the subprojects M05 - ``Numerically efficient multi scale material models for processes under thermal and chemical impact'' and F03 - ``Processes with chemical impact'' of the transregional Collaborative Research Center 136 ``Process Signatures'' with the project number 223500200 is gratefully acknowledged.

\clearpage

\appendix

\section{Appendix}
\label{sec:appendix}

%----------------------------------------------------------------------------------------------------------------------------------%
\subsection{Linearization}
\label{sec:app_linearization}

The linearization of $\gv$ and $\gT$ (Eqs.~(\ref{eqn:gv}) and (\ref{eqn:gT})) about a known state $\left( \vbnpe, \Tbnpe \right)$ reads:
\begin{align}
  L_{\gv} &               =  \gv   \left( \vbnpe, \, \Tbnpe, \, \varv \right)            \notag \\
          & \hspace{20mm} +  \dvgv \left( \vbnpe, \, \Tbnpe, \, \varv, \, \dvnpe \right) \notag  \\
          & \hspace{40mm} +  \dTgv \left( \vbnpe, \, \Tbnpe, \, \varv, \, \dTnpe \right) \stackrel{!}{=} 0 \qquad \forall \, \varv  \label{eqn:Lgv} \\
  L_{\gT} &               =  \gT   \left( \vbnpe, \, \Tbnpe, \, \varT \right)            \notag \\
          & \hspace{20mm} +  \dvgT \left( \vbnpe, \, \Tbnpe, \, \varT, \, \dvnpe \right) \notag  \\
          & \hspace{40mm} +  \dTgT \left( \vbnpe, \, \Tbnpe, \, \varT, \, \dTnpe \right) \stackrel{!}{=} 0 \qquad \forall \, \varT  \label{eqn:LgT}
\end{align}
Furthermore, the G\^{a}teaux-derivatives are defined as follows:
\begin{align}
   \dvgv \left( \vbnpe, \, \Tbnpe, \, \varv, \, \dvnpe \right)
  &\coloneq \td{}{\alpha} \Big[ \gv \left( \vbnpe + \alpha \, \dvnpe, \, \Tbnpe, \, \varv \right) \Big]_{\alpha = 0} \\
   \dTgv \left( \vbnpe, \, \Tbnpe, \, \varv, \, \dTnpe \right)
  &\coloneq \td{}{\alpha} \Big[ \gv \left( \vbnpe, \, \Tbnpe + \alpha \, \dTnpe, \, \varv \right) \Big]_{\alpha = 0} \\
   \dvgT \left( \vbnpe, \, \Tbnpe, \, \varT, \, \dvnpe \right)
  &\coloneq \td{}{\alpha} \Big[ \gT \left( \vbnpe + \alpha \, \dvnpe, \, \Tbnpe, \, \varT \right) \Big]_{\alpha = 0} \\
   \dTgT \left( \vbnpe, \, \Tbnpe, \, \varT, \, \dTnpe \right)
  &\coloneq \td{}{\alpha} \Big[ \gT \left( \vbnpe, \, \Tbnpe + \alpha \, \dTnpe, \, \varT \right) \Big]_{\alpha = 0}  
\end{align}

In detail, the linearization of $\gv$ with respect to $\vnpe$ reads:
\begin{align}
  \dvgv &= \td{}{\alpha}  \bigg[ -   \intO \jL \left( \E    \left( \alpha \right) \right) \cdot \grad{\varv} \dV \notag \\
        & \hspace{25mm}          - 2 \intO \jV \left( \Edot \left( \alpha \right) \right) \cdot \grad{\varv} \dV 
                          \bigg]_{\alpha = 0} \notag \\
         &=\hspace{5.5mm}  \bigg[ -   \intO \left( \djLdE    \cdot \pd{\E}{\alpha}    \right) \cdot \grad{\varv} \dV \notag \\
         & \hspace{25mm}          - 2 \intO \left( \djVdEdot \cdot \pd{\Edot}{\alpha} \right) \cdot \grad{\varv} \dV 
                          \bigg]_{\alpha = 0} \notag \\
        &=\hspace{10mm}          -   \intO \left( \djLdE    \cdot               \left( - \grad{\dvnpe} \right) \right) \cdot \grad{\varv} \dV \notag \\
        & \hspace{25mm}          - 2 \intO \left( \djVdEdot \cdot \frac{1}{\dt} \left( - \grad{\dvnpe} \right) \right) \cdot \grad{\varv} \dV
        \label{eqn:dvgv}
\end{align}

In detail, the linearization of $\gv$ with respect to $\Tnpe$ reads:
\begin{align}
  \dTgv &= \td{}{\alpha}  \bigg[ -   \intO \jS \left( \grad{ \Tnpe \left( \alpha \right)}        \right) \cdot \grad{\varv} \dV 
                          \bigg]_{\alpha = 0} \notag \\
        &=\hspace{5.8mm}  \bigg[ -   \intO     \left( \djSdgradT \cdot \pd{\grad{\Tnpe}}{\alpha} \right) \cdot \grad{\varv} \dV
                          \bigg]_{\alpha = 0} \notag \\
        &=\hspace{8.9mm}         -   \intO     \left( \djSdgradT  \cdot \grad{\dTnpe}            \right) \cdot \grad{\varv} \dV
        \label{eqn:dTgv}
\end{align}

In detail, the linearization of $\gT$ with respect to $\vnpe$ reads:
\begin{align}
  \dvgT &= \td{}{\alpha}  \bigg[ - \intO \pibarn  \jL \left( \E    \left( \alpha \right) \right) \cdot \grad{\varT} \dV \notag \\
        & \hspace{25mm}          - \intO \pibarn  \jV \left( \Edot \left( \alpha \right) \right) \cdot \grad{\varT} \dV \notag \\
        & \hspace{40mm}          - \intO \jL \left( \E    \left( \alpha \right) \right) \cdot \E \left( \alpha \right) \varT \, \dV \notag \\
        & \hspace{55mm}          - \intO \jV \left( \Edot \left( \alpha \right) \right) \cdot \E \left( \alpha \right) \varT \, \dV \notag \\
        & \hspace{70mm}          - \intO \jS                                            \cdot \E \left( \alpha \right) \varT \, \dV 
                          \bigg]_{\alpha = 0} \notag \\
                                              \notag \\                
        &=\hspace{5.5mm}  \bigg[ - \intO \left( \pibarn  \djLdE    \cdot \pd{\E}{\alpha}    \right) \cdot \grad{\varT} \dV \notag \\
        & \hspace{25mm}          - \intO \left( \pibarn  \djVdEdot \cdot \pd{\Edot}{\alpha} \right) \cdot \grad{\varT} \dV \notag \\
        & \hspace{40mm}          - \intO \left( \left(   \djLdE    \cdot \pd{\E}{\alpha}    \right) \cdot \E \left( \alpha \right) + \jL \left( \E    \left( \alpha \right) \right) \cdot \pd{\E}{\alpha} \right) \varT \, \dV \notag \\
        & \hspace{55mm}          - \intO \left( \left(   \djVdEdot \cdot \pd{\Edot}{\alpha} \right) \cdot \E \left( \alpha \right) + \jV \left( \Edot \left( \alpha \right) \right) \cdot \pd{\E}{\alpha} \right) \varT \, \dV \notag \\
        & \hspace{70mm}          - \intO \left(                                                                     \jS                                            \cdot \pd{\E}{\alpha} \right) \varT \, \dV
                          \bigg]_{\alpha = 0} \notag \\
                                              \notag \\                                          
        &=\hspace{10mm}          - \intO \left( \pibarn  \djLdE    \cdot               \left( - \grad{\dvnpe} \right) \right) \cdot \grad{\varT} \dV \notag \\
        & \hspace{25mm}          - \intO \left( \pibarn  \djVdEdot \cdot \frac{1}{\dt} \left( - \grad{\dvnpe} \right) \right) \cdot \grad{\varT} \dV \notag \\
        & \hspace{40mm}          - \intO 2 \, \jL \cdot \left( - \grad{\dvnpe} \right) \varT \, \dV \notag \\
        & \hspace{55mm}          - \intO \left( \djVdEdot \cdot \frac{1}{\dt} \, \E + \jV \right) \cdot \left( - \grad{\dvnpe} \right) \varT \, \dV \notag \\
        & \hspace{70mm}          - \intO   \, \jS \cdot \left( - \grad{\dvnpe} \right) \varT \, \dV
        \label{eqn:dvgT}
\end{align}

In detail, the linearization of $\gT$ with respect to $\Tnpe$ reads:
\begin{align}
  \dTgT &= \td{}{\alpha}  \bigg[ - \intO \pibarn  \jS \left( \grad{\Tnpe \left( \alpha \right) } \right) \cdot \grad{\varT} \dV \notag \\
        & \hspace{25mm}          - \intO          \qF \left( \grad{\Tnpe \left( \alpha \right) } \right) \cdot \grad{\varT} \dV \notag \\
        & \hspace{40mm}          + \intO  \rhoVbar \cTbar \, \thetadot_{n+1} \left( \alpha \right)              \, \varT \, \dV \notag \\
        & \hspace{55mm}          - \intO  \jS \left( \grad{\Tnpe \left( \alpha \right ) } \right) \cdot \E      \, \varT \, \dV        
                          \bigg]_{\alpha = 0} \notag \\
                                              \notag \\                                          
        &=\hspace{5.5mm}  \bigg[ - \intO \left( \pibarn  \djSdgradT \cdot \pd{\grad{\Tnpe}}{\alpha} \right) \cdot       \grad{\varT} \dV \notag \\
        & \hspace{25mm}          - \intO \left(          \dqFdgradT \cdot \pd{\grad{\Tnpe}}{\alpha} \right) \cdot       \grad{\varT} \dV \notag \\
        & \hspace{40mm}          + \intO        \rhoVbar \cTbar \, \frac{1}{\dt} \, \dTnpe                           \, \varT \,     \dV \notag \\
        & \hspace{55mm}          - \intO        \left( \djSdgradT \cdot \pd{\grad{\Tnpe}}{\alpha}   \right) \cdot \E \, \varT \,     \dV \notag
                          \bigg]_{\alpha = 0} \notag \\
                                              \notag \\                                          
        &=\hspace{10mm}          - \intO \left( \pibarn  \djSdgradT \cdot \grad{\dTnpe} \right) \cdot \grad{\varT} \dV \notag \\
        & \hspace{25mm}          - \intO \left(          \dqFdgradT \cdot \grad{\dTnpe} \right) \cdot \grad{\varT} \dV \notag \\
        & \hspace{40mm}          + \intO        \rhoVbar \cTbar \, \frac{1}{\dt} \, \dTnpe         \, \varT \,     \dV \notag \\
        & \hspace{55mm}          - \intO \left( \djSdgradT \cdot \grad{\dTnpe} \right) \cdot \E    \, \varT \,     \dV
        \label{eqn:dTgT}
\end{align}

In addition, the corresponding tangents in Eqs.~(\ref{eqn:dvgv})~-~(\ref{eqn:dTgT}) read:
\begin{align}
  \djLdE     &= \kEbarn \I                \\
  \djVdEdot  &= \epsnull \epsrbar \, \I   \\
  \djSdgradT &=-\kEbarn \, \alphabarn \I  \\
  \dqFdgradT &=-\kTbarn \I
\end{align}

%----------------------------------------------------------------------------------------------------------------------------------%
\subsection{Element vectors and matrices}
\label{sec:app_elemvecmat}

\begin{align}
  \kvv &= \hphantom{-}     \intOe \BvT \, \left( \djLdE \right)^e \, \Bv \, \dVe                                                                      \\
  \kvT &= -                \intOe \BvT \, \left( \djSdgradT \right)^e \, \BT \, \dVe                                                                  \\
  \kTv &= \hphantom{-}     \intOe \NTT \, \left( 2 \, \jL^e + \jS^e \right)^{\mathrm{T}} \, \Bv
                                + \BTT \, \pibar^e \left( \djLdE \right)^e \, \Bv \, \dVe                                                             \\
  \kTT &= -                \intOe \NTT \, \E^{e \mathrm{T}} \, \left( \djSdgradT \right)^e \, \BT \, \dVe                                      \notag \\
       &  \hspace{4.5mm} - \intOe \BTT \, \left( \pibar^e \left( \djSdgradT \right)^e + \left( \dqFdgradT \right)^e \, \right) \, \BT \, \dVe         \\
  \cvv &= \hspace{1.5mm} 2 \intOe \BvT \, \frac{1}{\dt} \left( \djVdEdot \right)^e \, \Bv \, \dVe                                                     \\
  \cTv &= \hphantom{-}     \intOe \NTT \, \left( \frac{1}{\dt} \left( \djVdEdot \right)^e \E^e + \, \jV^e \right)^{\mathrm{T}} \, \Bv                                                               
                                + \BTT \, \frac{1}{\dt} \, \pibar^e \left( \djVdEdot \right)^e \, \Bv \, \dVe                                         \\
  \cTT &= \hphantom{-}     \intOe \NTT \, \frac{1}{\dt} \, \rhoVbar^{\,e} \cTbar^{\,e} \, \NT \, \dVe                                                 \\
  \rv  &= -                \intOe \BvT \, \left( \j^e + \jV^e \right) \dVe                                                                            \\
  \rT  &= \hphantom{-}     \intOe \NTT \left(  \rhoVbar^{\,e} \cTbar^{\,e} \thetadot^e - \j^{e\mathrm{T}} \, \E^e - q^\ast \right) \dVe
                         - \intOe \BvT \, \q^e \, \dVe
\end{align}

\bibliographystyle{agsm}
\bibliography{literature}

\end{document}